\documentclass{aa}  

\usepackage{graphicx}
\usepackage[varg]{txfonts}
\usepackage{natbib}
\usepackage{soul}
\usepackage{caption}
\usepackage{subcaption}
\usepackage{multirow}
\usepackage{makecell}
\usepackage{array}
\usepackage{CJKutf8}
\usepackage[dvipsnames]{xcolor}

\usepackage{enumerate}
\usepackage[inline]{enumitem}

\usepackage{amssymb}
\newcommand{
\yes}{\ensuremath{\checkmark}}
\newcommand{\no}{\ensuremath{\times}}

\newcommand{\xb}{z} 
\newcommand{\xc}{w} 
\newcommand{\Gini}{Gini} 

\newcommand\sk{\mbox{\textit{skew}}}
\newcommand\kurt{\mbox{\textit{kurt}}}

\newcommand\Dh{\mbox{$D_{\rm H}$}}

\newcommand{\HI}{\ion{H}{I}}

\providecommand{\sorthelp}[1]{}

\begin{document}

   \title{Quantitative morphology of Galactic cirrus in deep optical imaging}

   \subtitle{Statistical structural analysis in a multi-wavelength perspective}

   \author{Qing~Liu~(\begin{CJK}{UTF8}{gbsn}刘青\end{CJK})\inst{1},
          Peter~G.~Martin\inst{2},
          Roberto~G.~Abraham\inst{3,4,7},
          Pieter van~Dokkum\inst{5,7},
          Henk~Hoekstra\inst{1},
          Juan~Mir\'{o}-Carretero\inst{6,1},
          William~P.~Bowman\inst{5,7},
          Steven~R.~Janssens\inst{7},
          Seery~Chen\inst{3,4,7},
          Deborah~Lokhorst\inst{8},
          Imad~Pasha\inst{5,7},
          Zili~Shen\inst{5}
          }

   \institute{Leiden Observatory, Leiden University, P.O. Box 9513, 2300 RA Leiden, The Netherlands\\
              \email{qliu@strw.leidenuniv.nl}
              \and
              Canadian Institute for Theoretical Astrophysics, University of Toronto, 60 St. George St., Toronto, ON M5S 3H8, Canada
              \and
              David A. Dunlap Department of Astronomy \& Astrophysics, University of Toronto, 50 St. George St., Toronto, ON M5S 3H4, Canada
              \and
              Dunlap Institute for Astronomy and Astrophysics, University of Toronto, Toronto, ON M5S 3H4, Canada
              \and
              Department of Astronomy, Yale University, New Haven, CT 06520, USA
              \and
              Departamento de Física de la Tierra y Astrofísica, Universidad Complutense de Madrid, E-28040 Madrid, Spain   
              \and
              Dragonfly Focused Research Organization, 150 Washington Avenue, Santa Fe, NM 87501, USA
              \and
              NRC Herzberg Astronomy \& Astrophysics Research Centre, 5071 West Saanich Road, Victoria, BC V9E 2E7, Canada
              }
              
\titlerunning{Quantitative morphology of galactic cirrus in deep optical imaging}
\authorrunning{Liu et al.}

\date{Received 9 July 2025; accepted 3 Nov 2025}

\abstract {Imaging of optical Galactic cirrus, the spatially resolved form of diffuse Galactic light, provides important insights into the properties of the diffuse interstellar medium (ISM) in the Milky Way. While previous investigations have focused mainly on the intensity characteristics of optical cirrus, their morphological properties remain largely unexplored. In this study, we employ several complementary statistical approaches -- local intensity statistics, angular power spectrum / $\Delta$-variance analysis, and wavelet scattering transform analysis -- to characterize the morphology of cirrus in deep optical imaging data. We place our investigation of optical cirrus into a multi-wavelength context by comparing the morphology of cirrus seen with the Dragonfly Telephoto Array to that seen with space-based facilities working at longer wavelengths (Herschel $250\,\mu m$, WISE $12\,\mu m$, and Planck radiance), as well as with structures seen in the DHIGLS HI column density map. Our statistical methods quantify the similarities and the differences of cirrus morphology in all these datasets. The morphology of cirrus at visible wavelengths resembles that of far-infrared cirrus more closely than that of mid-infrared cirrus; on small scales, anisotropies in the cosmic infrared background and systematics may lead to differences. Across all dust tracers, cirrus morphology can be well described by a power spectrum with a common power-law index $\gamma\sim-2.9$. We demonstrate quantitatively that optical cirrus exhibits filamentary, coherent structures across a broad range of angular scales. Our results offer promising avenues for linking the analysis of coherent structures in optical cirrus to the underlying physical processes in the ISM that shape them. Furthermore, we demonstrate that these morphological signatures can be leveraged to distinguish and disentangle cirrus from extragalactic light. 
}

\bibpunct{(}{)}{;}{a}{}{,} 


\maketitle
%

\section{Introduction}

The interstellar medium (ISM) in galaxies comprises a variety of components with distinct densities, kinematics, chemical compositions, and thermal phases -- from cold/warm neutral medium (CNM/WNM) to hot ionized gas, and from dense molecular clouds near the Galactic plane to diffuse structures at high latitudes. Interwoven in this multiphase ISM is a pervasive population of interstellar dust grains, which plays a crucial role in regulating star formation and chemical enrichment (e.g., \citealt{1998ApJ...501..643D}, \citealt{2007ApJ...657..810D}, \citealt{2016SAAS...43...85K}). Dust is involved in multiple radiative transfer processes; therefore, dust emission has been widely used as a diagnostic tool for probing the physical conditions within the ISM (e.g., \citealt{2001ApJ...551..807D}, \citealt{2003ARA&A..41..241D}, \citealt{2004ApJS..152..211Z}, \citealt{2011A&A...525A.103C}).

Diffuse radiation from dust in the Milky Way, in the optical historically referred to as (the reflected portion of) the diffuse Galactic light (DGL), is a prominent manifestation of the diffuse ISM. 
All-sky infrared (IR) mapping 
\citep{1984ApJ...278L..19L}
revealed characteristically wispy, filamentary structures in dust emission prominently at high Galactic latitudes, which became referred to as Galactic cirrus. Similar structures had already been seen in optical imaging as high latitude reflection nebula \citep{1976AJ.....81..954S}.
In this work we use the common term ``cirrus'' to denote this
spatially resolved form of the DGL, recognizing that the radiative mechanisms are distinct at different wavelengths.  

IR cirrus has been extensively investigated using observations from infrared missions such as the IR Astronomical Satellite (IRAS), the Diffuse Infrared Background Experiment (DIRBE) aboard the Cosmic Background Explorer (COBE), the Wide-field Infrared Survey Explorer (WISE), the Herschel Space Observatory (e.g., \citealt{1987A&A...184..269L}, \citealt{1988ApJ...330..964B}, \citealt{1998ApJ...508...74A}, \citealt{1999A&A...352..645Z}, \citealt{2007A&A...469..595M}, \citealt{2010A&A...518L.105M}, \citealt{2011MNRAS.412.1151B}, \citealt{2015ApJ...811...77S}, \citealt{2020MNRAS.492.5420S}). 
Far-infrared (FIR) observations probe thermal emission from characteristically larger dust grains, whereas mid-infrared (MIR) observations detect primarily non-equilibrium emission from small grains. 
These observations have provided characterizations of the spatial distribution and spectral energy distribution (SED) of the IR DGL, thereby informing our understanding of the variations in dust properties, including dust temperature, composition, and emissivity (\citealt{planck2011-7.12},\citealt{planck2013-XVII},\citealt{planck2013-p06b}).

The optical cirrus, or DGL, first revealed through early observations before the 90s (e.g., \citealt{1937ApJ....85..213E}, \citealt{1976ApJ...208...64L}, \citealt{1976AJ.....81..954S}, \citealt{1979A&A....78..253M}, \citealt{1989ApJ...346..773G}), and more recently through deep {charge-coupled device} (CCD) imaging surveys (e.g., \citealt{2008ApJ...679..497W}, \citealt{2013ApJ...767...80I}, \citealt{MivilleDeschene2016}, \citealt{2020A&A...644A..42R}, \citealt{2023ApJ...948....4Z}, \citealt{2024AJ....168...88Z}), originates predominantly from the scattering of the interstellar radiation field (ISRF) by large ($a>0.1~\mu m$) dust grains (e.g., \citealt{2001ApJ...548..296W}, \citealt{2003ApJ...598.1017D}, \citealt{2011A&A...525A.103C}).
Optical observations can thus offer unique insights into the physical properties of dust grains, scattering anisotropy, and the characteristics of the incident ISRF (\citealt{2004ASPC..309...77G}, \citealt{2012ApJ...744..129B}, \citealt{2013ApJ...767...80I}, \citealt{2023ApJ...948....4Z}, \citealt{2024AJ....168...88Z}). In addition, the higher angular resolution achievable 
makes observations of the optical cirrus particularly valuable for investigating the turbulent cascade processes in the ISM (\citealt{MivilleDeschene2016}).

Observations of optical cirrus have not yet been utilized extensively as a tool to study the diffuse ISM because, in the optically-thin conditions most easily modeled, the cirrus is very faint,
typically only a few percent as bright as the night sky.
Similar to many other low surface brightness phenomena, the photometry of optical cirrus is susceptible to various systematic effects, such as stray light from off-axis diffraction and the extended point-spread function (PSF) of sources in the image, improper sky subtraction, and flat-fielding errors. These systematics substantially suppress, if not completely remove, the cirrus signals.
Only recently have advancements in instrumental design and in data reduction tailored for low surface brightness science (e.g., \citealt{2014PASP..126...55A}, \citealt{2016ApJ...823..123T}, \citealt{2023ApJ...953....7L}, \citealt{2024MNRAS.528.4289W}, \citealt{2025A&A...697A...6C}) substantially overcome these challenges. These improvements have facilitated the imaging of optical Galactic cirrus with optimized observing and data reduction strategies (e.g., \citealt{MivilleDeschene2016}, \citealt{2020A&A...644A..42R}, \citealt{2023ApJ...948....4Z}, \citealt{2025ApJ...979..175L}), opening new opportunities to study the diffuse ISM.

Existing characterizations of optical Galactic cirrus have focused primarily on their photometric characteristics: surface brightness, colors, and correlations with their IR counterparts. 
``Pixel-by-pixel'' correlations, however, necessitate downgrading high-resolution data to the lower resolution of the pair and may be non-linear. Although such correlation encodes important insights on dust physical properties, the spatial coherence information about the hierarchical structures of cirrus is lost, because pixels of the same intensity are binned regardless of whether they are from structures with very different shapes and scales.

Statistical approaches to characterize the cirrus morphology remain rather limited but offer an intriguing path forward.
\cite{MivilleDeschene2016} found that the angular power spectrum of optical cirrus in deep imaging data from {the Canada France Hawaii Telescope} (CFHT) follows a power law with an index of $\gamma=-2.9$, consistent with Planck and WISE results in the same field. 
They found no break in the power spectrum, as an indication of energy dissipation, down to a physical scale of 0.01 pc.
\cite{2021MNRAS.508.5825M} studied the fractal properties of optical and IR cirrus in the {Sloan Digital Sky Survey} (SDSS) Stripe82 field and found a mean 2D fractal dimension of 1.69 and 1.38, respectively. They concluded that this difference cannot be attributed solely to differing angular resolutions of optical and IR data and might reflect intrinsic physical differences such as imposed by the scattering phase function. 

In this work, we provide characterizations of the morphology of optical Galactic cirrus using a suite of statistical approaches. We place our investigation of optical cirrus in a multi-wavelength context by comparing the morphology of optical cirrus with those of other dust tracers. 
Although cirrus maps from various tracers may appear visually similar, to what extent are they \textit{quantitatively} the same (or different)?
We address this by seeking ``distribution-to-distribution'' approaches that preserve and extract structural information across a range of scales.
Furthermore, observations in different wavelengths have varying beam sizes, sensitivity, reduction processes, and systematics. We correct these effects where feasible, and where corrections are not feasible, 
we discuss how these effects could result in different statistics at certain scales. 
The main objectives of this paper are therefore to:
\begin{itemize}
    \item Quantify the spatial coherence of optical Galactic cirrus across a range of angular scales.
    \smallskip
    \item Investigate the morphological similarities and differences among cirrus maps at different wavelengths, corresponding to different dust tracers.
    \smallskip
    \item Develop and evaluate statistical tools for exploratory and diagnostic data analysis in forthcoming deep imaging surveys.
\end{itemize}

The goals focus on developing a better understanding of cirrus for its own sake, but we have an additional objective. Galactic cirrus is often a source of foreground contamination on deep images, masking out other low surface brightness phenomena -- including ultra-diffuse galaxies (\citealt{2021ApJS..257...60Z}), intracluster light (ICL) (\citealt{2017ApJ...834...16M}, \citealt{2025A&A...697A..13K}), and tidal features (\citealt{2020MNRAS.498.2138B}, \citealt{2023A&A...671A.141M}) -- which can profoundly impact their detection and measurement. Existing techniques to distinguish are based on colors (\citealt{2020A&A...644A..42R}, \citealt{2023MNRAS.524.2797M}, \citealt{2023MNRAS.519.4735S}) or supplementary IR data (\citealt{2010MNRAS.409..102D}, \citealt{2016ApJ...825...20B}, \citealt{2017ApJ...834...16M})
More recently \citet{2025ApJ...979..175L} proposed a method combining morphological filtering with color constraints to distinguish faint diffuse galaxies from cirrus. However, this approach might not be optimal for extragalactic sources with angular scales comparable to cirrus (e.g., ICL) or with visually similar morphology (e.g., tidal tails), and many imaging surveys may not have high resolution IR coverage or multi-band photometry available at the same depth and/or resolution. Therefore, a single-band approach would be highly valuable to facilitate low surface brightness analyses. Another important objective of this paper is therefore to: 
\begin{itemize}
    \item Identify morphological clues of optical cirrus that can differentiate cirrus from other faint diffuse extragalactic emission.
\end{itemize}

The paper is organized as follows. 
Section~\ref{sec:data} describes the datasets employed, supplemented by Appendix~\ref{appendix:dragonfly_appendix} on the Dragonfly Telephoto Array.
Section~\ref{sec:method} outlines statistical methods that quantify the cirrus spatial coherence using local probability density
functions (PDFs), angular power spectra (supplemented by Appendix~\ref{appendix:supp_ps}) and its variants, and wavelet scattering transforms (WSTs, supplemented by Appendix~\ref{appendix:supp_wst}). 
Sections~\ref{sec:result_local} to \ref{sec:result_wst} present the principal results from application of these methods. 
Section~\ref{sec:discussion} explores
distinguishing between
extragalactic light (from tidal tails) and cirrus based on a single band and discusses the principles of the methodology, with supplementary material in Appendix~\ref{appendix:tidal}.
Finally, Section~\ref{sec:propsects} 
presents a summary and prospects for application in deep wide-field imaging surveys.

\section{Datasets}
\label{sec:data}
The subsections below describe the datasets used in this work. 
The main results are focused on optical imaging data obtained from the Dragonfly Telephoto Array (hereafter Dragonfly for short). To characterize the similarity and difference of dust morphology with different tracers, we used FIR data from Herschel, MIR data from WISE, and the radiance image from thermal dust modeling of Planck observations.
We also used {\HI} observations from DHIGLS as supplementary data. 
{The beam widths and pixel sizes of the datasets described below are summarized in Table \ref{tab:beampix}.}

\begin{table}[ht]
\centering
\caption{{Beam widths and pixel sizes of datasets used}}
\label{tab:beampix}
\begin{tabular}{lll}
\hline\hline
Dataset & Beam width & Pixel size\\
\hline
Dragonfly optical & 5\arcsec & 6\arcsec\\
Herschel 250 $\mu m$  & 18\arcsec & 6\arcsec\\
WISE 12 $\mu m$ reprocessed & 15\arcsec & 6\arcsec\\
Planck radiance & 5\arcmin & 1\arcmin\\
DHIGLS HI  & 55\arcsec & 18\arcsec\\
\hline
\end{tabular}
\end{table}

\begin{figure*}[!htbp]
\centering
  \resizebox{\hsize}{!}{\includegraphics{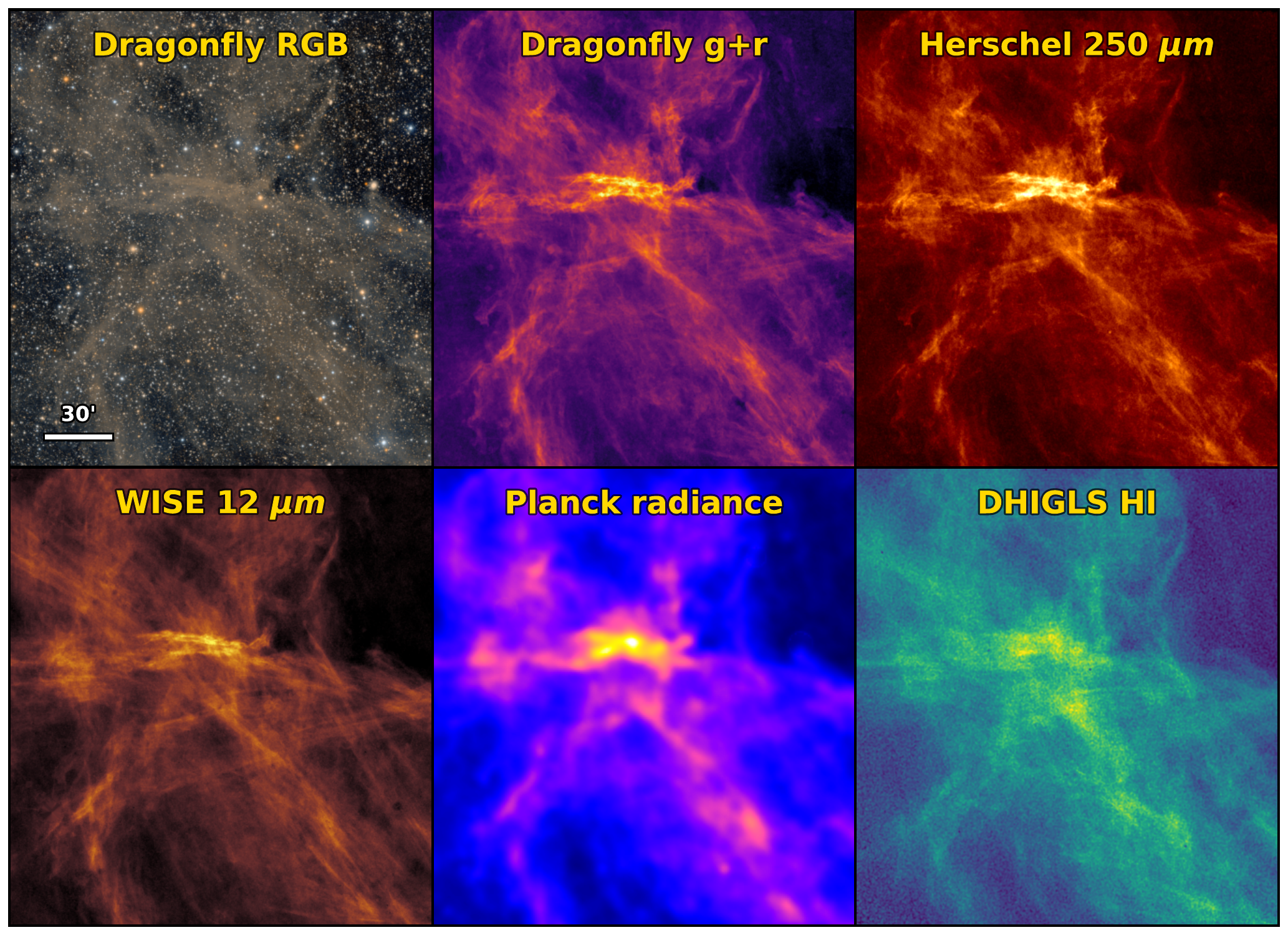}}
  \caption{Dragonfly RGB mosaic image of the Spider field at the top left and multi-wavelength images from different dust tracers used in this work. Top middle: Dragonfly combination of $g$ and $r$ after source removal, approximating diffuse radiation in the V band. Top right: Herschel 250~$\mu m$. Bottom left: WISE 12~$\mu m$. Bottom middle: Planck radiance. 
  Bottom right: DHIGLS {\HI} LVC column density. See text for details. The images are displayed on a linear scale with the same contrast between 0 and the 99.99\% quantile.}
\label{fig:spider}
\end{figure*}

\subsection{The Spider field}
The ``Spider'' field was chosen as the example for this work. This diffuse region of the ISM at intermediate Galactic latitude, centered at $(l,b)\sim(135^\circ, 40^\circ)$ and located at a distance of $\sim$320 pc and a height of $\sim$205 pc above the Galactic plane (\citealt{2020A&A...633A..51Z}, \citealt{2023ApJ...942...70M}), 
is part of the North Celestial Pole Loop (NCPL; \citealt{1991A&A...245..247M}, \citealt{2015ApJ...809..153M}, \citealt{2023ApJ...942...70M}). 
For the most part, the field is faint and optically thin (\citealt{2023ApJ...948....4Z}). It has been observed by several facilities at a range of wavelengths, making it an interesting target for ISM investigations and dust modeling.

\subsection{Dragonfly optical imaging}
At visible wavelengths, Galactic cirrus radiation mainly originates from the scattering of starlight by larger-size ($a>0.1~\mu m$) dust grains. Our primary dataset is deep optical imaging of Galactic cirrus obtained from Dragonfly. Dragonfly is a mosaic aperture telescope optimized for imaging diffuse extended emission (\citealt{2014PASP..126...55A}). Details about the telescope and data reduction are summarized in Appendix~\ref{appendix:dragonfly}. In particular, we adopt the background modeling recipes in \cite{2023ApJ...953....7L} to preserve the faint diffuse cirrus signal, which prevent it from being suppressed or smeared out by conventional sky subtraction methods. 

This work uses the same images of the Spider field presented in \cite{2023ApJ...953....7L}. 
The final coadd was a mosaic of six Dragonfly fields, with an average of 53 frames in $g$ and 64 frames in $r$ for each field passing quality checks. In the $g$/$r$ bands, the 3$\sigma$ surface brightness limit of the coadd is 28.5/28.3 mag$\,$arcsec$^{-2}$ on $10\arcsec \times 10 \arcsec$ scales, and 29.6/29.3 mag$\,$arcsec$^{-2}$ on $60\arcsec \times 60 \arcsec$ scales.\footnote{The surface brightness limits were measured on the cirrus-removed background; otherwise the sky RMS would be significantly overestimated due to confusion noise from cirrus. The calculation follows the method used in the appendix of \cite{2022ApJ...935..160K}.} 
The total field-of-view of the coadd is $4\fdg1 \times 3\fdg8$. An RGB composite image constructed from the $g$ and $r$-band coadds (R:$r$; G:($g$+$r$)/2; B:$g$) is displayed in the upper left panel of Figure~\ref{fig:spider}. 

We performed source removal following the procedures in \cite{2025ApJ...979..175L}, followed by a component separation process to yield the diffuse light from cirrus. This process, which aims to eliminate the contamination from faint extragalactic sources and residuals in the source removal, is based on cirrus morphology and SED constraints calibrated using the Planck thermal dust model \citep{planck2013-p06b}. 
The compact source removal and cirrus component separation are summarized in Appendix~\ref{appendix:diffuse}.
During this process, the images were converted to physical units in kJy$\,$sr$^{-1}$ with zero-points from Planck, assuming that the optical cirrus is the counterpart of Planck thermal emission. We then combined the decomposed $g$ and $r$ cirrus images into a 
brightness image approximating the V band\footnote{This diffuse $g+r$ image is constructed by $V$ = $0.435\,g + 0.565\,r - 0.016$ as in \cite{2025ApJ...979..175L}, which follows from the conversion from the SDSS $g$ and $r$ bands to the $V$ band (\citealt{2006A&A...460..339J}). Note that although $V$ is in the Vega system and the native Dragonfly images are in the AB system, the image was converted to physical units with zero-points from Planck (\citealt{2025ApJ...979..175L}) and the majority of this work is focused on the morphology rather than the absolute intensity.} and resampled to a pixel size of $6\arcsec$ to enhance the signal-to-noise ratio (S/N). {The image size is [2460x2272] pix$^2$.}
This optical image of the Spider field is displayed in the top middle panel of Fig.~\ref{fig:spider}.

\subsection{Herschel 250~$\mu m$}

FIR (sub-mm) observations probe thermal emission from interstellar dust. We used data from the Spectral and Photometric Imaging Receiver (SPIRE) instrument on the Herschel Space Observatory (\citealt{2010A&A...518L...3G}). SPIRE produces images in three bands: 250~$\mu m$, 350~$\mu m$, and 500~$\mu m$. The 250~$\mu m$ image was chosen because it has the highest S/N and spatial resolution. We retrieved the image from the Herschel Science Archive (HSA).\footnote{\url{https://archives.esac.esa.int/hsa/whsa/}} We used the Level 2.5 products created using HIPE12 pipeline version 14.0 (\citealt{2010ASPC..434..139O}). The SPIRE images have been zero-level corrected and calibrated in units of MJy$\,$sr$^{-1}$, with a beam FWHM of $18\arcsec$ on $6\arcsec$ pixels. {The image size is the same as the optical image.} 
The Herschel 250~$\mu m$ image is shown in the top right panel of Fig~\ref{fig:spider}. 

To reduce the impact of cosmic infrared background anisotropies (CIBA) on the morphology of dust emission, we first masked all bright point sources using the Herschel/SPIRE Point Source Catalog (HSPSC; \citealt{2017arXiv170600448S}).\footnote{The HSPSC is not complete, especially on top of highly structured backgrounds. We masked all sources above 30 mJy, which corresponds roughly to $50\, \%$ completeness in a field with low confusion noise (\citealt{2017arXiv170600448S}).} 
Next, we removed relatively faint sources. This was done by fitting a correlation between the intensities of Dragonfly $g+r$ data and Herschel 250~$\mu m$ data. The $g+r$ image was scaled to remove the diffuse background to create a residual image for the detection of faint compact sources. We then masked all sources above $S/N$ of 3, and replaced masked pixels with median filtering.
To mitigate the impact of fainter sources below the detection limit, we further performed a low-pass filtering 
through multiplying the Fourier transformed image by a Gaussian filter whose FWHM corresponds to twice that of the Herschel SPIRE 250~$\mu m$ beam. Effectively, this low-pass filtering smooths small-scale structures below this scale.

It should be noted that these conventional approaches are not optimal because the CIBA still has a residual contribution with power at low spatial frequencies (\citealt{viero2013}, \citealt{2017ApJ...839....7M}, \citealt{singh2022}).
More advanced decomposition techniques, such as by \cite{2024A&A...681A...1A},\footnote{The authors used wavelet phase harmonics (WPH), a similar approach to scattering transform described below that effectively extracts structural information using the wavelet-based moments to separate dust emission from CIBA, by assuming the true CIBA behind the cirrus dust emission has the same WPH statistics as a dust-free CIBA image.} 
ought to be preferred but are deferred to future work.

\subsection{WISE 12~$\mu m$}
The emission at MIR bands is dominated by the non-equilibrium emission from the smallest ($a<0.01~\mu m$) dust grains such as polycyclic aromatic hydrocarbons (PAHs; \citealt{2008ARA&A..46..289T}). We used the WISE 12~$\mu m$ (W3 band) map from the all-sky WISE 12~$\mu m$ atlas reprocessed by \cite{2014ApJ...781....5M},\footnote{To restore the emission at scales larger than 2$\degr$, the reprocessing used the Planck 857~GHz (350~$\mu m$) emission map, which is thermal rather than non-equilibrium emission.} which optimized the removal of compact sources and artifacts. The WISE intensities were converted from counts to MJy$\,$sr$^{-1}$ following the prescription of \cite{2012wise.rept....1C}, Sect.~4.4h.  

We further performed a source residual cleaning. First, we did a first visual inspection and masked the negative holes and bright sources in the map. We then followed the same approach as for the FIR data, i.e, we subtracted a diffuse component scaled based on the $g+r$ image and masked sources detected above S/N $>$ 3 in the residual image. The masked pixels were replaced with median filtering. 

The native FWHM of the WISE W3 band is $6\farcs5$ but the reprocessed all-sky WISE 12~$\mu m$ atlas  has been smoothed to $15\arcsec$. The cleaned WISE map was resampled to the finer pixel resolution of the Dragonfly image for comparison of structures on the same grid, which of course does not gain back the original WISE W3 resolution.
{The image size is again the same as the optical image.}
The WISE 12~$\mu m$ map of the Spider field is shown in the lower left panel of Fig~\ref{fig:spider}.

\subsection{Planck radiance}

The all-sky thermal dust model derived from Planck observations \citep{planck2013-p06b} was used as complementary data (see also \citealt{planck2016-XLVIII}). In particular, we used the product dust radiance map, defined as the integral of the model thermal emission:
$\mathcal{R} = \int I_\nu\,d\nu$.
The $\mathcal{R}$ map retrieved from the Planck Legacy Archive\footnote{The PR1-2013 thermal dust model from \url{https://pla.esac.esa.int/#maps}.} has a beam width of $5\arcmin$ in a HEALPix representation with pixel size approximately 1\farcm72 and for the Spider field was resampled to a pixel size of $1\arcmin$. {The Planck image size is [246x227] pix$^2$, amounting to the same angular extent as the optical image.} The map was visually inspected and two point-like sources were masked and refilled by smoothing. The radiance map is shown in the lower middle panel of Fig~\ref{fig:spider} in units of $\rm 10^{-7}\,Wm^{-2}sr^{-1}$. 

One major advantage of using the Planck thermal dust model is the reduction of the impact of the CIBA.
In \cite{2025ApJ...979..175L}, we also demonstrated the advantage of using $\mathcal{R}$ over other quantities like optical depth as the dust surrogate (see also discussion in \citealt{2023ApJ...953....7L}). In summary, $\mathcal{R}$ is a better representation of the total thermal emission observed by Planck and is less affected by optical depth effects in the illuminating radiation. 

\subsection{Different couplings to the interstellar radiation field}
\label{sec:isrf}

{The morphology in each of the products discussed above obviously depends on the amount of dust from which the radiation arises along the line of sight. But for each product there is also a different coupling to the ISRF. 
For high latitude fields like the Spider, the incident ISRF outside of the cirrus structures can be treated as uniform across the field. However, this ISRF could be affected internally by the optical depth of the cirrus itself, which is strongly wavelength dependent, thus potentially impacting the morphology.} 

{The cirrus in optical images arises from the scattering of light by the classical ``large'' grain component of the dust size distribution at particular optical wavelengths where the optical depth is small (though probably not completely negligible).}

{The FIR radiation considered is thermal radiation from the same large dust grains responsible for the scattered light. The SED can be modeled as a modified blackbody. In equilibrium, the radiance, the integral over this SED, is equal to the integral of the radiation absorbed from the local ISRF.  Most of the energy in the ISRF is in the optical and near infrared, where optical depth effects are not large. For more in-depth discussion, see \citet{2023ApJ...948....4Z}.
To the extent that the ISRF decreases, so does the radiance and the SED also shifts to slightly lower temperature. The emission in the 250~$\mu m$ band, chosen to sample the SED at wavelengths longward of the peak of the SED, would decrease but not non-linearly as for bands at the peak of the SED or shortward.}

{Thus it can be anticipated a priori that the optical-FIR coupling of our morphology-tracing products will be fairly robust. But any changes in properties of the large dust grains with environment across the field could induce some decoupling.}

{The MIR is non-equilibrium emission from small grains responding directly to the intensity of incident UV photons in the ISRF. Therefore, the morphology traced could be different not only because of changes in the grain size distribution (large vs.\ small) across the field but also because of changes in the shape of the ISRF arising from higher attenuation in the UV. For these fundamental reasons, some optical-MIR decoupling can be expected.}

\subsection{DHIGLS HI}
Gas and dust are known to be spatially correlated, especially at high Galactic latitudes \citep{planck2011-7.12}.   %
It is therefore interesting to investigate the morphology of neutral hydrogen ({\HI}) column density as an indirect tracer of dust. 
We used data for the DF field of the DHIGLS {\HI} survey \citep{2017ApJ...834..126B},\footnote{\url{https://www. cita.utoronto.ca/DHIGLS/}} 
which targeted intermediate-to-high Galactic latitude fields using the Synthesis Telescope at the Dominion Radio Astrophysical Observatory (DRAO).
The {\HI} emission can be distinguished by a variety of gas velocity components (VCs). When the emission is optically thin, the brightness temperature cube can be integrated over these component velocity ranges to obtain maps of the column density of {\HI} ($N_{{\HI}}$) for low-, intermediate-, and high-VCs (LVC, IVC, and HVC, respectively). 
Because IVC of the Spider field is faint and the HVC is negligible, we used the diffuse emission map of the LVC in units of $\rm 10^{19}\,cm^{-2}$. {The velocity range defining the {\HI} LVC emission is $30 > v >-15$ km s$^{-1}$ (row 2 of Table 2 in \citealt{2017ApJ...834..126B}).} The {\HI} image has a beam width of $55\arcsec$ and pixel size $18\arcsec$. {The image size is [820x758] pix$^2$, again amounting to the same angular extent as the optical image.}
The {\HI} map is shown in the lower right panel of Fig~\ref{fig:spider}.

\section{Methods} \label{sec:method}
{We employ complementary families of statistical methods for structural analysis on Galactic cirrus. The local intensity statistics approach (Sec. \ref{sec:method_stats}) is focused on local non-Gaussianity, while Fourier-domain approaches  (Sect.~\ref{sec:method_fourier}) quantify global coherence across scales. The scattering transform approach  (Sect.~\ref{sec:method_wst}) incorporates both local and global structural information; although physical interpretation of its wealth of outcomes can be subtle, benefits can arise from using summary statistics derived from this approach to build on insights obtained from other methods.} As a concrete example of how all these approaches can be applied together to interpret a panchromatic dataset, they are applied to investigate the Spider field in Sect.~\ref{sec:result_local}, Sect.~\ref{sec:result_fourier}, and Sect.~\ref{sec:result_wst} below.

\subsection{Local intensity statistics} \label{sec:method_stats}

A general approach to characterizing the ISM is to analyze the intensity statistics of the map or cube (e.g., 2D column densities/velocities, 3D densities). In particular, the probability density function (PDF) and its derived statistics encapsulate critical information regarding the physical processes that shape the ISM (\citealt{1999intu.conf..218N}, \citealt{Kowal2007}, \citealt{Burkhart2009}). Previous studies have identified correlations between the PDF and turbulence (\citealt{Burkhart2009}), self-gravity (\citealt{2015ApJ...808...48B}, \citealt{2018A&A...617A.125C}), stellar feedback (\citealt{2016ApJ...833..233B}), and astrochemistry (\citealt{2018ApJ...860..157B}), the first seemingly most relevant in the Spider field. In this analysis, we compute local PDFs of intensity maps across the field of view for various dust tracers and examine their properties. We derive statistics from the PDFs to characterize cirrus morphology and quantify the similarities and differences among maps.

\subsubsection{Probability density function} \label{sec:pdf}

Local PDFs encode structural information on specific physical scales, assuming uniform distance to the density field (\citealt{Burkhart2010}, \citealt{2018A&A...617A.125C}, \citealt{2018ApJ...860..157B}). Because cirrus structures span a wide range of scales, we employ a series of radii to extract coherent morphological information at different angular scales and examine its variation as a function of scale.
Given the broad dynamical range of intensities in the maps,
we compute the PDFs and their associated statistics using the logarithm of the intensity. This approach is also motivated by the fact that diffuse ISM morphology is governed largely by turbulence, where density fields driven by compressible isothermal turbulence typically follow a log-normal distribution (\citealt{1998PhRvE..58.4501P}).

At a given position $(x,y)$ on the $I_\nu$ map, we extract a representation of the underlying PDF using the pixel intensity values $I_{\nu, i}$ within a circular region centered at $(x,y)$ with radius of $r$. The PDFs are computed over on a grid of positions $\{x_i, y_i\}$ by moving the circular kernel across the map, where the grid spacing is equal to one-third of the kernel radius. To compute local PDFs and compare among different tracers, we normalize the intensities in the region, $\{I_{\nu, i} (x,y|r)\}$ by the local mean value so that the mean of the normalized intensities $\tilde{I}_{\nu}$ is equal to one. We then take the logarithm of $\tilde{I}_{\nu}$ as the input map $\xb=\log \,\tilde{I}_{\nu}$ 
(these PDFs are referred to as $p(\xb)d\xb$ below).\footnote{Because all intensity maps are calibrated in physical units, the occurrence of pixels with negative intensity values is negligible; otherwise, it would be necessary to add a low-level background value to the map before taking the logarithm of the map.}

To mitigate the influence of extreme outliers (e.g., due to inadequate removal of foreground/background sources), we apply a kernel density estimation (KDE) to the discrete values to produce a smooth empirical PDF $p({\xb = \log} \,\tilde{I}_{\nu}|x,y,r)$. The bandwidth of the estimator $h$ is determined using the empirical Scott's rule: $h=n^{-1/(d+4)}$, with $n$ the number of pixels used for evaluation and $d$ the number of dimensions.

\begin{figure*}[!htbp]
\centering
  \resizebox{0.98\hsize}{!}{\includegraphics{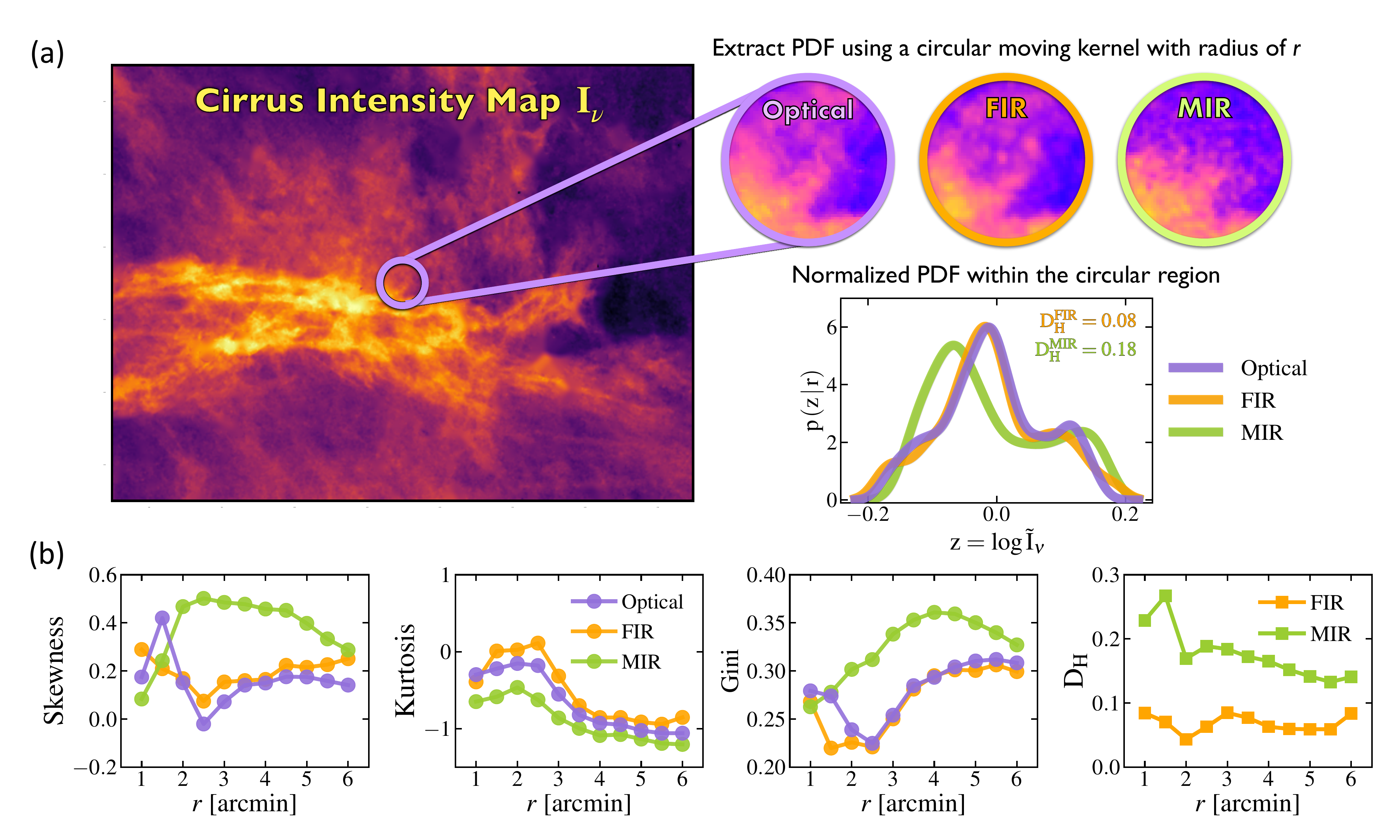}}
  \caption{Illustration of the extraction of local PDFs and PDF statistics. \textbf{(a)} Local PDF extracted from the intensity map within a circular region moving across the field. This example shows an extraction with radius of $r=3\arcmin$ at the same position on the optical (left), FIR (middle), and MIR (right) map. The lower right panel shows the KDE-smoothed PDFs of the logarithm of the normalized intensity in the circular regions (${\xb = \log} \,\tilde{I}_{\nu}$). The PDFs are used for computing statistics and distance metrics. {\Dh} indicates the distance between the PDF of the FIR/MIR data and that of the optical data within the particular subregion. \textbf{(b)} Skewness, kurtosis, and Gini coefficient of the local PDFs at the same position measured on the optical, FIR, and MIR cirrus data with varying kernel radius $r$. The rightmost panel shows the PDF distance of the FIR and MIR data relative to the optical as a function of kernel radius.}
\label{fig:pdf_demo}
\end{figure*}

Figure~\ref{fig:pdf_demo}a illustrates the extraction of local PDFs using a circular kernel with a radius of 3$\arcmin$. The right-hand circular panels display the zoomed-in regions at a fixed position in the cirrus maps $I_{\nu}$ for optical, FIR, and MIR. The lower right panel shows the KDE-smoothed PDFs corresponding to the intensity distributions within the circular region. The PDFs show clear deviation from a single Gaussian and display multiple components that represent substructures that likely correspond to those identified by the dendrogram (\citealt{2009Natur.457...63G}).
Both the 2D and 1D distributions indicate that, although the optical, FIR, and MIR maps appear visually similar on large scales (see Fig~\ref{fig:spider}), they exhibit local differences. In particular, optical cirrus more closely resembles the FIR emission than the MIR emission. This can be explained by the fact that dust scattering in the optical is mainly contributed by the same dust population that emits thermally in the FIR, while MIR emission originates from stochastic heating of ultra-small dust grains \citep{2007ApJ...657..810D}. 
We will discuss this further in Section~\ref{sec:result_local}. The greater similarity between optical and FIR in this local region is further corroborated by the much lower Hellinger distance {\Dh} between the PDFs (see Section~\ref{sec:pdf_stats}).

\subsubsection{PDF statistics} \label{sec:pdf_stats}
While the local PDFs derived from maps encode comprehensive information, it is often more efficient to characterize the shape of the PDF and quantify differences using summary statistics and distance metrics. These measures are scale-invariant and their spatial variations across the field can be visualized. We compute the following statistics and metrics based on the PDF $p(\xb)$:

\begin{itemize}
    \item \textbf{Skewness and Kurtosis}: 
    Skewness and kurtosis are the third- and fourth-order moments of the PDF. They serve to characterize the shape of the PDFs. Skewness measures the symmetry of the distribution around the center value, with positive skewness indicating an excess of high values and negative skewness indicating an excess of low values. Kurtosis is a measure of the ``peakiness'' of the distribution; a distribution that is flatter than a Gaussian yields positive kurtosis, whereas a more concentrated distribution yields negative kurtosis. Skewness and kurtosis have been found to be correlated with turbulent properties, particularly the sonic Mach number (\citealt{Burkhart2010}). 
    The PDF-weighted skewness and kurtosis are computed using the following equations:
    \begin{equation}
        \sk = \frac{1}{\sigma_\xb^3} \int (\xb - \bar{\xb})^3 \cdot p(\xb)\, d\xb\, ,
    \end{equation}
    \begin{equation}
       \kurt = \frac{1}{\sigma_\xb^4}\int (\xb - \bar{\xb})^4 \cdot p(\xb)\, d\xb - 3\, ,
    \end{equation}
    where $\sigma_\xb^2$ and $\bar{\xb}$ are the PDF-weighted variance and mean given by:
        $\sigma_\xb^2 = \int (\xb - \bar{\xb})^2 \cdot p(\xb)\, d\xb$
    and 
        $\bar{\xb} = \int \xb \cdot p(\xb)\, d\xb$.
\medskip
    \item \textbf{Gini coefficient}: 
    The Gini coefficient (``$\Gini$'') is a measure of inequality in a given set of data values.\footnote{Use of the Gini coefficient was first introduced in astronomy to quantify galaxy morphology by \citet{2003ApJ...588..218A} and later generalized.} 
    $\Gini$ ranges from 0 to 1, with a higher value indicating greater imbalance. 
    For a continuous PDF, $\Gini$ can be computed using the following equation (\citealt{gastwirth1972}):\footnote{This double integral‐based formulation is mathematically equivalent to the commonly used Lorenz curve-based definition \citep{dorfman1979}.}
    \begin{equation}
        \Gini = \frac{1}{2\bar{\xc}} \iint |\xc-\xc'| \cdot p(\xc)\, p(\xc')\,d\xc\, d\xc' \, ,
    \end{equation}
    where $p(\xc)\, d\xc$ is first derived from $p(\xb)\,d\xb$ by a min-max normalization along $\xb$ after clipping extreme outliers so that $\xc~\in~[0,1]$. 
    The calculation of $\Gini$ does not require a center, which makes it well-suited for characterizing complex morphologies. We compute $\Gini$ for each map as a function of scale using the extracted $p({\rm log}\,\tilde{I}_\nu)$ at each grid position.
\medskip
    \item \textbf{Hellinger Distance}: 
    When comparing PDFs from different maps with the same kernel at a given position, it is useful to have a measure that quantifies their similarities. One distance metric commonly used for PDFs is the Hellinger distance ({\Dh}). Compared to other distance measures such as the Kullback-Leibler Divergence,
    {\Dh} is less sensitive to outliers. The Hellinger distance between two PDFs $p(\xb)$ and $q(\xb)$ is defined as (\citealt{hellinger1909}):
    \begin{equation}
        D_{\rm H}=\frac{1}{\sqrt{2}} \left\{ \int \left[\sqrt{p(\xb)}-\sqrt{q(\xb)}\,\right]^2 d\xb \right\}^{\frac{1}{2}}.
    \end{equation}
    By definition, {\Dh} ranges from 0 to 1, with lower values indicating greater similarity between $p(\xb)$ and $q(\xb)$. However, it is noteworthy that perfect similarity in the PDFs does not necessarily imply a one-to-one correspondence of the maps at the pixel level nor in frequency space. 
    In our analysis, we choose the optical map as the baseline, i.e., we compute the distance of the local intensity PDFs extracted from each map relative to that of the optical map at corresponding positions. 
\end{itemize}

Figure~\ref{fig:pdf_demo}b illustrates the skewness, kurtosis, and $\Gini$ calculated from the local PDFs of the optical/FIR/MIR maps at the same position in Fig.~\ref{fig:pdf_demo}a as a function of kernel radius $r$ from $1\arcmin$ to $6\arcmin$. The right panel shows the Hellinger distance of the FIR/MIR data relative to the optical data. In this example region, the skewness and kurtosis between optical and FIR are marginally consistent above $r=2\arcmin$; the small-scale differences are probably due to the presence of CIBA in FIR data, beam effects, and residuals from source removal in both bands. On nearly all scales $r>2\arcmin$, $\Gini$ in the optical and FIR are remarkably consistent.
However, statistics of MIR data, in particular, skewness and $\Gini$, show very different trends from those of optical and FIR. The difference is also revealed by its larger PDF distance {\Dh} to the optical data than FIR at all scales.

\subsection{Fourier statistics} \label{sec:method_fourier}

While local intensity statistics serve as useful diagnostics, they are often influenced by local anomalies and instrumental effects. Thus, a widely adopted approach in ISM studies is to characterize the structures in Fourier space by analyzing how structures correlate across different scales. It is also inherently connected to turbulent processes in the ISM, as theoretical models of the turbulent cascade and energy transfer (e.g., the Kolmogorov theory) are formulated in Fourier space.

We employ three widely used statistical tools: (angular) power spectrum, $\Delta$-variance, and cross-power spectrum. It is worth noting that several other techniques -- such as bispectrum/bicoherence (\citealt{Burkhart2009}) and multi-fractal analysis (\citealt{2018MNRAS.481..509E}) -- also characterize structures in Fourier space and provide additional phase information. However, these methods could be computationally demanding for high-resolution, wide-field data. Therefore, in this work, we adopt the stated three methods owing to their clarity and computational simplicity.

\subsubsection{Power spectrum analysis} \label{sec:method_ps}

The power spectrum is a standard tool for studying the statistical properties of the diffuse ISM (e.g., \citealt{2002A&A...393..749M}). Both observations and simulations indicate that the power spectrum of the ISM column density is closely related to the underlying turbulent flow and dissipation processes in the ISM (\citealt{2007A&A...469..595M}). For an optically thin ISM, the 2D column density power spectrum is equivalent to that of the 3D density (\citealt{2002A&A...393..749M}). The slope of the power spectrum, as well as the presence of any characteristic scale, could therefore provide valuable insights into how turbulence regulates the density structures of the ISM. 

We computed the 1D power spectrum as follows. The 2D power spectrum is calculated as the square of the modulus of the 2D Fast Fourier Transform (FFT). To mitigate edge effects (i.e., Gibbs phenomena) resulting from the FFT applied to nonperiodic distributions, we perform an apodization using a split cosine-bell function. The shape parameters of the apodization kernel are tuned until the edge effects effectively disappear. This choice primarily affects the power at only the largest couple of scales. The 2D power spectrum is then 
azimuthally averaged in annuli at $k$ to yield the 1D power spectrum, $P(k)$, where $k$ denotes the spatial frequency (wavenumber).

The 1D power spectrum, 
$P(k)$ is modeled using the following formula:
\begin{equation} \label{eq:ps}
   P(k) = B(k) \times [A k^\gamma + C k^\beta] + D\,,
\end{equation}
which consists of three components:
\begin{enumerate}
    \item The power term of the dust radiation, $A k^\gamma$;
    \item The contribution from other astrophysical sources and noise in the foreground or background sky signals, $C k^\beta$;
    \item Instrumental noise and systematics at the CCD level, $D$.
\end{enumerate}
The first two terms are convolved with a Gaussian beam-related factor, $B(k)$ that can be finely adjusted in the fit. This model is adapted from the modeling in \cite{MivilleDeschene2016}, with the addition of the $D$ term to account for instrumental effects that are not scale-dependent. 

Previous studies have reported values of $\gamma$ ranging between -2 and -5 for ISM emission in various environments (\citealt{2003A&A...399..177K}, \citealt{2012A&ARv..20...55H}). Accordingly, we adopt a broad fitting range for $\gamma$ between -2 and -5. The second term is typically assumed to represent 
Poisson noise (i.e., $\beta=0$)
but $\beta$ could be negative due to residuals in source removal and/or non-white noise characteristics (e.g., 1/f noise-like signal from clustering of sources, for which $\beta\simeq -1$). Therefore we set the fitting range of $\beta$ between 0 and -2. The constant term $D$ is constrained with an upper limit set by the $P(k)$ value at the highest frequency in the fitting range. 

For the Planck radiance, the beam size is fixed at 5\arcmin. The single channel maps used for the modified black body fit are each contaminated by CIBA, which in principle would be characterized by different values of $\beta$ because of a partially uncorrelated set of galaxies dominating in each map. It is not clear how this contamination would be propagated through the non-linear operator, integration over the fitted model modified black body SED.
We simply adopted $\beta = 0$ given that it is a high-order effect. The free parameter $D$ is still present, because although the integral of the SED model might seem noise-free, another set of observations with an independent set of noises would produce a slightly different radiance ($D$ is a measure of the reproducibility of the radiance over repeated observations).

For the {\HI} DHIGLS LVC data, there is no cosmic background and so $C$ is zero.  Following \cite{2015ApJ...809..153M} and \citet{2017ApJ...834..126B}, the noise term $D$
is replaced by a noise template constructed from the power spectra of a set of emission-free channels and multiplied by a free parameter $\eta$
that should be near unity.  The beam is asymmetical but is taken to be cirular with FWHM 56\farcs8, which is an adequate approximation (see \citealt{2017ApJ...834..126B} for further details).

The fitting is performed after excluding the $P(k)$ values at the largest scale (i.e., smallest $k$) and at scales much smaller than the beam size (i.e., the FWHM of the PSF).  The smaller the upper value of $k$ is, the less sensitive the value of $\gamma$ is to other details of the model. The fitting range is thus restricted as for each map.

\subsubsection{$\Delta$-Variance analysis}  \label{sec:method_delvar}

The $\Delta$-variance method is a useful tool for characterizing the structure of the ISM. Originally introduced by \cite{1998A&A...336..697S} and refined by \cite{Ossenkopf2008}, this technique has been used to study the turbulence of molecular clouds regulated by various physical processes -- such as magnetic fields, shock waves, and gravity -- in both simulations (\citealt{2015MNRAS.451..196B}) and observational data (\citealt{2016ApJ...833..233B}, \citealt{2023MNRAS.524.1625D}), as well as {\HI} gas in nearby galaxies (\citealt{2021A&A...655A.101D}). In \cite{2025ApJ...979..175L}, this approach was used to quantify the amount of structure before and after the separation of optical cirrus.

The $\Delta$-variance method can be regarded as a variant of the power spectrum method; it measures the power of structures across a range of spatial scales by convolving the image with a series of kernels of increasing width. Compared to power spectrum analysis, $\Delta$-variance is more robust when handling observational data with irregular boundaries or missing pixels. 

We adopt the implementation of $\Delta$-variance calculation provided by the publicly available \texttt{TurbuStat} package (\citealt{2019AJ....158....1K}), which utilizes the formulation and kernel separation outlined in \cite{Ossenkopf2008}. Briefly, the implementation generates a series of Mexican hat (Ricker) kernels, separated into their core and outer components, convolves the intensity map with these kernels, and computes the weighted variance of the resulting map. As the kernel width varies, the $\Delta$-variance is a function of the spatial scale $L$:
\begin{equation}
    \sigma^2_\Delta(L) = \bigr \langle (I(x,y) \ast \odot_L)^2 \bigr \rangle _{x,y}\,,
\end{equation}
where the average is computed over the entire map, $\ast$ denotes convolution, and $\odot_L$ represents the kernel at scale $L$. The inverse variance map of the intensity map is used as a weight map in the spatial integration. By convention, $L$ is referred to as the ``lag''.

If the underlying density field exhibits a power-law behavior in its power spectrum, i.e., $P(k)\propto k^\gamma$, then $\sigma^2_\Delta(L)$ is expected to also show a power-law scaling:
\begin{equation} \label{Eq:delvar_power}
    \sigma^2_\Delta(L) \sim L^{\alpha}\,,
\end{equation}
where the power index $\alpha$ is related to that of the power spectrum via $\alpha=-\gamma-2$ (\citealt{1998A&A...336..697S}). We fit a power-law function to the main portions of the $\Delta$-variance spectra and examine this relationship.

\subsubsection{Cross-power spectrum analysis}  \label{sec:method_cps}

While the power spectrum measures the statistical properties of intensity fluctuations in the field, comparisons between datasets need to account for systematics, contamination by other sky signals, and instrumental noise effects first. The cross-power spectrum provides a complementary means to characterize the correlation of structures across spatial scales in two fields (\citealt{tristram2005}). By construction, systematics, contamination, and noise exclusive to a single dataset are suppressed in the cross-spectrum, thereby enhancing the S/N of genuine correlation. This method has been adopted widely in analysis of various astrophysical maps in cosmology, such as the cosmic microwave backgrounds, 21~cm intensity maps, and weak gravitational lensing fields (e.g., \citealt{2016MNRAS.460..434H}, \citealt{planck2016-l11A}), yielding more robust constraints to theoretical models than auto-power spectra alone.

For two given maps, $I_a$ and $I_b$, the 2D cross-power spectrum is defined as:
\begin{equation} \label{eq:cps}
    \mathit{CP}_{a \times b} ({\bf k}) = \langle Re\ [ \mathcal{I}_a ({\bf k}) \, \times \, \mathcal{I}_b^* ({\bf k}) ]\, \rangle\,,
\end{equation}
where $\mathcal{I}_a$ and $\mathcal{I}_b$ denote the discrete Fourier transforms of $I_a$ and $I_b$, respectively ($*$ denoting the complex conjugate), ${\bf k}$ is the wavevector, and $\langle \cdot \rangle$ represents averaging over radial binning of ${\bf k}$.  
Because the input maps are in real values and we aim to measure the correlations between common structures with the same spatial phase, we retain only the real part in the product of their transforms.
In practice, the 1D cross-power spectrum is then calculated by azimuthally averaging the 2D spectrum over radial bins (annuli) in Fourier space. Following the prescription in Sect.~\ref{sec:method_ps}, we apply an apodization using a split cosine-bell function to both maps.\footnote{Conventionally, the cross spectrum of two fields, A and B, is computed via their spherical harmonics expansions: $\mathcal{C}_l^{AB}=\frac{1}{2l+1}\sum\limits^{l}_{m=-l}a^A_{l,m}\,a^{B\,*}_{l,m}$, where $a_{l,m}$ are the spherical harmonic coefficients. This formalism is convenient for maps in \texttt{healpix} format. Although the results in Sect.~\ref{sec:result_cps} are calculated using the 1D spectrum following from Eq.~\ref{eq:cps}, we have also calculated using the above formalism and obtained consistent results.}

We further calculate the cross-correlation ratio, which is the 1D cross-power spectrum normalized by the auto-power spectra of each map:
\begin{equation} \label{eq:cpr}
    \xi_{a \times b} (k) = \frac{\mathit{CP}_{a \times b} (k)}{\sqrt{P_a(k)P_b(k)}}\,.
\end{equation}
where $P_a(k)$ and $P_b(k)$ are the auto‐power spectra of $I_a$ and $I_b$, respectively. At a given spatial frequency $k$, one would expect $\xi_{a \times b}$=1 for perfect correlation, 0 for no correlation, and negative values in the case of anti-correlation.

\subsection{Wavelet scattering transform (WST)} \label{sec:method_wst}

Although power spectrum-based approaches provide useful information about the energy partition across scales, they are insufficient to extract the non-Gaussian information. This is particularly important because many physical processes in the ISM are intrinsically non-Gaussian. Consequently, two fields may exhibit identical first- and second-order statistics in Fourier space while displaying markedly different morphologies. One novel statistical tool for quantifying the non-Gaussianity and morphological characteristics of a physical field is the wavelet scattering transform. 
The WST has been successfully applied to cosmology (e.g., \citealt{Cheng2020},  \citealt{2022MNRAS.513.1719G}, \citealt{2025arXiv250514400J}) and ISM astrophysics (e.g., \citealt{Allys2019}, \citealt{2021ApJ...910..122S}, \citealt{2023ApJ...947...74L}). Compared to conventional high-order statistics, the WST estimator offers advantages in robustness, rapid convergence, and stability against additive noise and deformations in observational data.

\subsubsection{Formalism to generate WST fields}

\begin{figure}[!htbp]
\centering
  \resizebox{0.95\hsize}{!}{\includegraphics{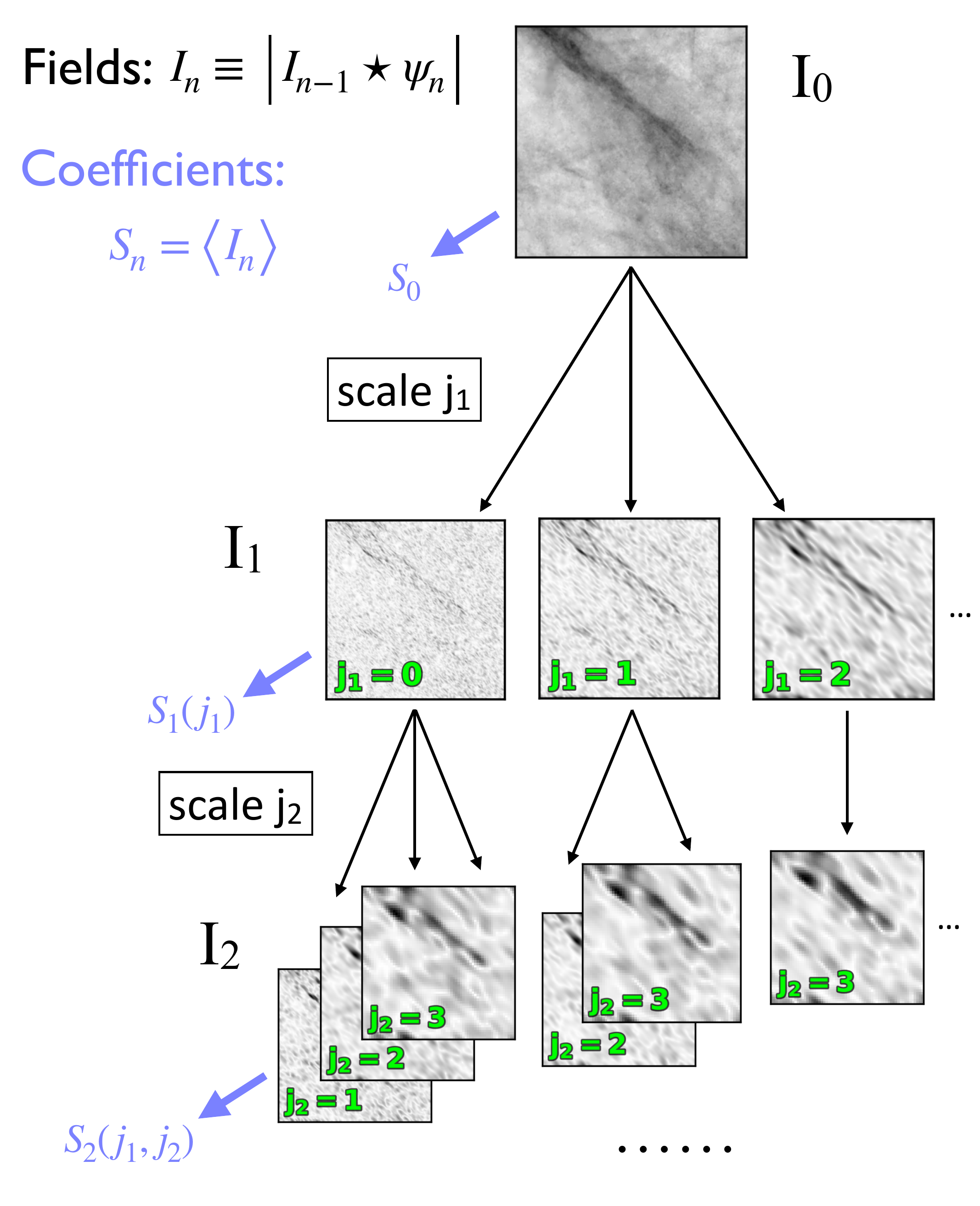}}
  \caption{{Illustration of wavelet scattering transform applied to cirrus images, adapted from Figure 4 of \cite{2021arXiv211201288C}}. The input field is ``scattered" through convolution with a bank of Morlet wavelets at different scales $j$ and orientations $\theta$ (not displayed for clarity), followed by a non-linear operation (modulus). The output fields are shown up to $j_2=3$. The scattering coefficient $S_n$ is computed by taking the spatial average of the output field. Only $j_2>j_1$ coefficients are used.
  }
\label{fig:wst_schematic}
\end{figure}

Here, we briefly summarize the formalism of the WST approach, which is illustrated in Figure~\ref{fig:wst_schematic}. 
For a 2D input field $I_0=I(x,y)$, the scattering transform computes a set of first-order fields $I_1=I_1(x,y)$ by convolving $I_0$ with a family of Morlet wavelets $\{\psi^{j,\theta}(x,y)\}$ and applying a modulus operation:
\begin{equation} \label{eq:I1}
    I_0 \rightarrow I_1^{j_1,\theta_1} \equiv \left| I_0 \ast \psi^{j_1,\theta_1}\right|\,,
\end{equation}
where $I_1 ^{j_1,\theta_1}$ represents a set of fields $\{I_1\}$ characterized by the scale index $j_1$ and the orientation index $\theta_1$ of the wavelet. The second-order fields $I_2=I_2(x,y)$ can be generated by applying the same process to $I_1$:
\begin{equation}\label{eq:I2}
    \{I_1\} \rightarrow I_2^{j_1,j_2,\theta_1,\theta_2} \equiv \left| I_1 \ast \psi^{j_2,\theta_2} \right| = \left|\, \left| I_0 \ast \psi^{j_1,\theta_1}\right| \ast \psi^{j_2,\theta_2}\,\right|\,,
\end{equation}
where $I_2^{j_1,j_2,\theta_1,\theta_2}$ represents a set of fields $\{I_2\}$ characterized by the wavelet indices ($j_1,\theta_1$) and ($j_2,\theta_2$). By recursion, the $n$-th order fields are generated by:
\begin{equation}\label{eq:In}
  \{I_{n-1}\} \rightarrow I_n^{j_1,\dotsc,j_n, \theta_1,\dotsc,\theta_n} \equiv \left| I_{n-1} \ast \psi^{j_n,\theta_n}\right|\,.
\end{equation}

By convention, the number of scales $J$ is defined in a dyadic sequence of $2^j$ for $0\leq j < J$, with the largest scale $2^{J}$ being smaller than the size of the field.\footnote{The scale index $j$ refers to the wavevector in Fourier space. Therefore, the index $j$ corresponds to a scale of $2^{j+1}$ in real space, and the largest scale (in pixels) is of $2^J$ set by $j=J-1$.} Because the wavelets scatter the energy into larger scales, only coefficients with $j_2 > j_1$ are physically meaningful. 

The number of orientations, $\Theta$, defines wavelets with position angles $\phi=\pi\theta/\Theta$ in the range $0\leq \phi < \pi$ generated by integers $\theta$ over the range $0\leq \theta < \Theta$. 

\subsubsection{Formalism to generate WST 
coefficients}

The set of $n$th-order WST coefficients, $S_n$, is computed by taking the mean amplitude of the $n$th-order fields:
\begin{equation}
    S_n = \, \bigr \langle I_n \bigr \rangle_{x,y}\,,
\end{equation}
where $\langle \cdot \rangle_{x,y}$ denotes spatial averaging over the field. 
These coefficients form a compact set of translation-invariant descriptors that capture the coherence of the structures among different scales and orientations. In practice, most of the information in a physical field is contained in the leading orders of the scattering coefficients, as the norm of the coefficients decays exponentially with increasing order. Therefore, as widely adopted in the literature (\citealt{2023ApJ...947...74L}, \citealt{2024A&A...681A...1A}), we focus on WST coefficients up to the second order. 

Because the fields $I_{n}$ averaged in the coefficients $S_n$ are computed from the previous-order fields $I_{n-1}$, $S_n$ is correlated with $S_{n-1}$ under the same index family. Therefore, it is usually convenient to normalize the coefficients according to $s_n = S_n/S_{n-1}$, which renders the coefficients dimensionless and scale-invariant.

\subsubsection{Intepretability}

Here we briefly describe the intepretability of the WST coefficients. Readers with deeper interests can refer to \cite{Allys2019} and \cite{2021arXiv211201288C} for further exposition. 

The first-order coefficients, $S_1$, provide information similar to that obtained from the power spectrum (or the two-point correlation function), as both quantify the power of density structures as a function of scale. The primary differences are that the power spectrum adopts an $\mathcal{L}^2$ norm and a Fourier kernel, while the WST uses an $\mathcal{L}^1$ norm and localized wavelet kernels. 

It is often more interesting to investigate the second-order coefficients $S_2$. When averaged over orientations (yielding isotropic statistics), $S_2 (j_1,j_2)$ characterizes the \textit{clustering} of patterns at a scale $j_1$ as a function of scale $j_2$, thereby encoding information about the coherence and interaction between scales. A larger value of $S_2(j_1,j_2)$ indicates that structures at scale $j_1$ occur more frequently on a scale $j_2$. 

In the case where anisotropy is also of interest, it is often useful to reduce the second-order coefficients by averaging only over certain orientations. In particular, we follow the reduction formalism in \cite{2023ApJ...947...74L}, which retains the angular difference dependence $\Delta\theta = \theta_2-\theta_1$ and averages over $\theta_1$:
\begin{equation} \label{eq:wst_coef_iso}
    S_2(j_1,j_2,\Delta \theta)=\langle S_2(j_1,j_2,\theta_2-\theta_1,\theta_1)\rangle_{\theta_1}\,,
\end{equation}
where $\langle \cdot \rangle_{\theta}$ denotes averaging over $\theta$. Large components with small $\Delta\theta\sim0$ delineate coherent structures with certain scales distributed along the same direction on larger scales (e.g., cirrus filaments), whereas large components with $\Delta \theta \sim 90 \degr$ correspond to structures predominantly extending in the perpendicular direction (e.g., sheets and patches). 

\subsubsection{Adopted summary statistics}

We examine only the two summary statistics proposed by \cite{2021arXiv211201288C}, which further reduce the second-order coefficients:
\begin{itemize}
    \item \textbf{Sparsity $s_{21}$}: 
    The sparsity is defined as the ratio between $S_2$ and $S_1$ coefficients averaged over orientations:
    \begin{equation} \label{eq:sparsity}
        s_{21} \equiv \langle S_2/S_1\rangle_{\theta_1,\theta_2}\,.
    \end{equation}
    This diagnostic measures the degree of deviation from Gaussian random fields. A higher $s_{21}$ indicates the structures being more localized, i.e., concentrated at certain positions in the field, whereas a lower $s_{21}$ suggests a more widespread distribution.
    \medskip
    \item \textbf{Linearity $s_{22}$}: 
    The linearity is defined as the ratio between the parallel ($\Delta\theta=0\degr$) and perpendicular ($\Delta\theta=90\degr$) components of the coefficients averaged over orientations:
    \begin{equation} \label{eq:linearity}
        s_{22} \equiv \langle S_2^{\parallel}/S_2^{\perp}\rangle_{\theta} = \langle S_2^{\Delta\theta=0}/S_2^{\Delta\theta=90\degr}\rangle_{\theta}\, ,
    \end{equation}
    which is a scale- and rotation-invariant measure of the shape of structures. Fields with $s_{22}<1$ tend to exhibit curvy, bubbly, or swirly patterns, while fields with $s_{22}>1$ display more linear, wispy, filamentary features.
\end{itemize}

Both summary statistics depend solely on the two scales $j_1$ and $j_2$. Because $S_2$ and $S_1$ at different scales are correlated, the sparsity and linearity may also exhibit correlations across various scale combinations. Consequently, in our analysis, we further compute the sparsity and linearity averaged over all scale combinations with $0\leq j_1 < j_2 <J$, denoted as $\tilde{s}_{21}$ and $\tilde{s}_{22}$, which are treated as scale-averaged statistics. 

\subsubsection{Computation}

The WST coefficients are computed using the publicly available code \texttt{scattering} (\citealt{Cheng2020}), which is developed based on the \texttt{Kymatio} package (\citealt{2018arXiv181211214A}). This implementation performs the wavelet convolution through FFT and supports GPU acceleration. 

The choice of the number of scales ($J$) and the number of orientations ($\Theta$) is a necessary input for WST analysis. In this work, we choose $J=\lfloor \text{log}_2 N-1 \rfloor $, where $N$ is the minimum dimension (in pixels) of the field and $\lfloor \cdot \rfloor$ denotes rounding to the nearest integer. Thus, the statistical description is restricted to scales that are smaller than or comparable to half the size of the field to ensure reliable estimation. Following \cite{RSB2020} and \cite{2021ApJ...910..122S}, we set $\Theta=8$ to decompose the half-circle into discrete directions, which provides as a reasonable trade-off between the compactness of the descriptor set\footnote{Although $\Theta$ can be arbitrarily large, too large $\Theta$ would lead to highly correlated coefficients especially for small-scale (low $J$) structures because of the pixelization, and thus introduce redundancy.} and smooth sampling of the directions. 

{We did not apply apodization in the computation because in comparison with Fourier power spectra the use of localized wavelets in WST analysis makes it less sensitive to non-periodic boundary conditions. However, we note that the cutoff of cirrus structures at the field edges might still bias the measured shape parameters (such as $s_{22}$) for the largest couple of scales. This effect, however, does not affect the relative comparisons between datasets in our analysis below.}

Note our analysis applies to the intensity image as $I_0$ rather than the logarithm of intensity. Therefore, when applying such analysis for studying the cirrus morphology, bright interlopers -- such as nearby galaxies or bright stars -- should be carefully removed to minimize their impact on the non-Gaussian statistics of the field.
This task, however, could be challenging in a crowded field (such as galaxy cluster or near the Galactic center) if the extended PSF wings dominate the sky background.

\begin{table*}[!htbp]
  \caption{{Summary of statistical methods for structural analysis of Galactic cirrus}}
  \label{table:method}
  \centering
  \renewcommand{\arraystretch}{1.0}
  \begin{tabular}{%
      >{\centering\arraybackslash}m{3.2cm}  
      >{\centering\arraybackslash}m{2.4cm}  
      >{\centering\arraybackslash}m{1.4cm}  
      >{\centering\arraybackslash}m{2cm}  
      >{\centering\arraybackslash}m{2cm}  
      >{\centering\arraybackslash}m{2cm}  
      >{\centering\arraybackslash}m{2cm}  
  }
    \hline\hline
    \textbf{Method}
      & \textbf{Diagnostics\tablefootmark{a}}
      & \makecell[c]{\textbf{Scales} \\\textbf{Probed}}
      & \makecell[c]{\textbf{Anomaly} \\\textbf{Insensitive?}\tablefootmark{b}}
      & \textbf{\makecell[c]{\textbf{Non‑Gaussian} \\\textbf{Information?}\tablefootmark{c}}}
      & \textbf{\makecell[c]{\textbf{Invariant} \\\textbf{Across Field?}\tablefootmark{d}}}
      & \textbf{Complexity\tablefootmark{e}}
      \\
    \hline

    Local intensity statistics
      & \makecell[c]{Moments\\PDF distance}
      & \makecell[c]{Local\\(tunable $r$)}
      & \no
      & \yes
      & \no
      & Moderate to high
    \\[1.0ex]

    \hline
    \medskip
    
    Power spectrum
      & 
      & 
      & 
      & 
      & 
      & \medskip Low
    \\[-1ex]

    $\Delta$‑Variance
      & Power‑law slope
      & \makecell[c]{Global\\(all $k$ or $L$)}
      & \yes
      & \no
      & \yes
      & Moderate
    \\[-1ex]

    Cross-power spectrum
      & 
      & 
      & 
      & 
      & 
      & Low
    \\[1.0ex]
    
    \hline
    
    Wavelet scattering transform
      & \makecell[c]{Scattering coeff.\\Summary statistics}
      & \makecell[c]{Global\\(all $J$)}
      & \yes
      & \yes
      & \yes
      & Moderate to high
    \\

    \hline
  \end{tabular}
  \tablefoot{%
    \tablefoottext{a}{Diagnostics used in this work for morphological characterization and comparison across dust tracers.}
    \tablefoottext{b}{Is the diagnostic robust against local anomalies such as residuals of source removal?}
    \tablefoottext{c}{Does the method retain non-Gaussian or equivalent local intermittency information?}
    \tablefoottext{d}{Is the diagnostic invariant across the field?}
    \tablefoottext{e}{Relative computation cost of each method, with the exact overhead depending on the setup. Note that computation for local intensity statistics scales with image size by {$\mathcal{O}(N)$} and therefore is inefficient for high resolution wide fields, while computation for WST scales by {$\mathcal{O}(N\log N)$} and can be accelerated by GPU. Although it appears that WST 
    is preferred over other methods, physical interpretation of the abundant ($10^2\sim10^3$) coefficients is not straightforward, and so benefits accrue from complementary insights from other methods.}
  }
\end{table*}

\subsection{Summary}

{The statistical methods adopted in this work are summarized in Table~\ref{table:method}. These methods cover a range of algorithmic complexity, describing the morphology of cirrus structures across angular scales in different angles, with their own strengths and limitations.}
In the following sections, we apply these methods to statistical characterizations of the morphology of optical cirrus.
For a multi-wavelength context, we compare these results with results obtained using alternative dust tracers, highlighting the quantitative similarities and differences relative to optical cirrus. Where relevant, we discuss the connection between the observed structures and the interplay of physical mechanisms and instrumental effects that shape them. 

\section{Local intensity statistics} \label{sec:result_local}

Here we show analysis based on the local intensity statistics described in Section~\ref{sec:method_stats} using Dragonfly optical, Herschel FIR, and WISE MIR data. 
{The motivation here is to look into the ``neighborhoods'' around each pixel as an extension of the pixel-by-pixel comparison.}
Planck and {\HI} maps are not used for comparing statistics given their significantly larger beam widths. 

However, the {\HI} map is used here to exclude (mask) regions with very low ISM column densities, because, unlike simulations, PDFs in low column density regions in observational data have higher uncertainties and may suffer from incomplete sampling at the faint end (\citealt{2015A&A...576L...1L}). We produce a mean {\HI} map by computing the local mean value using a moving circular region (with the same kernel scale). We then exclude low column density regions with $N_{\HI}$ below the first quartile of the $N_{\HI}$ distribution in the mean {\HI} map. For reference, this corresponds to $2.6\times10^{-18}cm^{-2}$ for $d_k=5\arcmin$. 

\subsection{{PDF moment maps: a visualization of where morphology changes}}

\begin{figure*}[!htpb]
\centering
  \resizebox{0.67\hsize}{!}{\includegraphics{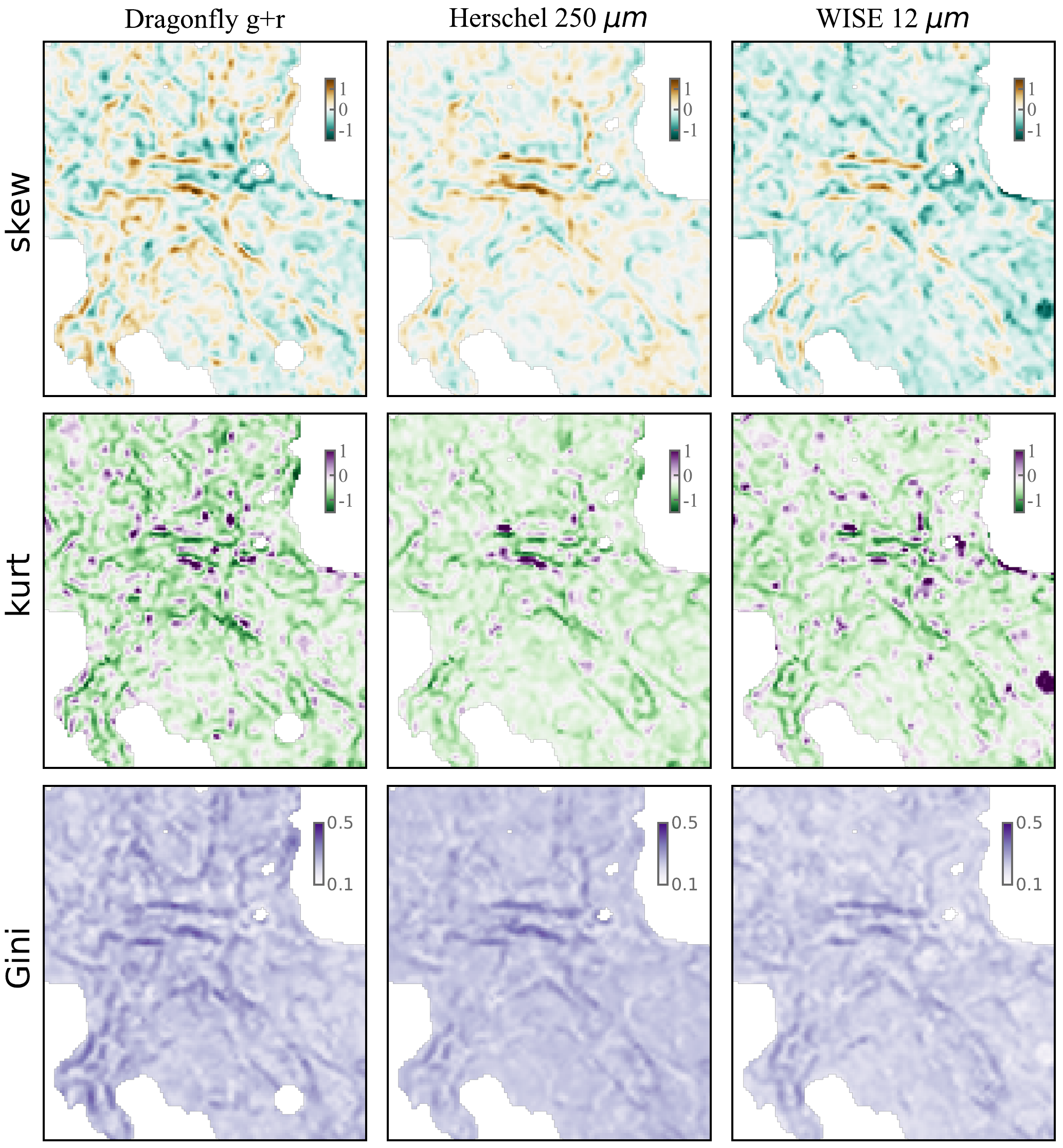}}
  \caption{Spatial distributions of local skewness, kurtosis, and $\Gini$ extracted from the Dragonfly optical (left column), Herschel FIR (middle column), and WISE MIR (right column) data, using a moving circular kernel with a kernel width of $d_k=10\arcmin$. These maps trace where cirrus morphology {changes}.}
\label{fig:pdf_stats_maps}
\end{figure*}

Figure~\ref{fig:pdf_stats_maps} shows the spatial distribution of skewness, kurtosis, and $\Gini$ extracted with 
a characteristic scale $d_k$ ($d_k=2 \times r$) of 10$\arcmin$, which correspond to $\sim$0.9 pc assuming a distance of 320~pc (\citealt{2020A&A...633A..51Z}). The maps were produced via bilinear interpolation using the grid values, as described in Sec~\ref{sec:method_stats}. 

In general, regions with positive skewness occur at the boundaries of sharp filaments and extend alongside bright structures, indicating strong discontinuities. Given the correlation between higher-order moments and turbulence (\citealt{Kowal2007}, \citealt{Burkhart2010}), these discontinuities likely correspond to the fronts of ISM turbulence -- possibly undergoing shearing or tidal flows. In the absence of energetic processes such as supernova explosions or stellar winds, the morphology is primarily shaped by intermittent turbulence and magnetic fields.

Positive kurtosis occurs more sporadically but is often co-spatial with positive skewness. Conversely, negative kurtosis regions are mainly found along the ridgelines of filaments. 

At small angular scales the skewness and kurtosis maps of FIR and MIR images exhibit several localized peaks. This is due to high-order PDF moments being vulnerable to extreme values or outliers, for example, residuals from imperfect source removal. The FIR image also tends to have higher values on average, which could be attributed to the presence of residuals of CIBA.

The spatial distribution of $\Gini$ is similar to that of skewness: the boundaries of cirrus filaments and patches exhibit higher $\Gini$, indicating greater density inequality in these regions, while patches and ridgelines have lower $\Gini$, reflecting a flatter intensity distribution. Indeed, $\Gini$ is correlated with skewness. However, it should be noted that $\Gini$ and skewness still trace different structural aspects because they respond differently to the same PDF.
For instance, a local region containing two symmetric peaks relative to the mean may exhibit near-zero skewness, but $\Gini$ can be high given its significant deviation from a uniform distribution.
The $\Gini$ maps of the FIR and MIR data present fewer local peaks at small angular scales, suggesting a much lower susceptibility to source residuals.

{In summary, the local PDF departs strongly from a Gaussian or a uniform distribution at the edges of sharp structures such as filaments, revealing local intermittency. The PDF moments thus serve as good indicators of where cirrus morphology changes.}

\subsection{{Which is closer to optical cirrus \textit{locally}: FIR or MIR?}}

\begin{figure*}[!htbp]
\centering
  \resizebox{0.925\hsize}{!}{\includegraphics{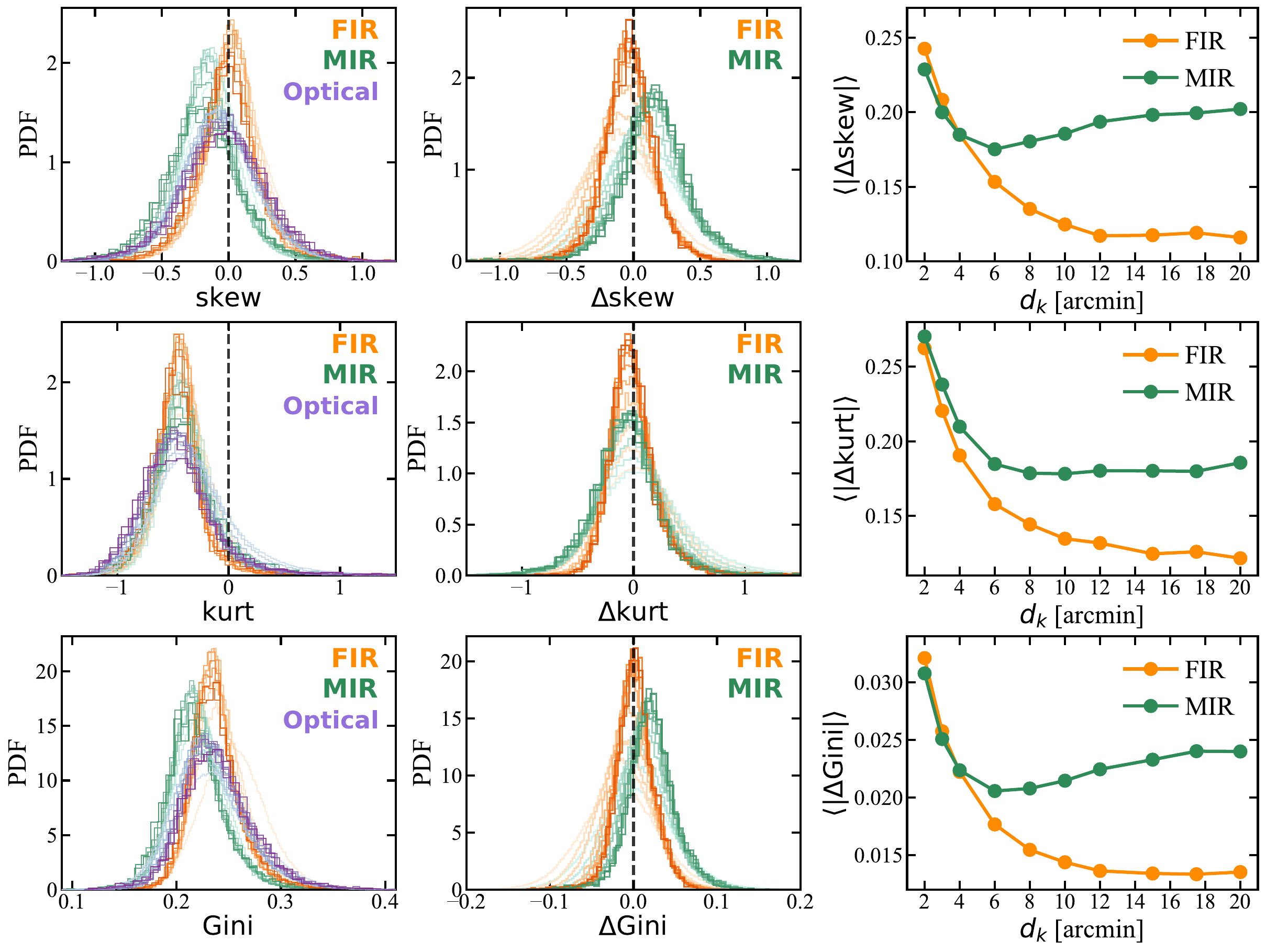}}
  \caption{\textbf{Left column}: Histograms of skewness, kurtosis, and $\Gini$ measured from the local PDFs in optical, FIR, and MIR data. Fainter histograms represent smaller kernel scales used for PDF extraction. 
  \textbf{Middle column}: Histograms showing the distributions of differences of the statistics between FIR or MIR and optical data. The difference is calculated by [optical-$\sf X$], where $\sf X$ stands for the band. \textbf{Right column}: Mean absolute difference 
  of statistics between FIR or MIR and optical as a function of kernel scale.}
\label{fig:pdf_stats_dist}
\end{figure*}

The left column of Figure~\ref{fig:pdf_stats_dist} shows the histograms of the distributions of skewness, kurtosis, and $\Gini$ across the field for scales $d_k$ ranging from $2\arcmin$ to $20\arcmin$. {The lower end of the range of kernel scale is chosen to be several times larger than the beam widths to ensure reliable statistics.
The upper end still retains good spatial resolution, while providing about sufficient points to sample the PDFs after masking the low ISM column density regions.} 
Fainter histograms represent smaller $d_k$.

For skewness, most values lie between -1 and 1. 
The skewness distribution of optical data is centered around zero, whereas the FIR data 
tend to be more positive and the MIR tend to be more negative.
Kurtosis values range between $-1.5$ to $1$ and peaks around $-0.5$. The optical data display a broader distribution than the FIR and MIR data, likely owing to the smaller beam width, which allows it to capture finer structures. 
The distributions of $\Gini$ for the three dust tracers are similar to those of skewness in a relative sense.

The middle column of Fig.~\ref{fig:pdf_stats_dist} shows the distributions of the differences between the statistics of the optical and FIR/MIR maps. For all statistics, the differences between optical and FIR (orange histograms) peak around zero. However, the differences in skewness and $\Gini$ between the MIR and optical maps (green histograms) are skewed toward positive values, indicating systematic offsets. The kurtosis difference relative to optical is broader for MIR than for FIR. 

These trends are further demonstrated in the right column of Fig.~\ref{fig:pdf_stats_dist}, which shows the 
mean absolute difference 
as a function of scale. {The mean of absolute difference quantifies the average discrepancy between the two maps.}
For $d_k<4\arcmin$, the differences between optical and FIR statistics are comparable to those between optical and MIR, while both follow a decreasing trend with increasing scale. This is likely due to residual CIB. For $d_k>4\arcmin$, the differences continue to drop at increasing scales for the FIR data, exhibiting more similar statistics to those of the optical data, whereas the MIR data present larger systematic offsets. These trends reinforce that the structures probed by optical and FIR are more similar than what MIR traces, 
{as anticipated in Sect.~\ref{sec:isrf}.}

\begin{figure*}[!htbp]
\centering
  \resizebox{0.85\hsize}{!}{\includegraphics{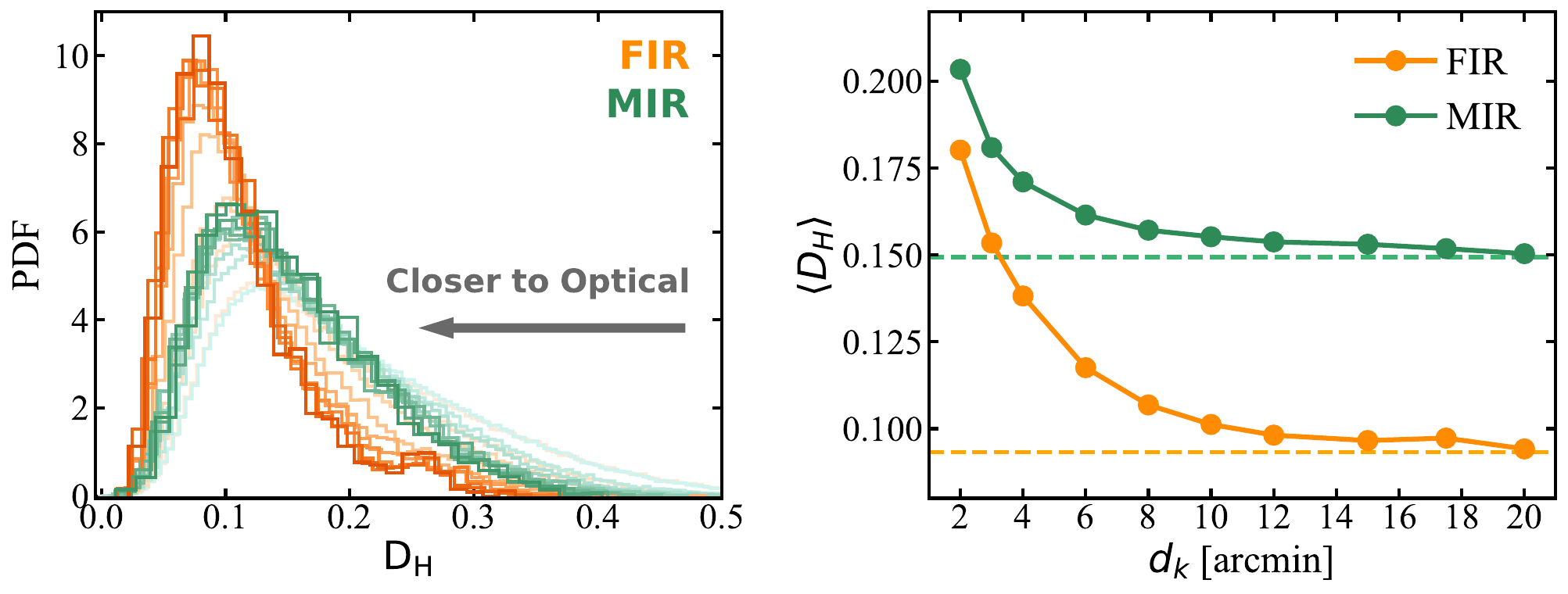}}
  \caption{\textbf{Left}: Histograms of the Hellinger distance ({\Dh}) between FIR or MIR and optical maps, which indicates the proximity of two cirrus maps. Smaller {\Dh} represents a closer local PDF. Histograms with lighter shades represent smaller kernel scales used for the PDF extraction. The PDF distance of FIR relative to optical is systematically lower than that of MIR. The dispersion is also much smaller. \textbf{Right}: Mean {\Dh} between FIR (orange) or MIR (green) and optical data as a function of kernel scale. The mean \Dh\ of FIR relative to optical is lower than MIR in all scales. The dashed lines show the asymptotic values at large scales.}
\label{fig:pdf_dist}
\end{figure*}

The greater similarity between FIR and optical can be further corroborated by the Hellinger distance \Dh. The left panel of Figure~\ref{fig:pdf_dist} shows the histograms of 
\Dh\ across the map extracted using various kernel scales, where lighter shades represent smaller scales. The distributions for FIR relative to optical are significantly narrower and peak at smaller values than those for MIR. The right panel of Fig.~\ref{fig:pdf_dist} plots the mean \Dh\
as a function of kernel scale, with the FIR distance being consistently smaller than the MIR distance at all scales. Because the PDF distance directly measures the proximity of local intensity distributions between two maps, these results indicate that the morphology of dust in optical imaging is closer to that traced by FIR than by MIR. This finding is consistent with the distinct origins of the diffuse light in the three bands, with the same characteristic dust size contributing to the optical and FIR but not the MIR (Sect.~\ref{sec:isrf}).

Furthermore, we note that the PDF distance approaches constant ($\sim$0.15 for MIR and $\sim$0.09 for FIR) on larger scales. Because the beam widths in FIR and MIR are several times smaller than the kernel scales, we attribute the increase in PDF distances at small scales to the presence of CIB residuals. We infer that if CIBA were optimally removed from the images, the PDF distance for the FIR data would shrink to a constant down to the beam width. At scales comparable to the beam width, PSF blurring smooths the density structures, and consequently smooths the intensity PDFs.

\subsection{$\mathcal{M}_s$ in diffuse ISM inferred from dust morphology} \label{sec:Mach_map}

The sonic Mach number, $\mathcal{M}_s$, is a measure of the compressibility of the medium. It is defined as the ratio of the local flow velocity to the sound speed of the medium. This parameter is crucial for hydrodynamics of astrophysical fluids, as the physics of compressible turbulence differs markedly from that of incompressible turbulence. Thus, investigating the variation of $\mathcal{M}_s$ in the ISM density field is of great interest.

$\mathcal{M}_s$ is known to correlate with statistical moments (variance, skewness, and kurtosis) of both observed column density distributions and simulated density fields (\citealt{Kowal2007}, \citealt{Burkhart2009}). The degree of non-Gaussian asymmetry in the PDF increases with $\mathcal{M}_s$. Although this correlation is strong for supersonic models, it is relatively weak for subsonic models where additional physical mechanisms may be at play; hence, caution is warranted when interpreting subsonic regions. 

While higher-order moments generally correlate with $\mathcal{M}_s$, we follow \citet{Burkhart2010} in using the fourth moment (kurtosis) to estimate $\mathcal{M}_s$ via the empirical relation: $\mathcal{M}_s\sim(\kurt+1.44)/1.05$, which exhibits an almost linear correlation for $\mathcal{M}_s$ in the range 1--8 and is less affected by boundary effects. This model is derived from simulations with Alfv{\'e}n Mach number $\mathcal{M}_A=0.7$ and the moments show little dependence on $\mathcal{M}_A$ (\citealt{Kowal2007}). We similarly assigned regions with very low kurtosis ($<-1$) with $\mathcal{M}_s=0$ to mark subsonic conditions whose behavior is not well characterized. We  masked a high-kurtosis region in the lower right of the field with a circular aperture, as it likely results from a residual of source removal.

\begin{figure}[!htbp]
\centering
  \resizebox{\hsize}{!}{\includegraphics{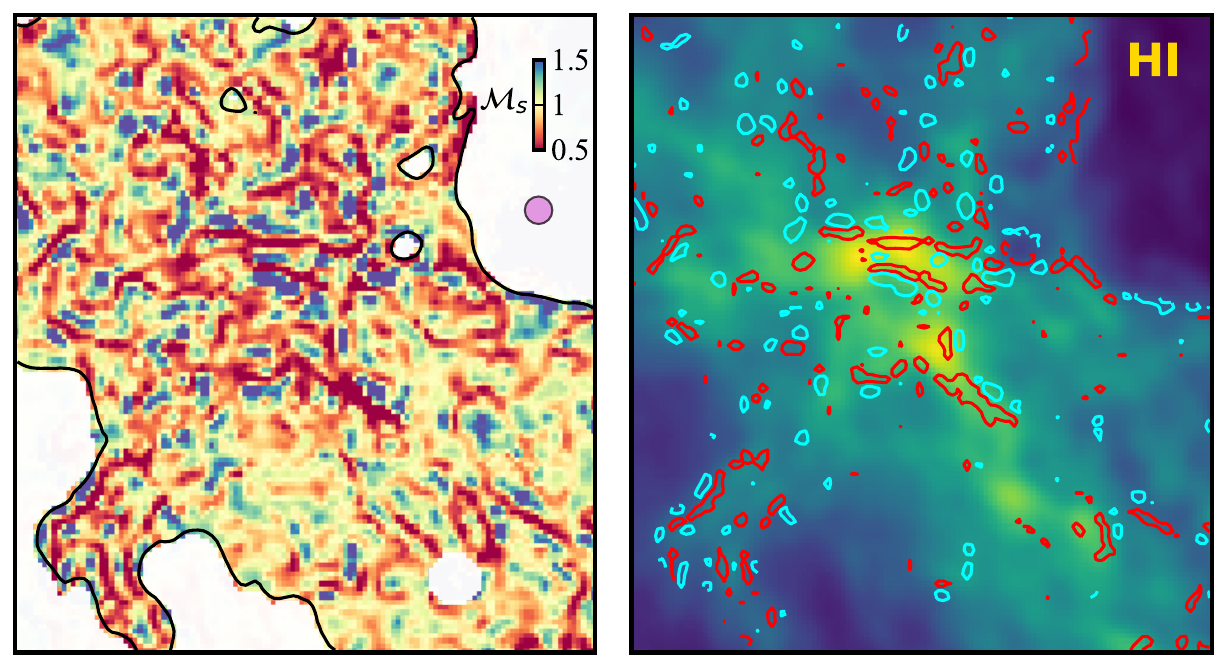}}
  \caption{\textbf{Left}: Maps of sonic Mach number inferred from optical cirrus map illustrated with a kernel scale of $d_k=10\arcmin$. 
  The circular kernel is illustrated with the purple disk. 
  \textbf{Right}: Mean {\HI} column density fields extracted with the same kernel in the upper panels. Regions with $\mathcal{M}_s<0.5$ and $>1.5$ in the upper panels are indicated by red and cyan contours, respectively, which aggreagate at edges of cirrus filaments or patches.}
\label{fig:Mach_maps}
\end{figure}

Figure~\ref{fig:Mach_maps} shows the $\mathcal{M}_s$ map derived from the local intensity PDF of optical cirrus maps using a kernel scales of $10\arcmin$. 
Most regions are subsonic ($\mathcal{M}_s<1$ or $kurt < -0.39$) or transonic ($\mathcal{M}_s\sim1$) -- see the kurtosis histograms in the middle of the left column in Fig.~\ref{fig:pdf_stats_dist}. For a scale of $5/10\arcmin$, $\sim$55/45\% of the map comprises regions with $\mathcal{M}_s<1$, and only $\sim$9/7\% of the map has $\mathcal{M}_s>1.5$ ($kurt > 
0.14$). These results are qualitatively consistent with those reported for the {\HI} in the Small Magellanic Cloud (\citealt{Burkhart2010}), where approximately 90\% of the areas in the Small Magellanic Cloud exhibit $\mathcal{M}_s<2$ based on PDF statistics. On average, $\mathcal{M}_s$ in the diffuse ISM is considerably lower than that found in star-forming regions, where shocks from stellar feedback can compress the medium and efficiently produce supersonic turbulence.
In the right panel we overlay contours of regions with $\mathcal{M}_s>1.5$ and $\mathcal{M}_s<0.5$ (in cyan and red, respectively) on the mean {\HI} column density map. 
These regions tend to cluster along or near the edges of the cirrus filaments and patches, which suggest transition of cirrus morphology from more compact structures to diffuse ones.

\section{Fourier statistics} \label{sec:result_fourier}

{Although local intensity statistics probe where the field becomes non-Gaussian and provide a localized picture of cirrus structures, a global picture of \textit{how} the structures are distributed across scales is missing. 
Structural analysis in Fourier space is better suited to extracting such information. Interpretation of the power-law slope can be directly linked to turbulence theories with supports from other investigations in the literature.}

In this section, we show the results of Fourier‑domain statistics, as described in Sec~\ref{sec:method_fourier}. 
We present the power spectrum and $\Delta$-variance of the optical cirrus in the example dataset. 
We also compute the results from other dust tracers for comparison with optical cirrus, including cross-correlations. 

\subsection{Power spectrum analysis} \label{sec:result_ps}

\subsubsection{A unified power spectrum across dust tracers} \label{sec:result_ps_plot}

The power spectra were computed following the procedures in Sec~\ref{sec:method_ps}. 
The fitting was performed using \texttt{Capfit},\footnote{\url{https://www-astro.physics.ox.ac.uk/~cappellari/software}} for robust non-linear least square fitting. The power indices of the signal term and noise term from the fitting and their estimated uncertainties are listed in Table~\ref{table:spec}. 
The ISM terms for the optical, FIR, and MIR maps have power index values ranging from $\gamma=-2.82$ to $-3.00$, which is comparable to values found using IRAS 100~$\mu m$ data (\citealt{1992AJ....103.1313G}, \citealt{2002A&A...393..749M}, \citealt{2007A&A...469..595M}) and using optical data from CFHT ($\gamma \sim -2.9$; \citealt{MivilleDeschene2016}). 
The power index for the Planck radiance lies near the lower (steeper) end of this range.
For the {\HI} LVC map,  $\gamma=-3.00$, which is also in accordance with the values reported \citet{2017ApJ...834..126B}.

The index of the sky contamination (noise) term, however, shows differences among the three maps in which it was included in Eq.~\ref{eq:ps}. The optical map has a nearly flat noise term ($\beta\sim -0.1$), whereas the FIR has a gray noise closer to $\beta\sim-1$. This difference is likely due to the presence of CIBA residuals and/or imperfect source removal in the FIR maps. We find a flatter dependence for the MIR ($\beta\sim -0.3$).  
\cite{MivilleDeschene2016} reported a flat noise term for WISE and a non-flat noise term for CFHT MegaCam. The difference in the optical might be attributed to differences in the sky area and source removal procedures. In any case, the variations in this noise term do not significantly affect the power index determined for the ISM component, which in our fitting is mainly dependent on the power at large scales.

\begin{table*}[!htbp] 
\caption{Power indices of the power spectrum and $\Delta$-variance slopes computed from maps of dust/gas tracers.}\label{table:spec}
\centering
\begin{tabular}{cccccc}
\hline \hline
\rule{0pt}{2ex} &  Optical & FIR & MIR & Planck Radiance & $N_{{\HI}}$\\
\hline
\rule{0pt}{2ex}$\gamma$ & $-2.88\pm0.01$ & $-2.82\pm0.01$ & $-3.00\pm0.01$ & $-2.98\pm0.09$ & $-3.00\pm0.05$ \\
$\beta$ & $-0.08\pm0.02$ & $-0.84\pm0.02$ & $-0.25\pm0.05$ & -- & -- \\
\hline
\rule{0pt}{2ex}$\alpha$ & $0.84\pm0.03$ & $0.70\pm0.01$ & $0.93\pm0.02$ & $0.72\pm0.04$ & $0.94\pm0.03$ \\
\hline
\end{tabular}
\tablefoot{%
    $\gamma$ describes the power term of the radiation from dust/gas, while $\beta$ describes the contribution from all other astrophysical sources and sky noise. $\alpha$ characterizes the $\Delta$-variance spectrum. See Eqs.~\ref{eq:ps} and \ref{Eq:delvar_power}.
    Theoretically, $\Delta$-variance slopes should be related to power indices via $\alpha=-\gamma-2$.  The listed uncertainties are from the model fitting and do not include systematic errors.
  }
\end{table*}

\begin{figure}[!htbp]
\centering
  \resizebox{\hsize}{!}{\includegraphics{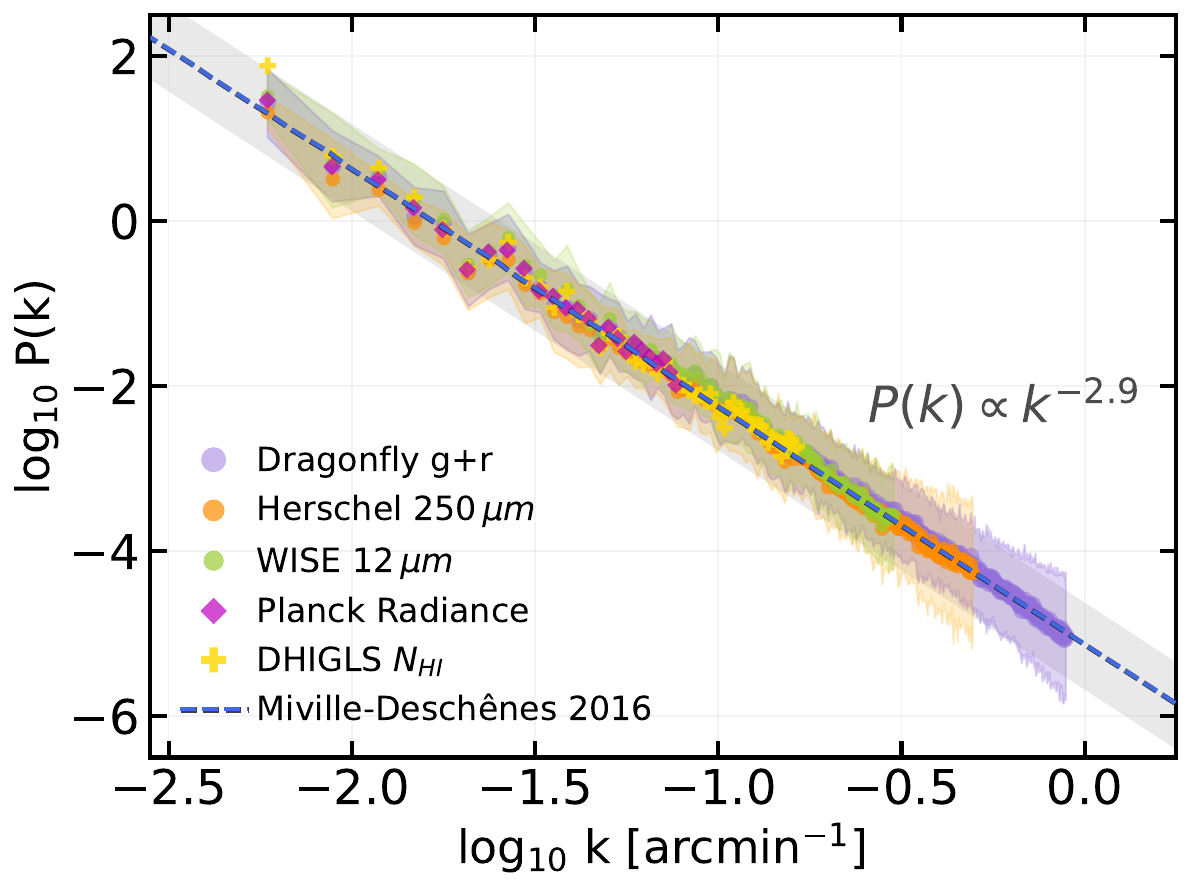}}
  \caption{Combined 1D power spectra of different dust/gas tracer maps. We only show the ISM components of the power spectra, i.e., the contaminating sky noise and instrumental noise components are subtracted.  
  The power spectra from different maps are similar, with a power index of $\gamma$ ranging from $-2.8$ to $-3.0$ (Table~\ref{table:spec}).
  Each power spectrum is displayed for a conservative range at higher $k$, excluding where the contaminating noise is most prominent and/or the cutoff by the beam is significant.
  The normalization is in arbitrary units, with the power-laws being aligned in the range where they overlap at lower $k$. 
  The power spectra are overplotted in order of increasing beam size (except FIR and MIR, guided by Fig.~\ref{fig:cps}), so that it can be appreciated that the spectrum for the Dragonfly optical (g+r) map extends to the highest $k$.
  The dashed blue line shows the result of \cite{MivilleDeschene2016} with $P(k)\propto k^{-2.9}$.}
\label{fig:ps}
\end{figure}

Figure~\ref{fig:ps} shows the combined 1D power spectra of maps of various dust tracers. We only show the ISM component (the $k^\gamma$ term in 
Eq.~\ref{eq:ps}), with other components subtracted from the total power spectra. The power spectra are vertically shifted to an arbitrary normalization. The overlapping ranges plotted are conservative relative to the available data (c.f., Fig.~\ref{fig:cps}), ensuring that the ISM model is not sensitive to details of the multi-component modeling.

In general, the power of structures decreases at smaller scales, following power-laws that are remarkably similar,
despite being derived from very different data products or wavelengths. 
The MIR map exhibits a similar power spectrum to optical and FIR even though it traces a different dust population. 
Also, the dust emission maps show power spectra consistent with that of {\HI} gas. 
This correspondence suggests that the power of density fluctuations on different structural scales, when averaged into 1D in Fourier space, is highly coherent in diffuse areas like Spider.\footnote{It should be noted, however, that such correspondence might not hold in optically thick regions such as dark nebulae or in areas that are locally heated, e.g., by shock waves or protostars.}
This is interesting considering that the tracer intensities are not always linearly correlated with each other, for example, the optical and FIR correlation becomes non-linear at high intensities. 

The fact that the power spectrum obtained from the Dragonfly optical imaging follows the same well-defined power law as other tracers indicates success in preserving cirrus light on large scales. By contrast, in Appendix~\ref{appendix:legacy_ps}, we show the power spectrum analysis on supplementary optical data from the DESI Legacy imaging survey (\citealt{2019AJ....157..168D}), where the background systematics suppress or remove cirrus light on large scales and thus bias the power spectrum measurements. 

\subsubsection{{Insights on turbulence in the diffuse ISM}} \label{sec:result_ps_discuss}

The shape of the power spectrum provides valuable information about turbulent flows. Previous studies revealed that the slope of the power spectrum correlates with $\mathcal{M}_s$ (\citealt{2005ApJ...630L..45K}, \citealt{Burkhart2009}). 
In Sect.~\ref{sec:Mach_map}, our results using local intensity statistics indicate that most regions in this field are subsonic or transonic ($\mathcal{M}_s<1.5$). A subsonic isothermal compressible turbulent flow is expected to exhibit a power spectrum of $\gamma=-11/3\approx-3.7$ in the column density distribution (\citealt{2014A&A...567A..16S}), which is steeper than the $\gamma\sim2.8-3$ slope we observed and those reported in other observations in the diffuse ISM (e.g., \citealt{MivilleDeschene2016}). In contrast, supersonic compressible turbulence can produce much shallower power spectra than the subsonic regime (\citealt{2005ApJ...630L..45K}). 
This discrepancy can be reconciled if the atomic gas in the local diffuse ISM is thermally bistable (i.e., non-isothermal), without requiring supersonic turbulence (\citealt{2014A&A...567A..16S}). \cite{MivilleDeschene2016} found that the 21 cm line emission in their data indicated thermally bistable {\HI}, and concluded that the $\gamma\sim-2.9$ spectra found in their results can be caused by thermal instability in {\HI}. The power spectra in our results, combined with the results in Sect.~\ref{sec:Mach_map}, support the scenario in which thermal instability could play a non-negligible role in shaping the cirrus structures at intermediate to high Galactic latitudes.\footnote{In subsonic/transonic and mildly supersonic regimes, the slope of the power spectrum also shows dependence on the strength of the magnetic field, where super-Alfv{\'e}nic turbulence ($\mathcal{M}_A>1$) exhibits flatter power spectra than sub-Alfv{\'e}nic turbulence ($\mathcal{M}_A<1$; \citealt{Burkhart2010}).  Consequently, magnetic fields further complicate the formation and regulation of the density structures.}

In addition to the slope, the presence of any characteristic scale would provide useful information. One would expect a loss of power below certain  scales if the turbulent energy is dissipated at small spatial scales (\citealt{2016A&A...589A..24N}). Such an imprint of dissipation was not observed down to 0.01 pc in the analysis of \cite{MivilleDeschene2016} using deep CFHT imaging of a high latitude region. Owing to the large pixel size of the Dragonfly, we are not capable of detecting such a break in the power spectrum at these small scales. However, the recently launched \textit{Euclid} mission, with its unprecedented resolution of $0\farcs1$ and superb sensitivity at low surface brightness, will shed new light on the turbulent nature of the ISM.

\subsection{$\Delta$-variance analysis} \label{sec:result_delta}

\subsubsection{{An alternative view of coherence in cirrus structures across scales}} \label{sec:result_delta_plot}

$\Delta$-variance is a variant of the power spectrum that is more robust at field edges and bad/missing pixels. We computed $\Delta$-variance as a function of angular scale as described in Sect.~\ref{sec:method_delvar} using Dragonfly optical, WISE MIR, Herschel FIR, Planck radiance, and DHIGLS $N_{{\HI}}$ data.

\begin{figure}[!htbp]
\centering
  \resizebox{\hsize}{!}{\includegraphics{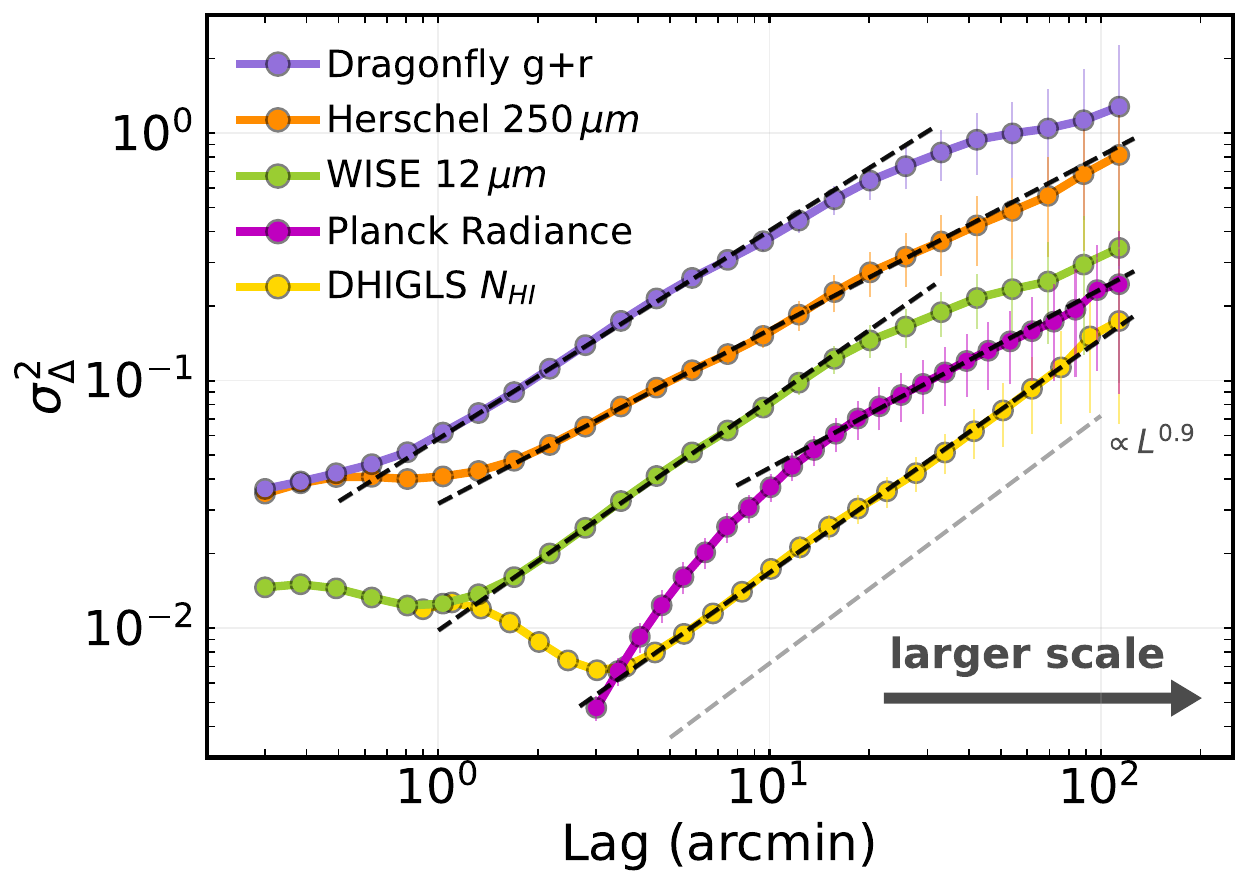}}
  \caption{
  $\Delta$-variance spectra of different dust tracer maps.
  The ``lag'' ($L$) corresponds to the structure scale and $\sigma_\Delta^2$ quantifies the variance of density fluctuations. 
  WISE, Planck, and {\HI} results are multiplied by a factor of $10^4$, 5, and $5\times10^{-3}$, respectively, for this visualization. 
  A similar coherence, characterized by a power-law scaling  $\sigma_\Delta^2 \propto L^\alpha$,  is revealed by the different dust tracers over a wide range of scales.
  A linear model is fit (dashed black lines) to find the slope, in the range $\alpha=0.7-1$. 
  At very small and large scales, the coherence is altered by map-specific systematics.} 
\label{fig:delvar}
\end{figure}

Figure~\ref{fig:delvar} shows the $\Delta$-variance spectra of the various cirrus maps. 
In general, $\Delta$-variance increases with the structure scale over a wide range, largely following power laws. This is a hallmark of the self-similarity of the density structures, indicative of the strong coherence between structures at different scales and the fractal nature of cirrus (\citealt{2021MNRAS.508.5825M}). Similar to power spectra, such power-law behavior is expected from interstellar turbulence in the ISM. The linear portions of the spectra were fit with a power law (Eq.~\ref{Eq:delvar_power}), as indicated by the dashed black lines. The fitted power-law slopes ($\alpha$) are listed in Table~\ref{table:spec}, which ranges from 0.7 to 0.94. Recall that theoretically the power index of the power spectrum is related to the slope of the $\Delta$-variance spectrum via $\gamma=-\alpha-2$; hence, the structural coherence revealed by $\Delta$-variance has indeed a quantitative likeness to that derived from the power spectrum analysis.

\subsubsection{{What breaks the structural coherence at small and very large scales?}} 

{The $\Delta$-variance spectra of various dust tracers follow a simple power law over a wide scale range, indicative of structural coherence across scales. However, they are affected by several factors at small and very large scales.}

On small scales ($\lesssim 1\arcmin$), the optical, FIR, and MIR spectra show clear flattening. The flattening is more pronounced and occurs at slightly larger scales in FIR and MIR than in optical, likely due to the combined beam effects (including filtering and PSF blurring) and the presence of CIBA residuals in these datasets.  In the simple case of white noise in a map, we would expect a spectrum with $\alpha = -2$, and as the beam cuts off the real sky signals in the $\Delta$-variance spectrum, a transition will be made to this steep noise-related spectrum.
The {\HI} map also shows an up-bending at small scales comparable to its beam, because of the transition to the $\Delta$-variance spectrum of the noise map from the cirrus-free end channels.
In contrast, the Planck radiance map shows steepening on scales smaller than or comparable to the Planck beam; it is derived from thermal dust modeling and the reproducibility noise is very small. 

On very large scales ($\gtrsim 0\fdg6$), $\Delta$-variance spectra of molecular clouds often display a plateau or turnover, presumably caused by wind feedback (\citealt{Ossenkopf2008}). The spectra of the diffuse Spider region do not reveal any clear characteristic scale. 
However, the MIR and optical spectra exhibit a small degree of flattening compared to other tracers. The flattening in the MIR map might result from imperfect gradient correction during the stacking process used to subtract zodiacal light and faint off-axis moon-glow (\citealt{2014ApJ...781....5M}). Alternatively, it could be a result of WISE 12~$\mu m$ tracing emission from ultra-small, stochastically heated dust grains that are distributed slightly differently from large grains. 

The flattening observed in the optical map cannot be explained by differences in dust compositions. But the flattening could stem from imperfect flat-fielding or subtle background mismatch in the mosaicking process. Boundary effects due to the finite size of the field could also influence the largest couple of scales (\citealt{2001A&A...366..636B}). 
There are other considerations with astrophysical origins. 
Optical cirrus can suffer from attenuation effects in the regime where it starts to transition from optically thin to optically thick (\citealt{2023MNRAS.524.2797M}). Several studies have revealed non-linearity in the optical-FIR correspondence in high column density regions (\citealt{2020A&A...644A..42R}, \citealt{2023MNRAS.524.2797M}, \citealt{2023ApJ...948....4Z}). However, it remains to be demonstrated that this would account for the flattening on large scales. 
This discrepancy between optical and FIR large-scale structures might relate to differences in the fractal dimension of optical cirrus in Stripe82 compared to its FIR counterpart reported by \cite{2021MNRAS.508.5825M}. 
Finally, we cannot dismiss the possibility that astrophysical sources other than dust -- in particular the cosmic optical background (COB) (e.g., \citealt{2007ApJ...666..663B}, \citealt{2017NatCo...815003Z}, \citealt{2024ApJ...972...95P}) -- also contribute faint emission on large scales. Further analysis with a larger dataset and refined data processing will be helpful in understanding the patterns observed in the optical $\Delta$-variance spectra at these scales.

\subsection{Cross-power spectra analysis} \label{sec:result_cps}

\subsubsection{{Structural correlation decoupled from systematics}} \label{sec:result_cps_plot}

A cross-power spectrum analysis enables the decoupling of map-specific systematics. We show the angular cross spectrum results computed following Sect.~\ref{sec:method_cps} using Dragonfly optical, WISE MIR, and Herschel FIR images. This cross-correlation analysis complements the comparisons of optical/FIR/MIR data based on local intensity statistics in Sect.~\ref{sec:result_local}.

We first plot the auto-power spectra for the optical, FIR, and MIR images as a consistency check; these spectra are in accordance with those shown in Fig.~\ref{fig:ps}. The main difference here is that we show the gross measurements, i.e., we do not separate the ISM components ($k^\gamma$ terms) by subtracting other components. At small scales (high $k$), below about three to four times the beam widths, the spectra deviate from a single power law due to CIBA residuals, beam effects, etc. 
The colored tick marks at the bottom of the plot denote twice the beam width for each map. 

\begin{figure}[!htbp]
\centering
  \resizebox{\hsize}{!}{\includegraphics{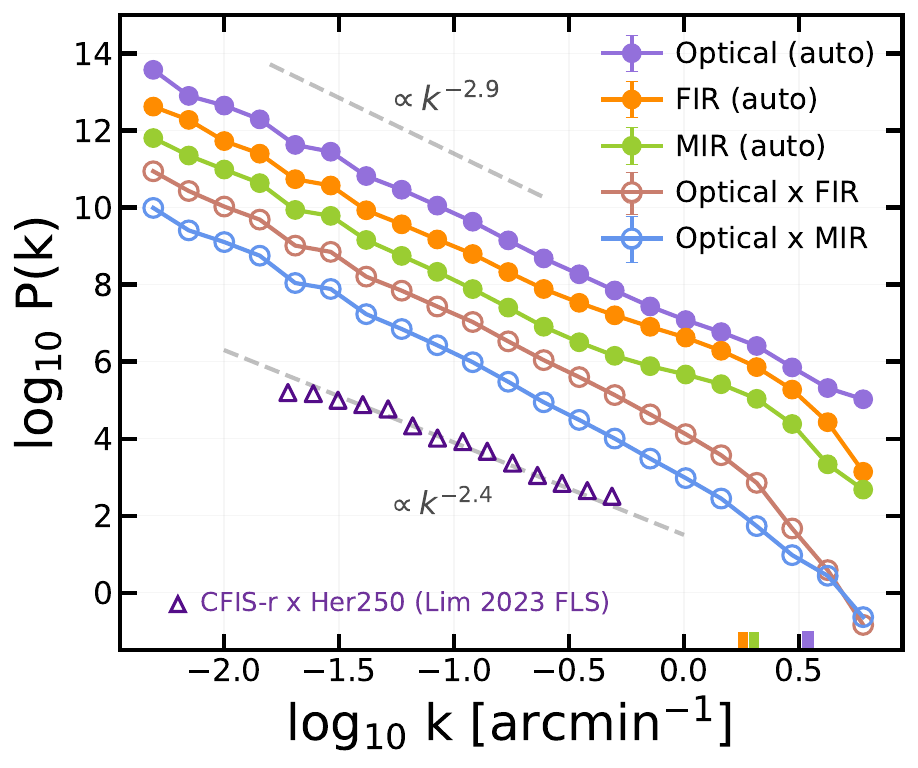}}
  \caption{Combined 1D cross-power spectra and power spectra of optical, MIR, and FIR cirrus maps. For visualization, the spectra are vertically offset, with the y-axis in arbitrary units. The colored ticks above the x-axis indicate twice the beam width of each map with the corresponding color. 
  The median error bars are indicated in the upper right for each spectrum. At larger spatial scales (lower $k$), the cross-power spectra between optical and FIR or MIR follow a power law similar to the auto-power spectra, indicating strong structural correlation. At small scales comparable to or below the beam width, the power actually declines because the non-cirrus systematics are uncorrelated between the maps.
  Purple triangles show the cross-power spectrum between the CFIS-$r$ image and the Herschel 250~$\mu m$ image of a cirrus-rich field (``FLS'') measured by \cite{2023MNRAS.525.1443L} (see text). }
\label{fig:cps}
\end{figure}

The 1D cross-power spectra between the optical image and the FIR or MIR images are shown as open markers in Fig.~\ref{fig:cps}. These measurements can be well described by a power law with an index of $\gamma=-2.87$ and $-2.95$, respectively, close to that of the auto-power spectra, demonstrating a strong correlation between structures in the optical and FIR or MIR data. These correlations persist down to scales comparable to the FIR or MIR beam widths; below these scales (higher $k$), density fluctuations become de-correlated due to differing beam effects, pixel window functions, noise properties, and other map-specific systematics. As a result, the power index measured using this approach is less sensitive to systematics and decoupled from the degeneracies with other components specific to a single map (c.f., Eq.~\ref{eq:ps}). 

\cite{2023MNRAS.525.1443L} conducted cross-correlation analyses between the Canada-France Imaging Survey (CFIS) optical images and the Herschel SPIRE images. In their work, the goal was to study the correlation between the COB and the CIB, where cirrus is a major contaminant. We plot their cross-spectrum between the CFIS $r$-band image and the Herschel 250~$\mu m$ map in the ``FLS'' field (their Fig.~13), which contains a substantial amount of cirrus (see their Fig.~5). Their cross-power spectrum is marginally shallower than ours, plausibly due to an inherent COB-CIB correlation.  Note that the CIBA has a power-law dependence $P(k) \propto k^{-1 \text{\,–\,} -1.5}$, much shallower than the Galactic component (\citealt{viero2013}), supported by their measurements in fields with minimal cirrus contamination. Therefore, it could be expected that our cross-power spectra focusing on the Galactic cirrus component, combined with efforts to mitigate extragalactic contributions by removal of point sources, yield a steeper power‐law dependence.

\subsubsection{{Which is closer to optical cirrus \textit{globally}: FIR or MIR?}}\label{sec:result_cps_ratio}

\begin{figure}[!htbp]
\centering
  \resizebox{\hsize}{!}{\includegraphics{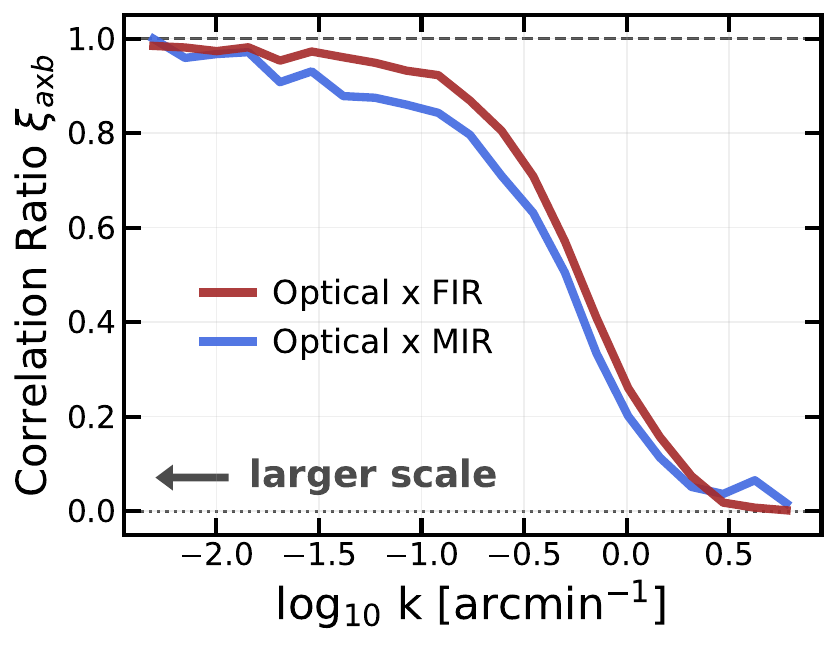}}
  \caption{Correlation ratio between the optical and FIR (red) / MIR (blue) cirrus maps as a function of spatial frequencies. For both paired maps, the correlation approaches unity at large scales, indicating strong correlation of dust structures across wavelengths. At small scales, the correlation falls to zero because of uncorrelated contamination, noise, and map-specific systematics. Notably, the FIR-optical correlation exceeds the MIR-optical correlation over a broad range of angular scales, reflecting the greater similarity between dust morphology traced by thermal emission and that traced by scattered light.}
\label{fig:cps_cr}
\end{figure}

In Figure~\ref{fig:cps_cr}, we show the correlation ratio $\xi_{a \times b}$ between the optical and FIR or MIR data computed using Eq.~\ref{eq:cpr}. For both paired maps, the correlation approaches unity at large spatial scales (low $k$), indicating the concordance of dust structures traced by different radiative mechanisms. At small scales, the correlation declines toward zero due to uncorrelated noise and systematics inherent to each map. It is worth noting that the FIR–optical correlation remains higher than the MIR–optical correlation across a wide range of scales. This result, from a global perspective, complements our findings in Sect.~\ref{sec:result_local} at local scales (note that the smallest kernel scale used there is $d_k=2\arcmin$) indicating that optical cirrus morphology more closely resembles that of FIR cirrus than MIR cirrus.  This quantifies what was anticipated in Sect.~\ref{sec:isrf}.

\section{Quantitative morphology of cirrus with wavelet scattering transforms}
\label{sec:result_wst}

{Local intensity statistics probe local non-Gaussianity but ignore multi-scale coupling, while the angular power-spectrum (and its variants) describe global coherence across scales but discard phase and non-Gaussian information. Is there a statistical approach covering both strengths? 
With this motivation, we explore the potential of the wavelet scattering transform. One key hurdle is that the physical interpretation of the full set of WST coefficients is not straightforward. However, we take advantage 
of the results in Sect.~\ref{sec:result_local} and Sect.~\ref{sec:result_fourier} as a basis for
extending our understanding, and 
use summary statistics that encapsulate the rich morphological information encoded in the coefficients.}

Below we show as morphological measures the summary statistics computed from the second-order WST coefficients using optical and IR maps. The results for the Planck radiance map and the {\HI} map are presented in Appendix~\ref{appendix:wst_Planck+HI}. We then explore the patterns of WST coefficients and summary statistics by dividing the field into subregions to study ensemble properties. Finally, as an application we show that mock cirrus can be generated using WST outputs from observations.

\subsection{WST statistics as morphological measures}
\label{sec:result_wst_optical_IR}
We start with the two summary statistics, sparsity $s_{21}$ and linearity $s_{22}$, defined by Eqs.~\ref{eq:sparsity} and \ref{eq:linearity}
and derived from the WST coefficients computed from the Dragonfly optical, Herschel FIR, and WISE MIR maps using $J=10$ and $\Theta=8$. These statistics are dimensionless and invariant under translation and rotation.
Relative to the power spectrum and $\Delta$-variance results in Sect.~\ref{sec:result_fourier}, the WST statistics incorporate additional information about the non-Gaussianity of the structures. Relative to the results based on local intensities in Sect.~\ref{sec:result_local}, they provide a more compact description of structure coherence across different scales. 

\begin{figure}[!htbp]
\centering
  \resizebox{\hsize}{!}{\includegraphics{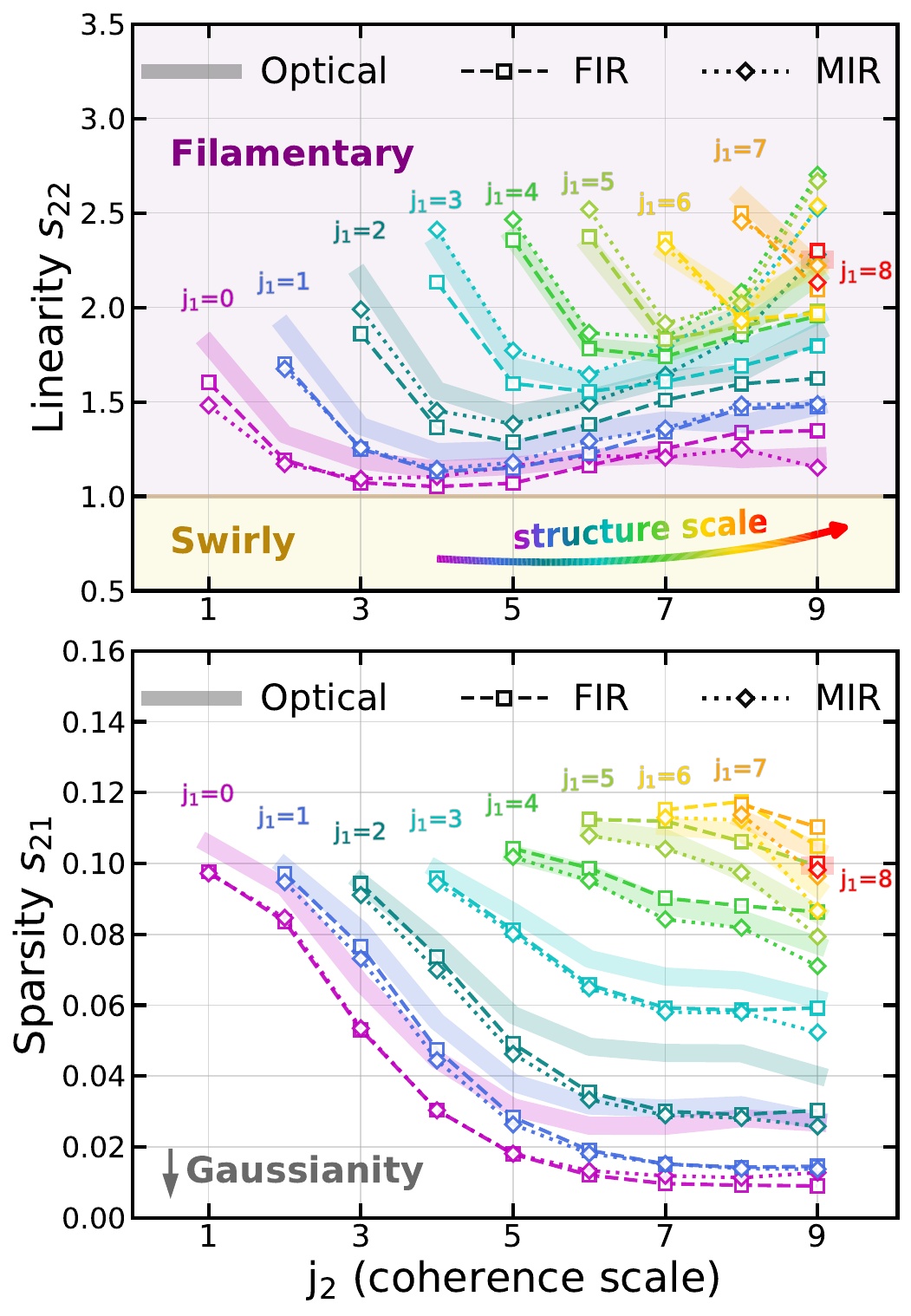}}
  \caption{Linearity $s_{22}$ and sparsity $s_{21}$ of optical, FIR, and MIR cirrus maps with different ($j_1$, $j_2$) combinations, which are summary statistics quantifying the non-Gaussianity of density fluctuations in the field. The structure scale is color coded by $j_1$. The coherence scale $j_2$ characterizes the clustering on a scale $j_2$ of $j_1$-scale patterns. 
  Only $j_2>j_1$ terms are shown because $j_2 \leq j_1$ terms are not physically significant. 
  Values of $s_{22}>1$ correspond to thin, filamentary structures, wheras values of$s_{22}<1$ correspond to swirly, curvy structures. Lower $s_{21}$ indicates that the structures are more widely spread, whereas $s_{21}=0$ is the expectation for a Gaussian random field. In general, $s_{22}$ and $s_{21}$ of optical, FIR, and MIR cirrus maps have similar patterns except for small-scale structures.}
\label{fig:wst_stats}
\end{figure}

Figure~\ref{fig:wst_stats} illustrates the hierarchical patterns of $s_{22}$ (upper panel) and $s_{21}$ (lower panel) obtained from the three maps for different ($j_1,j_2$) combinations. Here $j_1$ represents the structure scale (indicated by different colors) and $j_2$ (plotted on the x-axis) represents the coherence scale that characterizes the assembly or clustering of $j_1$-scale structures separated by $j_2$-scale. We display only terms with $j_2 > j_1$ because the terms with $j_2 \leq j_1$ are mainly determined by the properties of wavelets and thus do not carry significant physical information about the density field.

Across all scale combinations, the structures consistently exhibit $s_{22}>1$, indicating that the cirrus morphology is predominantly thin and filamentary. 
At a given coherence scale $j_2$, larger structures (i.e., those with higher $j_1$) generally exhibit higher linearity compared to smaller structures (with the exception of the largest scale, $j_1=8$). This suggests that cirrus morphology appears to be more elongated on larger scales, which can be interpreted as a consequence of the structure growth driven by the cascade of interstellar turbulence. 
Sparsity values are low, ranging from 0.01 to 0.12, indicating that cirrus is mostly diffuse and widespread across the field. 
At a given coherence scale $j_2$, larger structures exhibit higher sparsity (except for the largest scale).  This suggests that large structures are more ``localized'', i.e., occupying specific portions of the field.

Overall, both the FIR map and MIR map display trends highly consistent with the optical map, reinforcing the overall similarity in dust morphology. However, deviations from the optical data occur at specific scale combinations. 
For example, for both FIR and MIR maps, small-scale structures ($j_1\leq 2$) are slightly less filamentary at comparable coherence scales ($j_2=j_1+1$), and are more evenly distributed (lower sparsity) at large coherence scales ($j_2 \geq 6$). 
Additionally, for intermediate-large structures ($j_1=5$ and $6$) the FIR data appear to have slightly higher sparsity on larger coherence scales, while the MIR data show more deviation to higher linearity than the optical at the largest coherence scales. 
Note that FIR and MIR data have larger beam widths {(18\arcsec and 15\arcsec, respectively, in Table \ref{tab:beampix}). This would} smooth structures with $j_1=0$ {that correspond to $2^{j_1+1}=2$ pixels}, which would lower both linearity and sparsity. However, this factor alone could not explain the observed difference at larger scales, where the linearity is higher. We attribute these differences to residuals of {extragalactic emission}. For example, the CIBA can leave large-scale imprints across a wide range of structure scales (\citealt{2024A&A...681A...1A}). The presence of COB and/or its distinction from CIB may also contribute to the differences in these hierarchical patterns between optical and IR. {We note that in a different field, other diffuse extragalactic emission such as ICL and intragroup light may also contribute, if not well deblended from the cirrus foreground (see discussion in Sect.~\ref{sec:discuss_exgal}).}

Comparisons with Planck and {\HI} maps are provided in Appendix~\ref{appendix:wst_Planck+HI}. In general, the WST statistics computed from the Planck radiance map show trends similar to those of the optical map, while the {\HI} map exhibits deviations. We speculate that the deviations may result from a combination of the differences in data processing and the possible decoupling between gas and large dust grains; further analysis with larger datasets is needed to determine whether these deviations persist and to assess their physical implications.

One application of this method is the evaluation of statistical similarity between two fields. To be considered realistic, mock cirrus images, for example produced by generative models, should exhibit WST statistics with patterns comparable to what is observed. Further applications are discussed in the following sections. 

\subsection{A continuous representation of changing cirrus morphology}
\label{sec:result_wst_morph}

In this section, we explore the ensemble properties and variations of the cirrus morphology across the field. To achieve this, we divide the field into small subregions, each with a size of $45\arcmin \times 40\arcmin$, as shown in Figure~\ref{fig:spider_patch} in Appendix~\ref{appendix:spider_patch}. This division respects a compromise between covering a sufficiently large area for structural analysis and obtaining a reasonable number of samples. We computed the WST coefficients for all subregions $\{I_k\}$ (with $k=1$ to 30) using $J=7$ and $\Theta=8$.

Subsequently we calculated the sparsity $s_{21}$ and linearity $s_{22}$ for each subregion at different ($j_1$, $j_2$). The patterns are consistent with the patterns shown in Fig.~\ref{fig:wst_stats}.
The statistics $s_{21}$ and $s_{22}$ calculated for each subregion also show correlations at several scale combinations (see below). 

\begin{figure}[!tbp]
\centering
  \resizebox{\hsize}{!}{\includegraphics{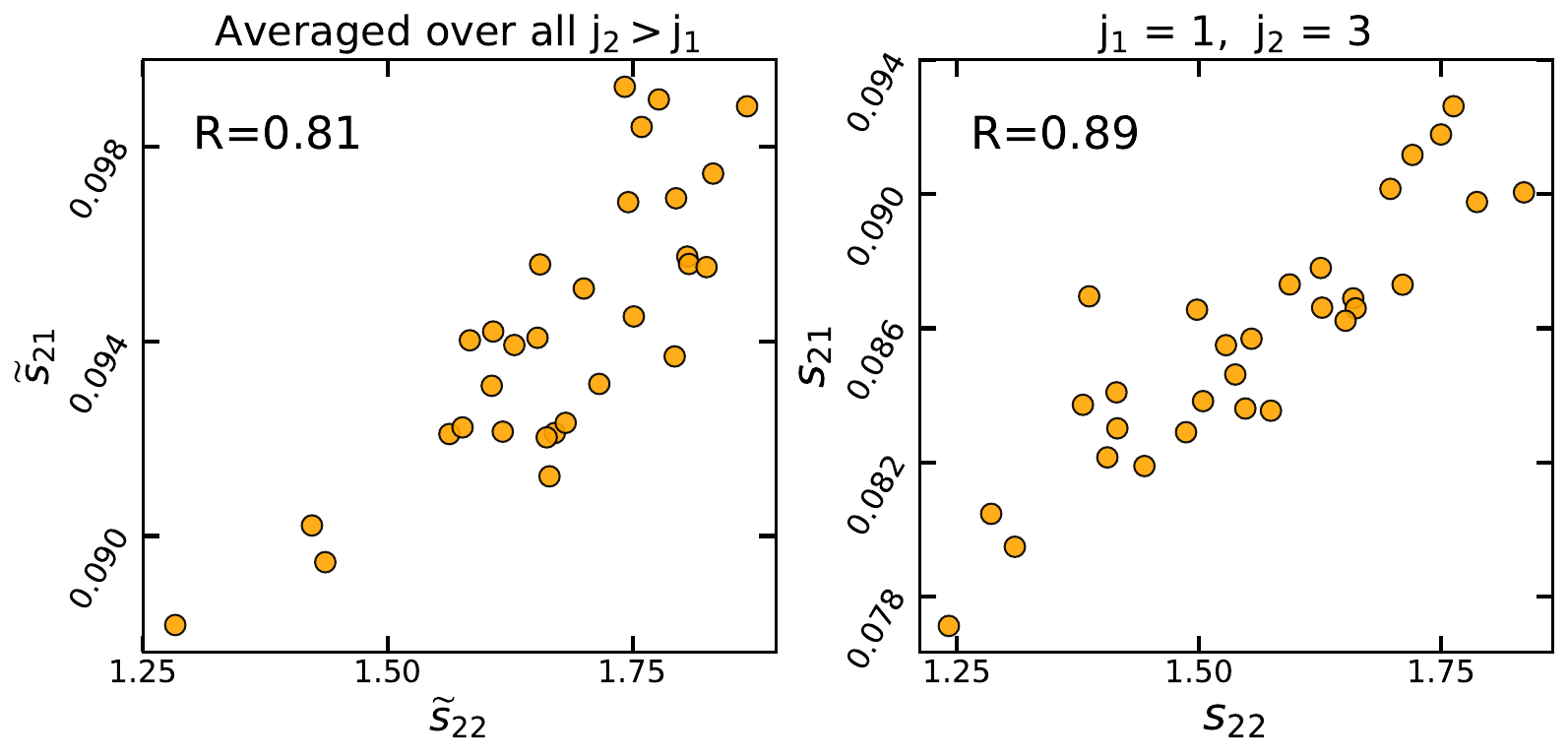}}
  \caption{Scatter plot of sparsity vs. linearity for subregions of cirrus in the example Spider field. Each marker represents a subregion within which the statistics have been evaluated. The left panel shows the distribution of reduced statistics, $\tilde{s}_{21}$ and $\tilde{s}_{22}$, after a weighted average over by all scale combinations with $j_2 > j_1$. The right panel shows $s_{21}$ vs. $s_{22}$ for one of the combinations with strong correlations ($j_1=1$, $j_2=3$). In both cases, the sparsity and linearity show strong correlations quantified by high Pearson correlation coefficients. Furthermore, some scale combinations are more relevant than others in describing the continuous changes in cirrus morphology.}
\label{fig:lin_vs_spa}
\end{figure}

We further calculated the normalized statistics, denoted as $\tilde{s}_{21}$ and $\tilde{s}_{22}$, by averaging over all scale combinations with $j_2 > j_1$, with the average weighted by the inverse variance of the corresponding distribution among subregions. The left panel of Fig.~\ref{fig:lin_vs_spa} shows the distribution of $\tilde{s}_{21}$ and $\tilde{s}_{22}$ for all of the subregions, revealing a clear correlation between these statistics. The Pearson correlation coefficient $R$ for $\tilde{s}_{21}$ versus $\tilde{s}_{22}$ is $R=0.81$, indicating a strong correlation among subregions. We note that this scale-averaged correlation is mainly driven by specific scale combinations; in particular, the correlations are strong for small coherence scales ($j_2\leq2$) or structures with comparable scales ($j_2-j_1\leq3$), but are weak or absent at larger scales ($j_1\geq 4$). 
For illustrative purposes, in the right panel of Fig.~\ref{fig:lin_vs_spa}, we show the scatter plot of $s_{21}$ and $s_{22}$ for $j_1=1,\,j_2=3$ , which has a Pearson correlation coefficient $R$ of 0.89, indicating a strong correlation in play at this scale combination, stronger than for the averages plotted in the left panel. It can therefore be inferred that not all scales are relevant to describe the cirrus morphology
with some scale combinations showing stronger correlations than others.

\begin{figure}[!tbp]
\centering
  \resizebox{\hsize}{!}{\includegraphics{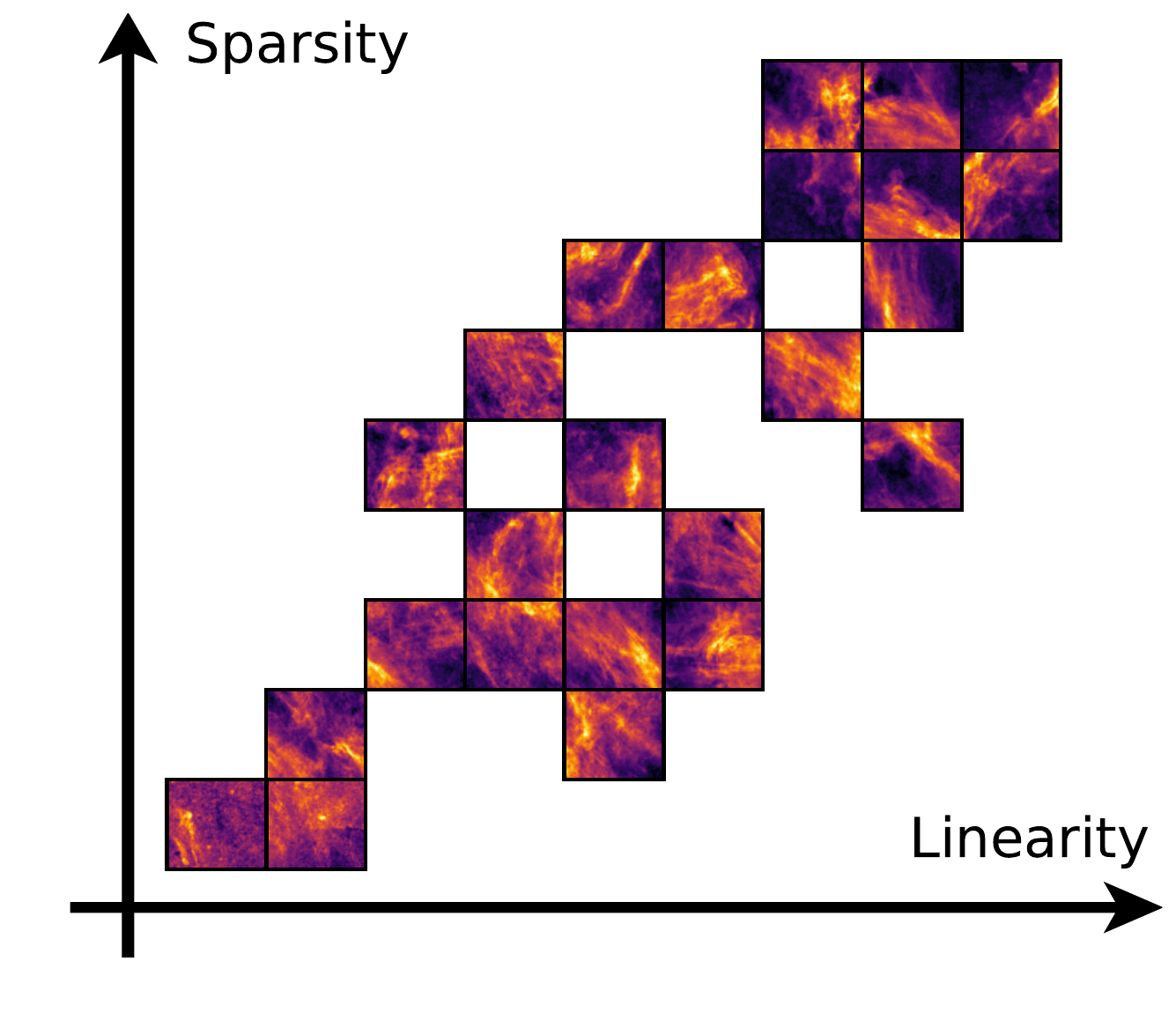}}
  \caption{Illustration of sparsity vs. linearity for subregions in the example field showing the cutout images, averaged over by all scale combinations with $j_2 > j_1$. Note that the image cutouts are sorted by their positions on the left panel of Fig.~\ref{fig:lin_vs_spa}, so the original layout does not strictly hold. In general, the patterns of the cirrus present a discernible trend varying with the two WST summary statistics, showing a regulation in cirrus morphology.}
\label{fig:lin_vs_spa_mosaic}
\end{figure}

In Fig.~\ref{fig:lin_vs_spa_mosaic}, we display the subregions sorted by their positions on the left scatter plot of Fig.~\ref{fig:lin_vs_spa}. Although the scatter plot layout is not strictly preserved, this representation effectively traces the correlation and visualizes the variation in cirrus morphology. In general, there is a discernible trend from the lower left to the upper right of the plot. This correlation between sparsity and linearity is a quantification of the visual impression that when cirrus structures are more concentrated, they tend to appear more filamentary, and conversely, when structures are more evenly distributed, they appear more diffuse. 
This morphological trend could be a result of the interplay between local ISM conditions and physical mechanisms.
For example, \cite{2022ApJ...937...81T} found that CNM exhibits more small-scale structures than WNM using data from DHIGLS.
Likewise, \cite{2023ApJ...947...74L} found that the fraction of CNM in {\HI} is correlated with small-scale linearity, indicating that CNM is more populated by {\HI} filaments. Future studies with larger datasets and multiple tracers may further elucidate the connection between cirrus morphology and physical conditions in the ISM.

From another perspective, this representation from WST analysis provides a two-dimensional characterization of the non-Gaussianity of each image, organizing various cirrus morphologies in a continuous and interpretable manner without requiring training or classification. Such representation is similar to the output of deep learning methods such as convolutional neural networks (CNNs). In fact, one interesting aspect of this approach is that it offers insights into the ``blackbox'' embedded in CNNs. As discussed in \cite{2021arXiv211201288C}, many elements of the WST are analogous to, or simplification of, features characteristic of a CNN. In effect, the WST bridges the gap between traditional power spectrum methods and CNNs. Using pre-defined convolutional kernels, the WST can be viewed as a non-trainable shallow CNN (\citealt{2021arXiv211201288C}), which circumvents potential overfitting issues while simplifying the training process, architecture, and interpretable bottleneck associated with CNNs.

\subsection{Application: statistical synthesis of mock cirrus}
\label{sec:result_wst_syn}

One interesting application is generation of mock cirrus fields. Such synthesis can be useful for testing detection and identification algorithms for extragalactic low surface brightness phenomena as well as providing estimates of bias on their measurements, where cirrus becomes one of the major sources of confusion.

We demonstrate using the same subregions of optical cirrus observations as shown in Fig.~\ref{fig:spider_patch}, Appendix~\ref{appendix:spider_patch}. For computational efficiency, we binned the subregion [4x4] and cropped the binned image into the same dimensions in height and width. This leads to a [100x100] $\rm pix^2$ image cutout corresponding to $40\arcmin \times 40 \arcmin$. To avoid artifacts caused by the FFT run on non-periodic density fields, we folded the pixel values at the boundaries to make the image twice as large as the original cutout and become periodic at the boundaries. As described in detail in Appendix~\ref{appendix:wst_syn}, the image synthesis is performed to match the scattering covariance computed from WST run on these cutouts.\footnote{The computation time for a [100x100] cutout is around 30s with T4 GPU provided by Colab and around 2 min with CPU on a laptop.} {A synthesis can be performed with WST coefficients only, but we use scattering covariance so that cross-scale structure couplings are reproduced as well.}
In brief, the scattering covariance is obtained by doing a cross-correlation of WST map outputs (wavelet-convolved footprints), which further captures scale interactions across different wavelets. 
{We adopt the calculation implemented in the \texttt{scattering} package (\citealt{Cheng2020, Cheng2023}).}

\begin{figure}[!htbp]
\centering
    \resizebox{\hsize}{!}{\includegraphics{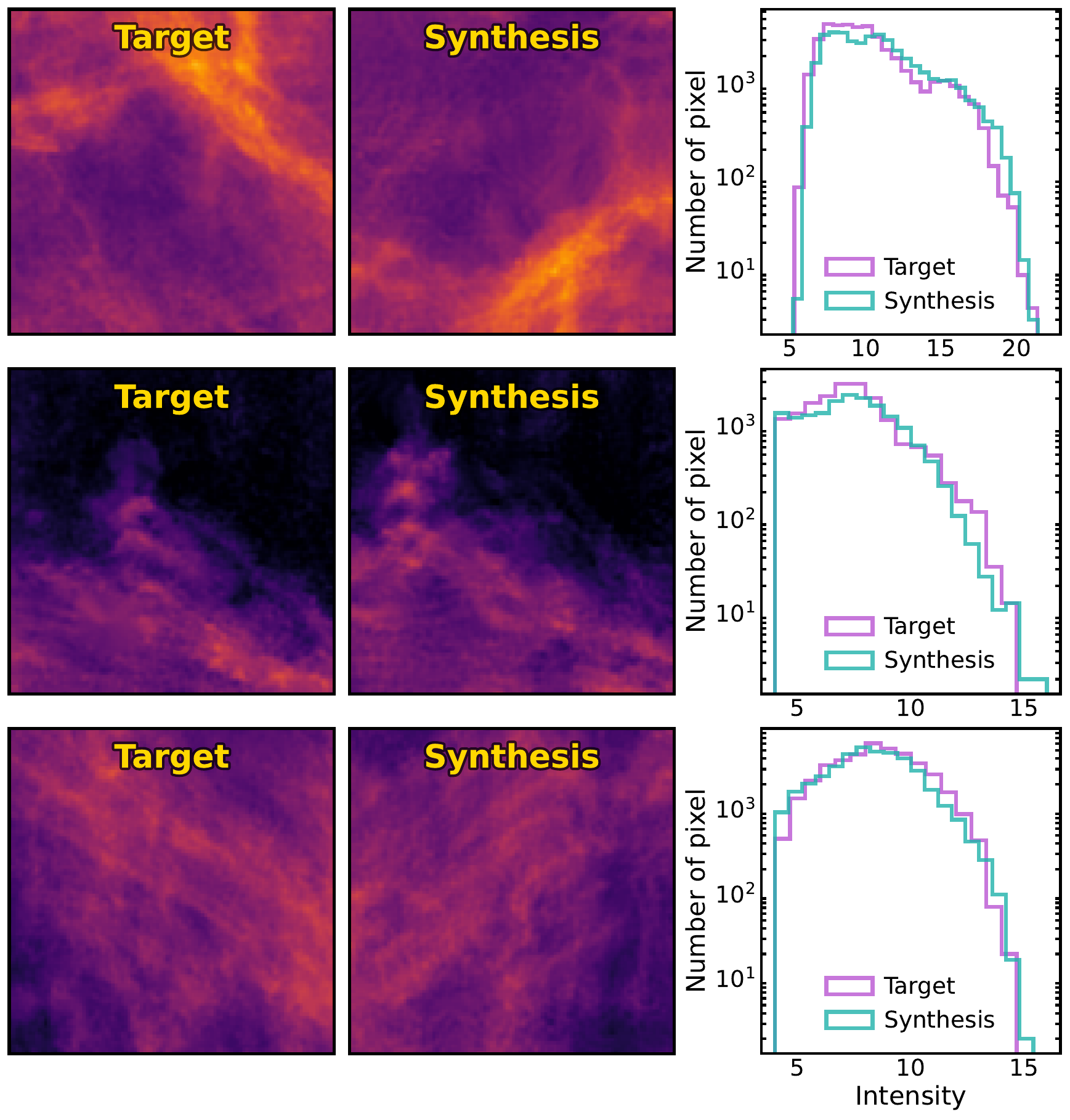}}
    \caption{Mock cirrus using scattering covariance computed from $40\arcmin \times 40 \arcmin$ subregions of the example dataset (from top to bottom: \#13, \#20, and \#21 in Fig.~\ref{fig:spider_patch}). The images are moderately filtered at high frequency to remove minor artifacts involved in the FFT process. The right panels show the histograms of the intensity distributions of target and mock cirrus.}
\label{fig:spider_syn}
\end{figure}

Here we demonstrate one-to-one synthesis by generating a mock cirrus field from a single cutout. An illustration of the image synthesis is presented in Figure~\ref{fig:spider_syn}, where the mock cirrus is displayed with the data from which it is generated side-by-side. We show three examples of realizations from top to bottom with different cirrus morphologies.
The visual similarity between the mock and the data is striking. In addition, the right panels show the histograms of the intensity distributions of the data and the mock cirrus, indicating good correspondence in intensity PDFs. By construction, the non-Gaussian statistics of the field characterized by the WST match up to the second order, which can be extended to higher orders at the cost of computational time.  

Although generative cirrus fields can be useful for image simulation tests, it should be noted that these synthesized images are purely phenomenological, i.e., no physical inference is involved and no new information is produced. The quality of the synthesis depends on several factors including the initial condition, the optimization, the parameterization, the data quality (e.g., surface brightness limits), and data processing effects (e.g., sky modeling, removal of foreground/background sources). Nevertheless, the fact that the mock cirrus fields look visually similar to the data strongly demonstrates the power of the WST approach in characterizing the cirrus morphology.

\section{Can extragalactic light be distinguished from cirrus via morphology?} \label{sec:discussion}

{We have demonstrated how a complementary suite of morphological diagnostics can robustly characterize the filamentary, non‑Gaussian signatures of optical Galactic cirrus. An interesting application is whether these signatures can be used to distinguish extragalactic light from cirrus. Of the various phenomena, we choose tidal features for demonstration because cirrus is widely recognized as significant confusion to tidal features given their visually similar wispy morphology. If one of our statistical tools can reliably differentiate in this challenging regime, it will hold potential for other emissions such as galactic halos and ICL.}

\subsection{{Identification of tidal features injected into cirrus-rich sky areas}}
\label{sec:discuss_tidal_inj}

\cite{2020A&A...644A..42R} showed that it is possible to use color differences to distinguish extragalactic stellar populations from cirrus. It is interesting to explore the feasibility of distinguishing between tidal tails and cirrus based solely on their morphological characteristics. 
The underlying motivation lies in the different physics, tidal structures being molded by gravitational interactions, while cirrus structures are primarily shaped by turbulence, thermal instability, and (modulated by) magnetic fields. As tidal tails orbit the host galaxy, they tend to present different curvatures (e.g., arcs, loops, and pretzel-like patterns) compared to those manifested by cirrus, which can be distinguished, though subjectively, by trained human eyes (e.g., \citealt{2025A&A...701A.182M}). 
The following proof-of-concept experiment aims to identify the presence of tidal structures in cirrus-rich regions by leveraging differences in the statistical properties of the two types of features. 
Here we experiment with a single-band WST analysis using the $g+r$ image, which does not require supplementary datasets.

\begin{figure}[!htbp]
\centering
  \resizebox{\hsize}{!}{\includegraphics{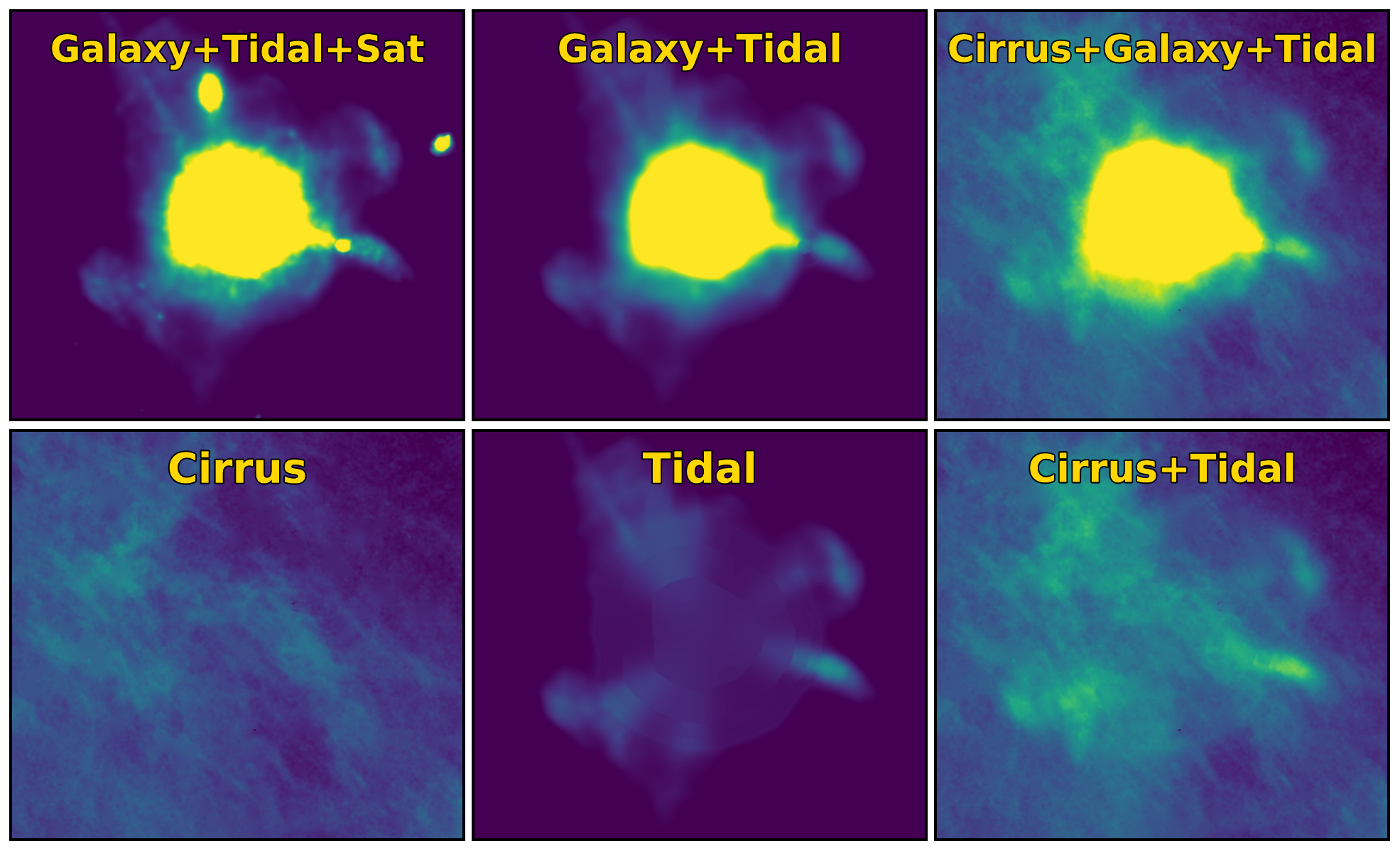}}
  \caption{Illustration of tidal tail injection into cirrus patches (see Appendix~\ref{appendix:inj_tidal}). \textbf{Top}: (from left to right) (a) simulated galaxy from TNG 50 with tidal features (for DECam). (b) same galaxy with satellites and density peaks removed, after convolution to Dragonfly PSF. 
  (c) galaxy with tidal tails injected into the cirrus subregion.
  \textbf{Bottom}: (from left to right)
  (d) example subregion of cirrus (\#22 in Fig.~\ref{fig:spider_patch}). 
  (e) tidal tails, excluding the central galaxy. 
  (g) tidal tails injected into the cirrus subregion.
  }
\label{fig:tidal_inj}
\end{figure}

We performed an injection test using realistic simulated tidal tail structures, which is illustrated in Fig.~\ref{fig:tidal_inj}. The simulated tidal structures were generated by \cite{Miro2025}. Details about the generation of mock galaxies and the injection process are described in Appendix~\ref{appendix:inj_tidal}. WST coefficients were computed for all subregions with tidal feature injections using the same scale indices ($J=7$, $\Theta=8$). Instead of reducing the coefficients into summary statistics as in Sec~\ref{sec:method_wst}, we examined the full set of the second-order coefficients -- after averaging over $\theta_1$ (Eq.~\ref{eq:wst_coef_iso}) -- yielding a total of $J\,(J-1)\,\Theta/2=168$ coefficients. We then employed dimensionality reduction techniques to extract the features that may distinguish subregions with tidal tails. The following investigations are motivated by \citealt{2021ApJ...910..122S}, who incorporated WST with dimensionality reduction to classify ISM structures in magneto-hydrodynamical simulations. Specifically, we applied principal component analysis (PCA) and linear discriminant analysis (LDA) on the WST coefficients.

We begin with PCA, a classic technique that finds the orthogonal components in a low-dimensional space that explain maximal variance. 
The input consists of the $N=30$ subregions from the original cirrus field, each described by 168 variables,\footnote{We did not include subregions with tidal injection in the training set of PCA because in reality tidal features occur much less frequently. Therefore, the WST-PCA space in cirrus-rich sky areas should be dominated by statistical properties of cirrus and we want to look for outliers.} which are scaled to have zero mean and unit variance.
This WST-PCA analysis allows the WST coefficients to be described by the first few PCA components, with one, two, and three components accounting for 68\%, 84\%, and 90\% of the variance, respectively, which indicates that the statistical properties of cirrus morphology can be effectively represented in a low-dimension space. 

\begin{figure}[!htbp]
\centering
  \resizebox{0.95\hsize}{!}{\includegraphics{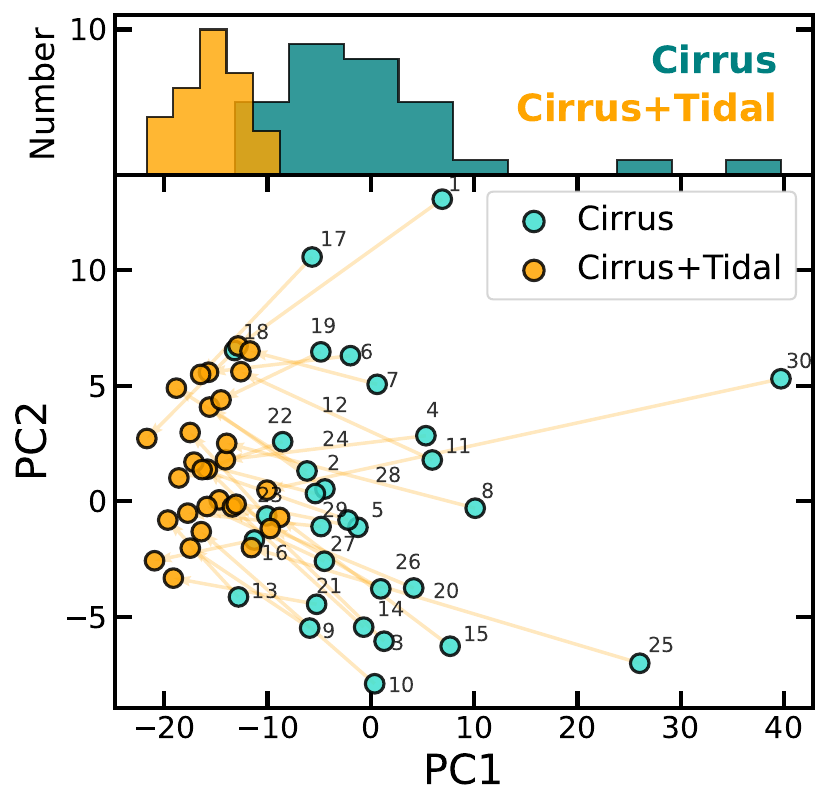}}
  \caption{Projection of the WST coefficients onto principal components PC1 and PC2 for
  subregions with (orange) and without (cyan) tidal tail injection. The top panel shows the histogram with respect to PC1. 
  The PCA is trained with cirrus-only subregions. 
  Even with only one component, the histograms for subregions with and without tidal tails show clustering with different peaks, although there is some overlap. 
  This supports the feasibility of using dimension reduction on WST coefficients to distinguish tidal features from cirrus.}
\label{fig:wst_pca_tidal}
\end{figure}

We then projected the WST coefficients of subregions with and without tidal tail injections onto those principal components. Figure~\ref{fig:wst_pca_tidal} shows the projection onto the first two principal components, PC1 and PC2. Each data point represents a subregion -- with or without tidal injection -- labeled according to Fig.~\ref{fig:spider_patch}. The two outliers (\#25 and \#30) correspond to regions with very little cirrus, which we designate as ``cirrus-free''.
In this WST-PCA space, subregions with tidal tails are shifted uniformly to the left in PC1 but up and down in PC2.
This is not unexpected, because PCA does not aim to maximize separation between groups.
The top histograms with respect to PC1, which explains 68\% of the variance, shows a distinct separation of peaks in an unsupervised manner.
There is some overlap of the histograms (including \#25 and \#30 with tidal tails added). 
Increasing the surface brightness of the tidal features would shift the points further to the left, making them more distinguishable.
%
One potential application, to search for tidal feature candidates entangled with cirrus in large surveys, is to examine the tail/outliers of this distribution in the WST-PCA space.
In general, this supports the feasibility of using dimension reduction on WST coefficients to distinguish tidal features from cirrus, or at least to establish the presence of tidal features despite the contamination by cirrus. 

We now turn to supervised learning to optimize the separation between patches with and without tidal injections. LDA is a classic machine learning technique 
that identifies hyperplanes in the $N-d$ space that best separate the predefined classes. Unlike PCA, LDA components need not be orthogonal. Although more advanced techniques exist, we used LDA here for exploratory purposes, noting that future analyses may improve with alternative algorithms. 
The input data now consist of $N=60$ subregions -- with and without tidal tail injections\footnote{This is to mimic the situation where we have already confirmed tidal features in some sky areas despite cirrus contamination, and we want to use their statistical properties to predict unknowns in other sky areas.} -- each characterized by 168 WST coefficients. 
We split the input into training and test sets using five-fold cross-validation.\footnote{The input was divided into five batches, each including 20\% of the data. Because there are few cirrus-free subregions in this sample, we kept both as flagged outliers in the training set. The training was performed over five rounds. In each round, four batches were used for training and the rest for testing. The performance score was averaged over five rounds.} The input data were standard-scaled similarly as with WST-PCA. We combined multiple LDA classifiers across the folds through weighted voting based on accuracy to ensure robustness in the classification.

\begin{figure}[!htbp]
\centering
  \resizebox{0.95\hsize}{!}{\includegraphics{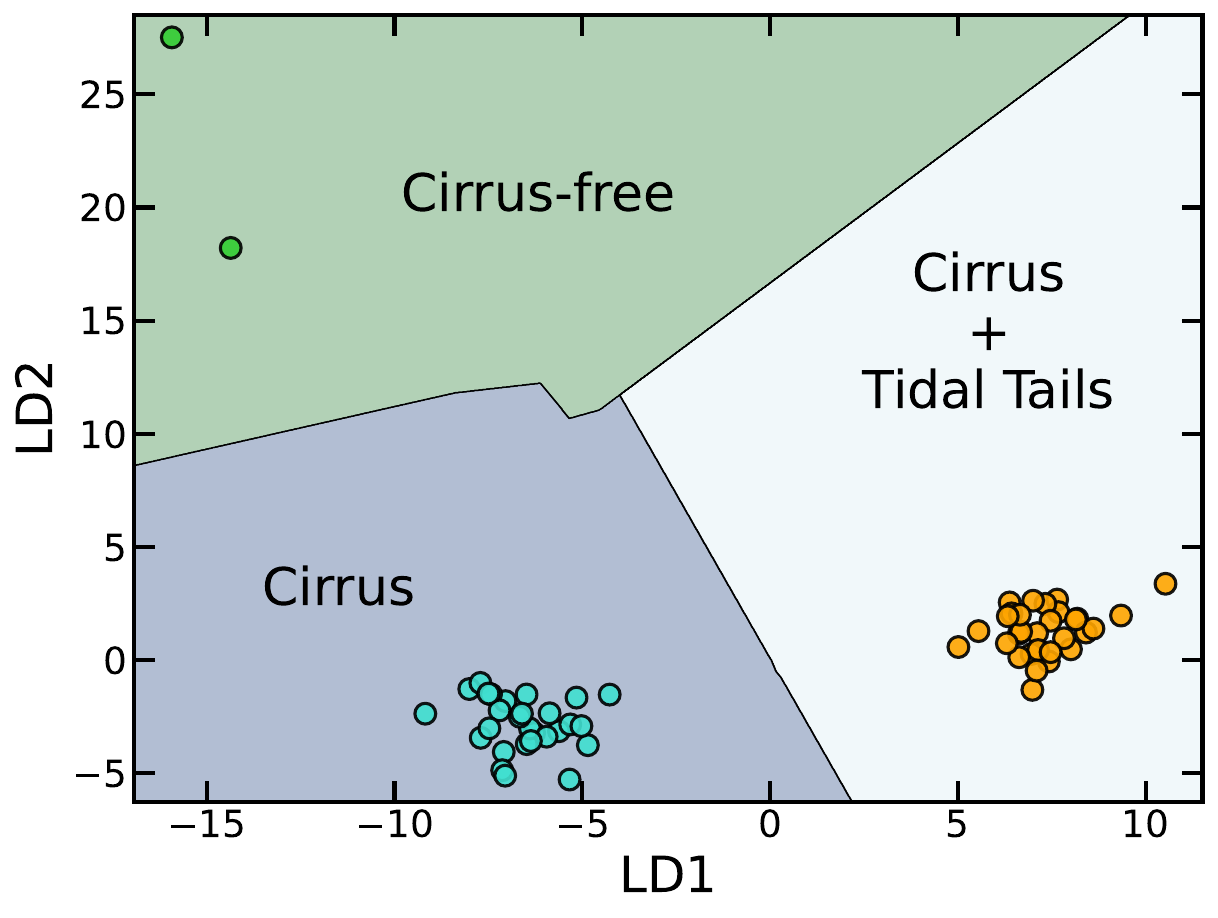}}
    \caption{LDA projection of the WST coefficients of subregions with and without tidal tail injections. Trained with labeled subregions, WST-LDA is able to find a good division in low-dimensional space between the categories. The good separation indicates that it is possible to identify the presence of tidal structures in cirrus-rich areas from their single-band morphology.}
\label{fig:wst_lda_tidal}
\end{figure}

Figure~\ref{fig:wst_lda_tidal} illustrates the WST-LDA results, which clearly distinguish the different groups. The accuracy and precision of the classification from cross-validation are 0.98 and 0.97, respectively. Note that this performance is based on a single injection case. We tested with different realizations of simulated tidal tails and obtained comparable results. Whether the performance is effective for a large number of tidal features remains to be tested, which requires more systematic investigation and is deferred to future work. We also tested the WST-LDA analysis using isotropic coefficients only (with J(J-1)=21 variables), which still yielded reasonably good performance with clear boundaries between categories. However, the ratio of inter- and intra-group variance (i.e., the Calinski-Harabasz index) decreased remarkably. Therefore, retaining anisotropic information in the WST is preferable for the classification. 

These results suggest that although cirrus and tidal features may appear visually quite similar, the statistical properties of their morphologies are subtly different. This distinction likely arises because the two types of structures are shaped by fundamentally different physical mechanisms. An intuitive explanation is that tidal features are more localized (as they are associated with the host galaxy) and become sparser on large scales, whereas cirrus appears more filamentary on large scales, because ISM physics operates over a broader range of scales than tidal stripping. We examined the patterns of WST statistics of pure tidal features and confirmed these trends (motivating this work; see Appendix~\ref{appendix:wst_tidal}). 
Combining morphological measures with color information holds
promise for identifying the presence of tidal features entangled with cirrus in deep wide-field surveys.

\subsection{Separation of extragalactic light and DGL via morphology} \label{sec:discuss_exgal}

\begin{figure*}[!htbp]
\centering
  \resizebox{0.95\hsize}{!}{\includegraphics{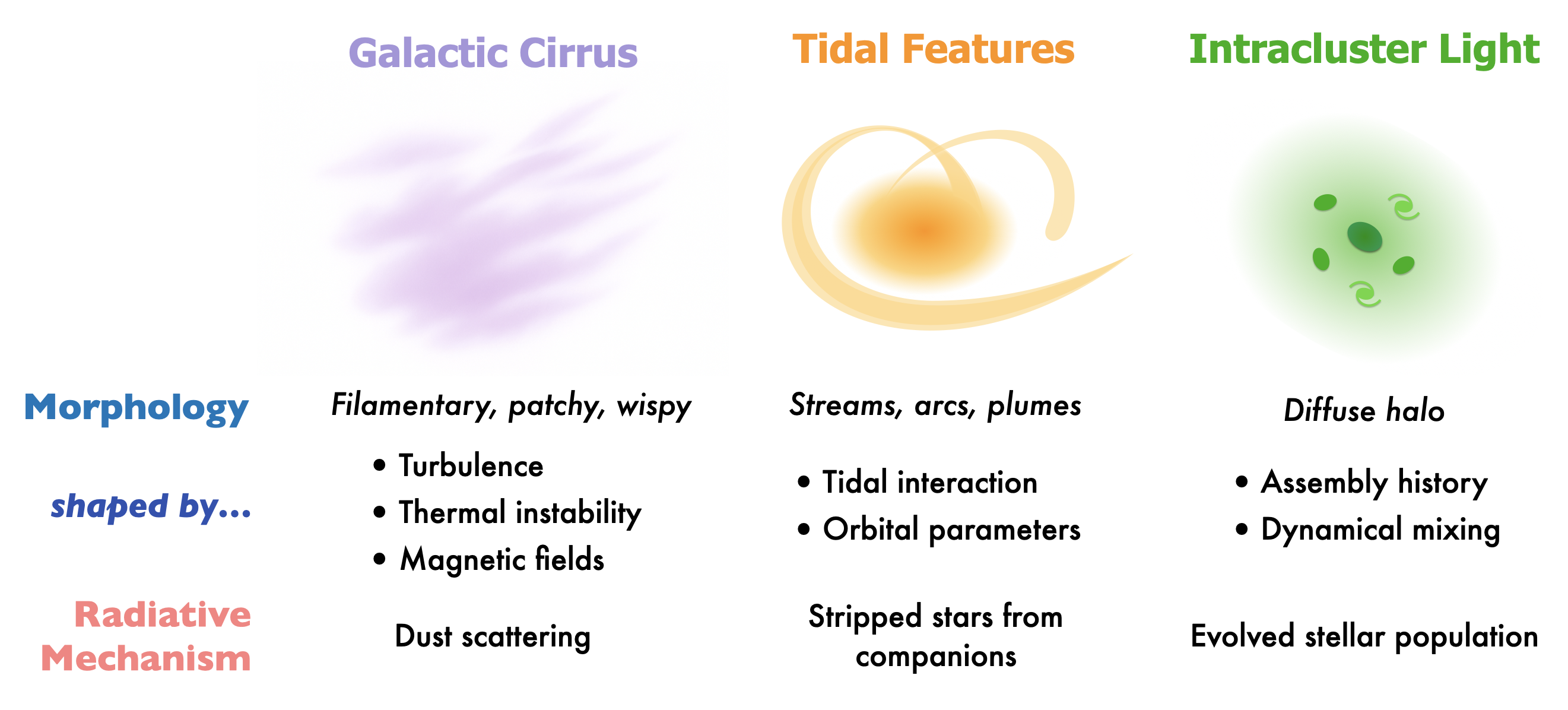}}
  \caption{Schematic summary of the typical morphology, underlying morphological drivers, and radiative mechanisms 
  for optical Galactic cirrus, tidal features, and intracluster light.
  The differences in their morphologies and SED shapes together will provide critical information to break the confusion and separate cirrus and diffuse extragalactic components in the combined diffuse light.}
\label{fig:schematic_summary}
\end{figure*}

Sect.~\ref{sec:discuss_tidal_inj} focuses on the detection challenge of extragalactic light in the presence of  cirrus contamination. Component separation of extended emission for low surface brightness studies is a much more challenging task for advanced deep imaging surveys, but is essential for unbiased characterization of the target signal. 
Addressing this issue is beyond the scope of this paper.  However, we outline possible morphology-based approaches for isolating the Galactic component from the integrated diffuse light.

Through training, a neural network can process subtle morphological cues that differentiate cirrus from extragalactic light, enabling an automated workflow scalable to large sky areas. 
However, one major obstacle is the lack of labeled training data.
\cite{2023MNRAS.519.4735S} developed a deep learning framework for the segmentation of cirrus filaments using annotated SDSS Stripe82 data. Their method can greatly facilitate the cataloging of cirrus, which is very time-consuming using manual annotation. The separation task, however, requires pixel-level ``ground truth" maps indicating the fractional contribution from each component, which is rarely the case for faint diffuse light such as cirrus or tidal tails where they are intertwined. In fact, optical imaging data of cirrus itself has been very limited. Wide-field imaging surveys optimized for low surface brightness covering the high Galactic latitude sky, such as the Dragonfly Ultra Wide Survey (DFUWS) and the Euclid Wide Survey, will provide valuable datasets for training. These can be combined with the synthesis method (see Sect.~\ref{sec:result_wst_syn}) for the purpose of data augmentation to mitigate overfitting.

An alternative approach is statistical modeling. {Statistical modeling enables component separation in a robust, unsupervised manner (e.g., \citealt{2023arXiv230516189S}, \citealt{2025ApJ...993....4L}).} In this paper we present statistical characterizations of optical Galactic cirrus, and we experiment with those signatures to identify the presence of tidal features among cirrus in Sect.~\ref{sec:discuss_tidal_inj}.
Going one step forward, one could use them for component separation by assuming that the statistics of optical cirrus that is entangled with extragalactic light are identical to those without extragalactic light.
A successful methodology can 
seen in the work by\cite{2024A&A...681A...1A}, who used WPH statistics to separate dust emission and CIB in FIR data. Similar algorithms using WPH or WST covariance (Appendix~\ref{appendix:wst_syn}) can be employed on optical images. Note that other diffuse light components, such as scattered light from the wide-angle PSF, should be carefully subtracted in advance. It is also possible to also include COB in the modeling in this framework by using ``flat-sky'' images. The main advantage of such approach is that there is no need for labeled training data. Aside from greater interpretability, the methodology is also more robust than deep learning in terms of domain transfer (from survey to survey) or the risk of overfitting. The drawbacks, however, include the longer convergence time and the requirements in systematics control (i.e., pre-processing). Therefore, such methods are best applied on a case-by-case basis.

The fundamental basis for any separation arises from the difference in the morphologies of Galactic cirrus and extragalactic light, which are driven by distinct physical mechanisms. 
Figure~\ref{fig:schematic_summary} presents a schematic comparison of optical Galactic cirrus and two typical low surface brightness emissions of  extragalactic origin, tidal features and ICL, summarizing their characteristic morphologies and the underlying mechanisms that shape them. We also indicate differences in the radiative sources.
While the investigation in the present work has focused on morphology alone, different radiative processes would lead to distinct SEDs characterized by multi-band photometry. Incorporating this complementary information with morphological signatures will further benefit the separation of cirrus and extragalactic light.

\section{Summary and prospects for application in deep wide-field imaging surveys}
\label{sec:propsects}

In this paper we applied multiple statistical approaches to characterize the morphology of optical Galactic cirrus. Although this morphological information is useful in itself, it is often most interesting to address the following: 
\begin{itemize}
\item What astrophysical insights can cirrus morphology provide?
\item Why does it matter to other low surface brightness science?
\end{itemize}
Here we summarize lessons learned from the characterization of cirrus morphology, and discuss prospects for its practical applications with modern deep wide-field imaging surveys in the context of both ISM studies and extragalactic investigations.

\begin{enumerate}[left=0pt, itemsep=0pt, leftmargin=10pt, labelindent=10pt, listparindent=10pt, labelsep=5pt]
\item \textbf{Local intensity statistics}:
PDF-based approaches have been widely applied to simulations and {\HI} observations to study turbulence in the ISM. Moments and metrics such as skewness, kurtosis, and $\Gini$ can used to probe where the morphology changes. However, few investigations have been extended to the optical regime. With next-generation deep imaging surveys, statistics of dust-scattered light derived from observations may be compared with those in state-of-the-art hydrodynamical ISM simulations at exceedingly high resolution to constrain dust prescriptions. Vice versa, if correlations between observables and ISM physical parameters (such as $\mathcal{M}_S$) exist and can be calibrated, they will provide a new avenue for inferring the latter. 

The greater similarity between optical and FIR cirrus shown in Sect.~\ref{sec:result_local} is expected from their related origins involving larger dust particles, while MIR cirrus exhibits differences due to its distinct origin as non-equilibrium emission for smaller dust particles like PAHs. Therefore, caution should be taken when using MIR as a dust surrogate to remove cirrus from optical images; FIR supplementary data would be preferred where available. Conversely, the PDF distance can be used as a metric to identify regions of strong PAH excitation versus thermal emission from cold dust. 

We also note that if extragalactic light is of interest at scales much larger than the Herschel beam width, such as in the case of ICL, it is possible to do a component separation by minimizing the local PDF distance of the optical cirrus to that of FIR over a range of spatial scales and across the sky. It is important that forefround/background systematics in both maps are removed, which could be demanding. 

\medskip

\item \textbf{Fourier statistics}:
Although power spectrum analysis, along with its variants, is a conventional tool, it has not been fully exploited over a wide sky area at visible wavelengths. The angular power spectrum of the optical cirrus in our example dataset follows a $\gamma\simeq -2.9$ power law, similar to other dust tracers, revealing coherent structures. Investigating the spatial variations of the cirrus power spectrum (and its variants) to study their dependence on the environment and ISM properties (e.g., dust temperature, CNM fraction), will be interesting. 

Second, the fact that the power spectra across dust tracers have consistent power indices indicates that the normalizations of power spectra can be used to calibrate the zero-point of the diffuse light background contributed by cirrus in optical images at the CCD chip level, with no need to dramatically downsample the optical data to, e.g., the Planck beam width. Another useful application is that one can determine whether the cirrus signal is suppressed by sky background subtraction, and if so, at what scales. This is clear in the illustration shown in Appendix~\ref{appendix:legacy_ps}, which presents the result for the same Spider sky area in the Legacy survey imaging. 

Furthermore, as discussed in Sect.~\ref{sec:result_ps}, whether higher-resolution data would reveal a characteristic scale will be interesting. The superb resolution and sensitivity of \textit{Euclid} and \textit{Roman} will render them the ideal facilities for this kind of study in the future.

\medskip

\item \textbf{WST analysis}:  This approach enables the extraction of higher-order, non-Gaussian information that is missing in power spectrum analyses, thereby enhancing our understanding of the complex, multi-scale cirrus structures. As stated in Sect.~\ref{sec:result_wst}, it is not yet well understood why \textit{some} cirrus is highly filamentary and how its morphology correlates with ISM properties. Therefore, it will be interesting to explore the correlation between ISM characteristics and morphological signatures, for example, by extending the analysis in \cite{2023ApJ...947...74L} to the optical regime. 

WST statistics could also be useful when comparing the morphology in different observations and that in simulations, where co-spatial correspondence does not exist. Furthermore, mock cirrus can be generated using wavelet-based covariance, as shown in Sect.~\ref{sec:result_wst_syn}. This can be useful for estimating cirrus contamination of extragalactic signals, as well as algorithmic improvement of detection and measurement pipelines directed at targets of interest. 

In Section \ref{sec:discuss_tidal_inj}, we demonstrated the potential of using WST coefficients to distinguish extragalactic light from cirrus. This is particularly useful when color information is not available. For example, in the Euclid Wide Survey, cirrus in near-infrared (NIR) may not be comparably significant to cirrus in the optical due to a combination of the lower sensitivity in Euclid NIR bands (\citealt{2025A&A...697A...6C}) and the fact that dust-scattered light is intrinsically fainter in the NIR in the absence of large grains (\citealt{2004ApJS..152..211Z}). Where colors are available, they will provide additional information for distinguishing extragalactic light from cirrus. 
\end{enumerate}

In summary, the morphology of Galactic cirrus in optical bands contains a wealth of information beyond measurements of intensity. Just as the morphology of a galaxy tells a tale of its physical properties and evolutionary history, the morphology of cirrus encodes insights about ISM properties and physical mechanisms that shape its structures, thereby rendering its morphology subtly different from extragalactic light. This information has been largely neglected. However, deep wide-field imaging surveys with low surface brightness optimizations should be capable of utilizing this information, advancing both ISM studies and extragalactic science.

\begin{acknowledgements}
Q.L. would like to thank Sihao Cheng, Wei Zhang, Yunning Zhao, Javier Roman, Jean-Charles Cuillandre, Nina Hatch, and Marc-Antoine Miville-Desch{\^e}nes for helpful discussions. Q.L. is supported by the Oort Postdoctoral Fellowship. The research of R.G.A. and P.G.M. is supported by grants from the Natural Sciences and Engineering Research Council of Canada. The Dunlap Institute is funded through an endowment established by the David A. Dunlap family and the University of Toronto.

This research has made use of data from the Planck Legacy Archive that provides online access to all official data products generated by the Planck mission. 
This research has made use of the APASS database, located at the AAVSO website. Funding for APASS has been provided by the Robert Martin Ayers Sciences Fund. The authors thank the excellent technical staff at the New Mexico Skies Observatory where Dragonfly is sited. 

This research has made use of data from the Herschel Space Observatory through the ESA Herschel Science Archive. Herschel is an ESA space observatory with science instruments provided by European-led Principal Investigator consortia and with important participation from NASA.

This research has made use of data products from the Wide-field Infrared Survey Explorer, which is a joint project of the University of California, Los Angeles, and the Jet Propulsion Laboratory/California Institute of Technology, funded by the National Aeronautics and Space Administration.

This research has made use of data from the Legacy Surveys. The Legacy Surveys consist of three individual and complementary projects: the Dark Energy Camera Legacy Survey (DECaLS; Proposal ID \#2014B-0404; PIs: David Schlegel and Arjun Dey), the Beijing-Arizona Sky Survey (BASS; NOAO Prop. ID \#2015A-0801; PIs: Zhou Xu and Xiaohui Fan), and the Mayall z-band Legacy Survey (MzLS; Prop. ID \#2016A-0453; PI: Arjun Dey). DECaLS, BASS and MzLS together include data obtained, respectively, at the Blanco telescope, Cerro Tololo Inter-American Observatory, NSF’s NOIRLab; the Bok telescope, Steward Observatory, University of Arizona; and the Mayall telescope, Kitt Peak National Observatory, NOIRLab. Pipeline processing and analyses of the data were supported by NOIRLab and the Lawrence Berkeley National Laboratory (LBNL). The Legacy Surveys project is honored to be permitted to conduct astronomical research on Iolkam Du’ag (Kitt Peak), a mountain with particular significance to the Tohono O’odham Nation.

NOIRLab is operated by the Association of Universities for Research in Astronomy (AURA) under a cooperative agreement with the National Science Foundation. LBNL is managed by the Regents of the University of California under contract to the U.S. Department of Energy.

This project used data obtained with the Dark Energy Camera (DECam), which was constructed by the Dark Energy Survey (DES) collaboration. Funding for the DES Projects has been provided by the U.S. Department of Energy, the U.S. National Science Foundation, the Ministry of Science and Education of Spain, the Science and Technology Facilities Council of the United Kingdom, the Higher Education Funding Council for England, the National Center for Supercomputing Applications at the University of Illinois at Urbana-Champaign, the Kavli Institute of Cosmological Physics at the University of Chicago, Center for Cosmology and Astro-Particle Physics at the Ohio State University, the Mitchell Institute for Fundamental Physics and Astronomy at Texas A\&M University, Financiadora de Estudos e Projetos, Fundacao Carlos Chagas Filho de Amparo, Financiadora de Estudos e Projetos, Fundacao Carlos Chagas Filho de Amparo a Pesquisa do Estado do Rio de Janeiro, Conselho Nacional de Desenvolvimento Cientifico e Tecnologico and the Ministerio da Ciencia, Tecnologia e Inovacao, the Deutsche Forschungsgemeinschaft and the Collaborating Institutions in the Dark Energy Survey. The Collaborating Institutions are Argonne National Laboratory, the University of California at Santa Cruz, the University of Cambridge, Centro de Investigaciones Energeticas, Medioambientales y Tecnologicas-Madrid, the University of Chicago, University College London, the DES-Brazil Consortium, the University of Edinburgh, the Eidgenossische Technische Hochschule (ETH) Zurich, Fermi National Accelerator Laboratory, the University of Illinois at Urbana-Champaign, the Institut de Ciencies de l’Espai (IEEC/CSIC), the Institut de Fisica d’Altes Energies, Lawrence Berkeley National Laboratory, the Ludwig Maximilians Universitat Munchen and the associated Excellence Cluster Universe, the University of Michigan, NSF’s NOIRLab, the University of Nottingham, the Ohio State University, the University of Pennsylvania, the University of Portsmouth, SLAC National Accelerator Laboratory, Stanford University, the University of Sussex, and Texas A\&M University.

BASS is a key project of the Telescope Access Program (TAP), which has been funded by the National Astronomical Observatories of China, the Chinese Academy of Sciences (the Strategic Priority Research Program “The Emergence of Cosmological Structures” Grant \# XDB09000000), and the Special Fund for Astronomy from the Ministry of Finance. The BASS is also supported by the External Cooperation Program of Chinese Academy of Sciences (Grant \# 114A11KYSB20160057), and Chinese National Natural Science Foundation (Grant \# 12120101003, \# 11433005).

The Legacy Survey team makes use of data products from the Near-Earth Object Wide-field Infrared Survey Explorer (NEOWISE), which is a project of the Jet Propulsion Laboratory/California Institute of Technology. NEOWISE is funded by the National Aeronautics and Space Administration.

The Legacy Surveys imaging of the DESI footprint is supported by the Director, Office of Science, Office of High Energy Physics of the U.S. Department of Energy under Contract No. DE-AC02-05CH1123, by the National Energy Research Scientific Computing Center, a DOE Office of Science User Facility under the same contract; and by the U.S. National Science Foundation, Division of Astronomical Sciences under Contract No. AST-0950945 to NOAO.
\end{acknowledgements}

\bibliographystyle{aa} 
\bibliography{bibtex} 
      
\begin{appendix}

\section{Acquisition, reduction, and post-processing of optical imaging data}
\label{appendix:dragonfly_appendix}

\subsection{The Dragonfly telephoto array}
\label{appendix:dragonfly}

Dragonfly is a distributed aperture telescope composed of 48 Canon 400 mm f /2.8 IS II USM-L telephoto lenses, together equivalent to a 1-m f/0.39 refractor. Each telephoto lens is equipped with a Santa Barbara Imaging Group (SBIG) CCD camera, with a field of view of $2\fdg6 \times 1
\fdg9$ and a pixel scale of $2\farcs84$ pix$^{-1}$. The cameras take exposures with Sloan $g$- and $r$- filters. By design, Dragonfly is optimized for low surface brightness science. Readers are referred to \cite{2014PASP..126...55A} and \cite{2020ApJ...894..119D} for detailed descriptions of the configuration of Dragonfly. 

The general strategy of data acquisition and the data reduction workflow of Dragonfly is described in \cite{2020ApJ...894..119D}, summarized as follows. 
Briefly, Dragonfly takes 10-minute science exposures with the 48 lenses with large dithers and performs initial quality checks. The mean FWHM at the New Mexico Skies Observatory under good conditions is $\approx$ $5\arcsec$ and the native pixel size is $2\farcs85$. The exposures are reduced using the upgraded Dragonfly data reduction pipeline \texttt{DFReduce} (Bowman et al. in prep). The pipeline performs dark-subtraction, flat-correction (using twilight flats), astrometric solutions, photometric calibration, and quality checks on individual frames. Pixel area maps are used to correct the distortion by airmass difference given the large field-of-view. 
For cirrus-rich fields as used in this work, we adopt sky subtraction and combination techniques specifically developed for the preservation of the faint diffuse light from cirrus. This is critical, because improper background modeling would oversubtract cirrus, which leaves significant bias in the image coadd and dramatically alters the radiometric precision of the signal. Details about the principles and procedures are described in \cite{2023ApJ...953....7L}. 
In brief, we adopt Planck thermal dust models \citep{planck2013-p06b} as priors for the background modeling of individual frames, and combine them with control in their background consistency. Demonstrations of improvement in the sky modeling can be found in Figure 2 and Appendix C of \cite{2023ApJ...953....7L}.

\subsection{Component separation of Galactic cirrus}
\label{appendix:diffuse}

We performed source modeling followed by a post-processing of the optical images to decompose the faint diffuse light from cirrus. The source models were built using the \texttt{mrf} package (\citealt{2020PASP..132g4503V}) developed for modeling \textit{compact} sources in low surface brightness imaging. In brief, \texttt{mrf} makes use of high-resolution imaging data and finds a matching kernel between the low-resolution image (here the Dragonfly data) and the high-resolution image. The flux models are built by convolving the high-resolution image with the kernel, which is then subtracted from the low-resolution image to leave out extended light in
the image. We used imaging data from the DESI Legacy imaging survey DR9 (\citealt{2019AJ....157..168D}) as the high-resolution source images. The scattered light in the extended PSF wings was incorporated into the modeling following the method described in \cite{2022ApJ...925..219L}. Although an iterative process was implemented by removing the local cirrus signal, we caution that the presence of cirrus could still affect the determination of the PSF wing on large scales. The core parts of compact sources are masked and refilled using the algorithm in \cite{2024PASP..136c4503V}.

After the compact source removal, we applied a cirrus decomposition technique to the residual $g$ and $r$ images following the procedures described in \cite{2025ApJ...979..175L}. The technique is based on two assumptions: (1) cirrus exhibits mostly filamentary or patchy morphology extending on scales larger than extragalactic sources, and (2) cirrus, which results from dust scattering in the Milky Way, is well constrained in its SED compared to that of extragalactic sources depending on the evolutionary tracks of stellar populations. This process involves a spatial filtering using a Rolling Hough Transform in combination with color constraints of cirrus calibrated using the Planck thermal dust model. This processing aims to remove any faint blob that does not agree with the cirrus SED (i.e., extragalactic sources), as well as residuals from compact source removal. Interested readers can refer to \cite{2025ApJ...979..175L} for further descriptions of the principles, procedures, and demonstrations.

It is noteworthy that the cirrus decomposition in this work adopted a simplified optical depth-dependent color model by using a piecewise linear model as described in \cite{2025ApJ...979..175L}. In general, optically-thick regions of cirrus appear redder than optically-thin regions and this effect is non-linear (\citealt{2023MNRAS.524.2797M}). Although the piecewise linear model accounts for the overall reddening at high optical depths, it is a simplified model and does not account for the variation of optical depth at small scales. Therefore, the presumption of color constraint in the decomposition does not hold strictly in regions with potentially higher optical depth (see discussion in \citealt{2025ApJ...979..175L}), specifically, the central region of the Spider field (subregion \#18, Fig.~\ref{fig:spider_patch}).
This could lead to a potential difference in the contrast of the cirrus structures at the center of the field relative to the rest of the regions, compared with other dust tracers. This is partially mitigated by using the $g+r$ image to average out the effect, compared to using individual bands. Because of the limited area of potentially optically-thick cirrus regions, we defer the investigation tackling the optical depth effect to future analysis with a larger dataset.

\section{Supplementary results of power spectrum analysis}\label{appendix:supp_ps}

\subsection{Results of the HOTT radiance map}
\label{appendix:hott_ps}

Thermal dust emission can also be modeled with modified blackbody parameterization using the high-quality Herschel PACS/SPIRE 160, 250, 350, and 500~$\mu m$ images. This has been derived by the HOTT (Herschel Optimized Tau and Temperature) analysis (\citealt{singh2022}),\footnote{\url{https://www.cita.utoronto.ca/HOTT/}} which provides much higher-resolution (compared to Planck) dust optical depth, temperature, and radiance maps for a suite of fields including the Spider field.
We made use of the HOTT maps 
for supplementary analysis. The HOTT radiance map has a beam width of $36\arcsec$, which produces a larger dynamical range in spatial frequencies across this field compared to Planck.

\begin{figure}[!htbp]
\centering
\resizebox{0.78\hsize}{!}{\includegraphics{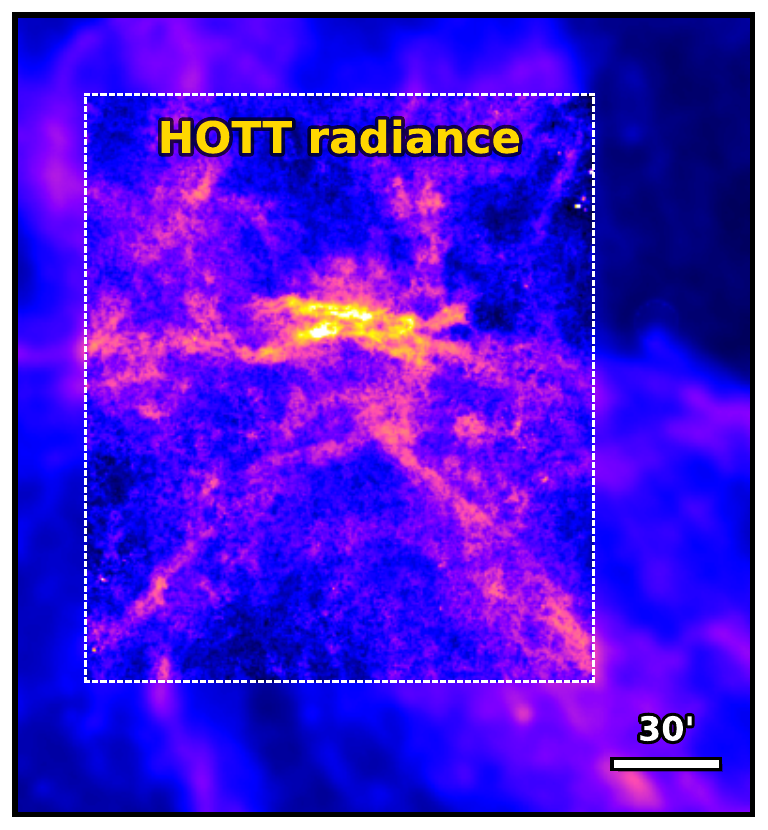}}
  \caption{HOTT radiance map derived from Herschel data. The map is cropped to avoid the noisy regions and cleaned to remove bright sources. The Planck radiance map from thermal dust modeling is shown in the background.}
\label{fig:hott_img}
\end{figure}

The HOTT radiance map also include sources and noisy pixels in faint diffuse regions. Therefore, we performed a cleaning on the map to mitigate the impact from sources and the CIBA by masking sources detected above S/N threshold of 2 with mean fitted temperature below/above the 5\%/95\% quantile and pixels with $\chi^2$ of fitting above 10. The criteria are empirical based on that the dust temperature
is relatively uniform in these fields \citep{singh2022} and emission other than thermal dust emission typically yields bad fitting. The masked map was convolved using a Gaussian kernel with a size of twice the HOTT beam width to fill in the masked pixels. In addition, we cropped out the map to avoid the diffuse noisy regions. This process mitigates the contribution of bright sources at large spatial frequencies, however, a more delicate component separation will be needed to fully remove those contaminations. The cropped and cleaned HOTT radiance map is displayed in Figure~\ref{fig:hott_img}, with the Planck radiance map shown in the background. 

\begin{figure}[!htbp]
\centering
  \resizebox{0.95\hsize}{!}{\includegraphics{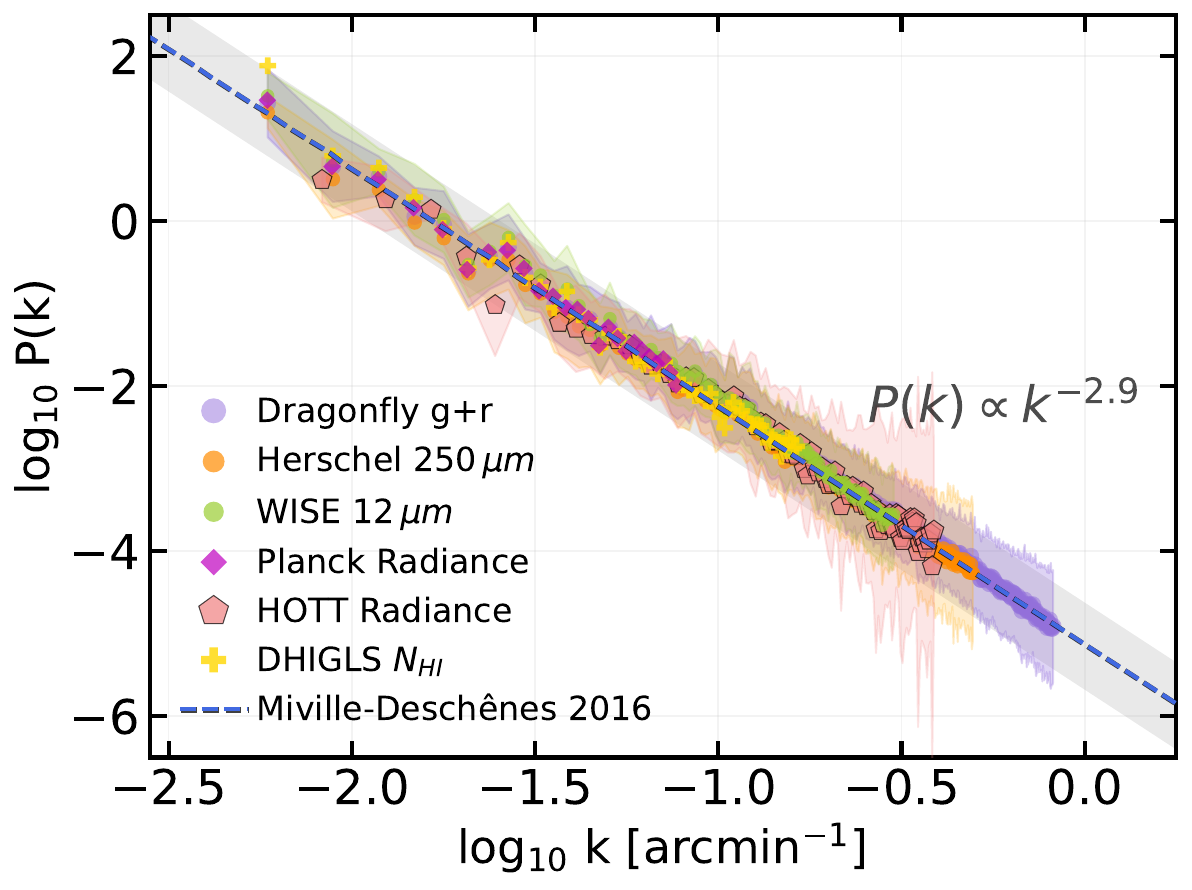}}
  \caption{Combined 1D power spectra of different dust tracer maps as Fig~\ref{fig:ps} added with the result of HOTT radiance map (pink markers). Only the ISM components are shown. The normalization is in arbitrary unit. The power spectrum of the HOTT map follows the same power law as other dust tracers. The larger uncertainties at high frequencies are likely due to the CIBA.}
\label{fig:ps_hott}
\end{figure}

We performed the power spectrum analysis on the HOTT radiance map following Sect.~\ref{sec:result_ps}. 
We adopted Eq.~\ref{eq:ps} for the power spectrum modeling due to the presence of (faint) sources and CIBA, yielding $\gamma=-2.9$ and $\beta=-1.1$. The power spectrum of the HOTT radiance map is shown in Fig.~\ref{fig:ps_hott}, with the contamination noise and instrumental components subtracted. The power index of the cirrus component is similar to that of Planck radiance and other dust tracers, while the contamination component has a similar power index to that of Herschel 250~$\mu m$ FIR image, indicating that the same components exist consistently in Herschel data (Table \ref{table:spec}). The uncertainties of the measured power spectrum are larger likely due to residuals in the source removal and CIBA.  However, the HOTT map extends the power spectrum of the radiance to larger spatial frequencies than could be reached using Planck radiance given the large Planck beam.

\subsection{Power spectrum measured in the DESI Legacy Imaging Survey}
\label{appendix:legacy_ps}

\begin{figure}[!htbp]
\centering
\resizebox{0.75\hsize}{!}{\includegraphics{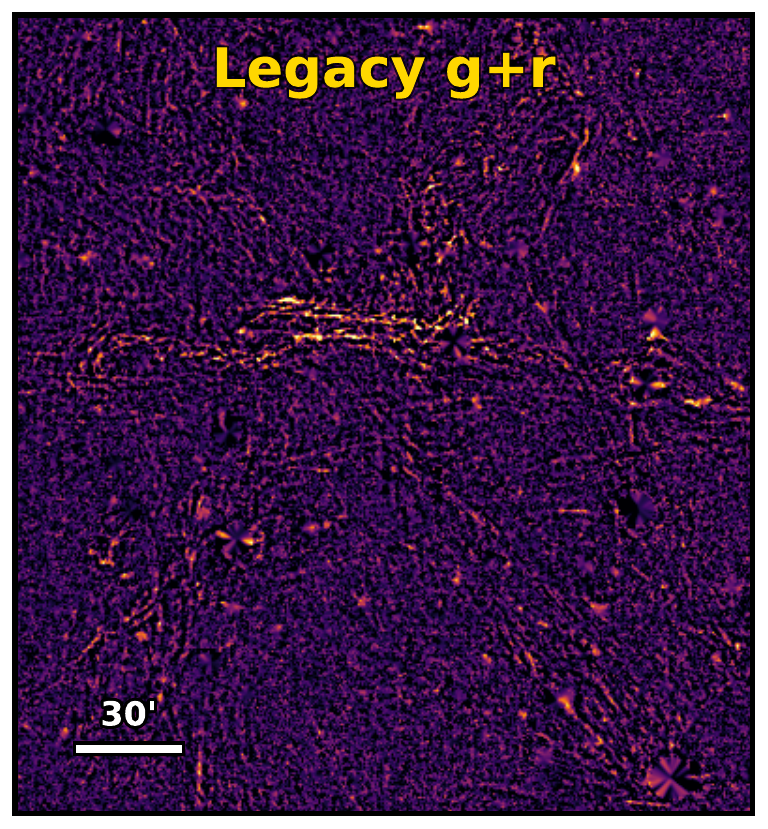}}
  \caption{DESI Legacy g+r image of the field at 6$\arcsec$ resolution with sources removed. The intensity scale is linear and stretched to the 95\% quantile to enhance the cirrus signal. Signals on large scales are largely removed by sky background subtraction.}
\label{fig:legacy_img}
\end{figure}

Here we show how sky background removal can oversubtract the cirrus signal in optical imaging and bias the structural statistics. We use the imaging data from the Data Release 10 of the DESI Legacy Imaging survey (\citealt{2019AJ....157..168D}) obtained with the Dark Energy Camera (DECam). As pointed out by \cite{2024AJ....168...88Z}, it is necessary to carefully mask the foreground and background sources to separate their contribution from cirrus signals. Therefore, we first run a local background subtraction to remove the diffuse light for compact source detection. All detected sources above signal-to-noise ratio of 2 were masked. In addition, saturated stars are masked using an empirical law according to their magnitudes, whose masks are manually enlarged after visual inspection to cover the extended PSF wings. The mask map was then applied to the original image and refilled following the recipe in \cite{2025ApJ...979..175L}. The g and r images are then combined into a g+r image in a similar way as Dragonfly g+r and resampled to 6$\arcsec$ resolution. This reprocessed image is displayed in Figure~\ref{fig:legacy_img}. 

It is clear by inspection that cirrus light in the Legacy survey image is suppressed. As discussed in \cite{2023ApJ...953....7L} (see their Figure 1 for a zoomed-in inspection), this is because the sky subtraction in the Legacy pipeline high-pass filters the image on spatial scales comparable to about half of the CCD chip size in DECam. This effectively eliminates the investigation of cirrus light on larger scales, in particular, its zero-point, and thus may bias the color measurements of cirrus. Such bias could partially explain the bluer mean color in the Legacy survey compared to other observations (\citealt{2024AJ....168...88Z}). It is noteworthy that such bias could not be resolved by binning adjacent pixels.

\begin{figure}[!htbp]
\centering
  \resizebox{0.8\hsize}{!}{\includegraphics{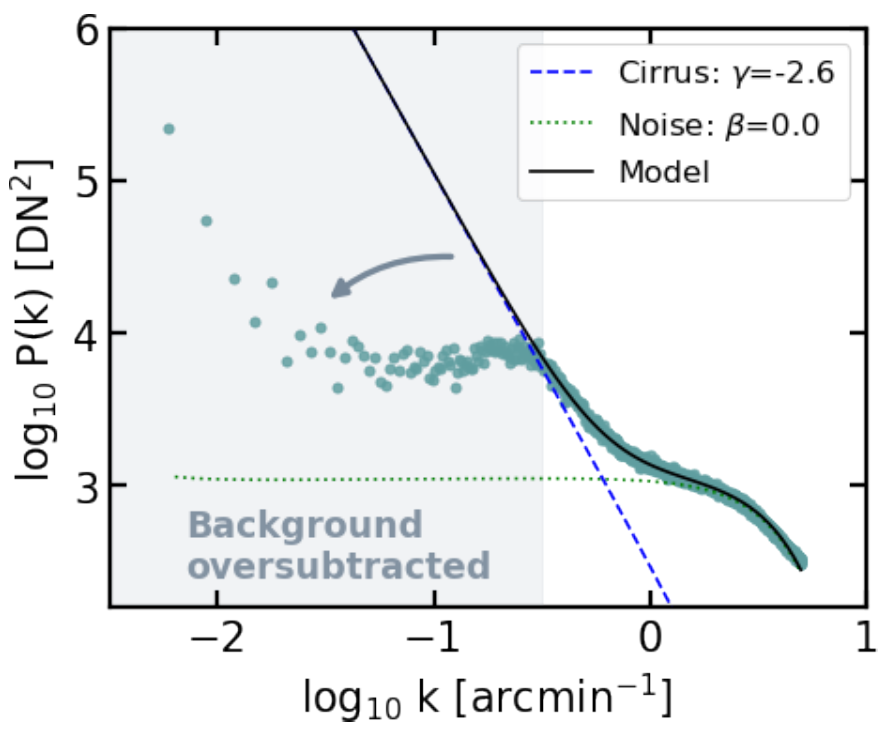}}
  \caption{1D power spectrum of the Legacy g+r image of the Spider field. The blue dashed line, dotted green line, and black line show the fitted cirrus component, the contamination noise component, and the combined model, respectively. The cirrus signal at large scales is significantly suppressed by sky background removal affecting the power spectrum at the low spatial frequencies in the shaded area.} 
\label{fig:legacy_ps}
\end{figure}

In the context of this work, the structures of cirrus on large scales are significantly smeared out or removed. This can be confirmed by the measured power spectrum shown in Figure~\ref{fig:legacy_ps}, where its shape is flattened at low spatial frequencies by the sky background removal. We performed power spectrum modeling following Eq.~\ref{eq:ps} 
for $k$ higher than where it flattens, yielding a power index of -2.6 for the cirrus signal. At even larger scales, the power spectrum rises again for reasons unknown.
Our speculation is that some structural information remains or it could be some weird field edge effect. 

For future analysis on cirrus using deep optical imaging surveys over a wide sky area, caution should be taken on the sky background modeling aiming for the preservation of the cirrus light (see further discussion in \citealt{2023ApJ...953....7L}). On the other hand, power spectrum analysis as a tool can be used to evaluate whether sky background removal has affected the cirrus structures. In this example, it indicates that cirrus light on scales above about $3\arcmin$ in the Legacy imaging data is affected.

\section{Supplementary results of WST analysis}\label{appendix:supp_wst}

\subsection{Comparison of WST statistics with Planck and {\HI} maps}
\label{appendix:wst_Planck+HI}

In this section, we compare WST results of the optical map with Planck and {\HI} maps. The Planck radiance map has a beam width of 
$\sim 5\arcmin$ 
with a pixel size of $1\arcmin$, and the DHIGLS $N_{{\HI}}$ map has a beam width of $\sim 1\arcmin$ with a pixel size of $18\arcsec$.
To match the beam and pixel grid for comparison, we convolved the optical cirrus map to the corresponding beam widths of the two maps, and resampled them to the same grid. We then computed the summary statistics of the beam-matched, downsampled optical maps matching the Planck or {\HI} map.

\begin{figure*}[!htbp]
\centering
  \resizebox{0.8\hsize}{!}{\includegraphics{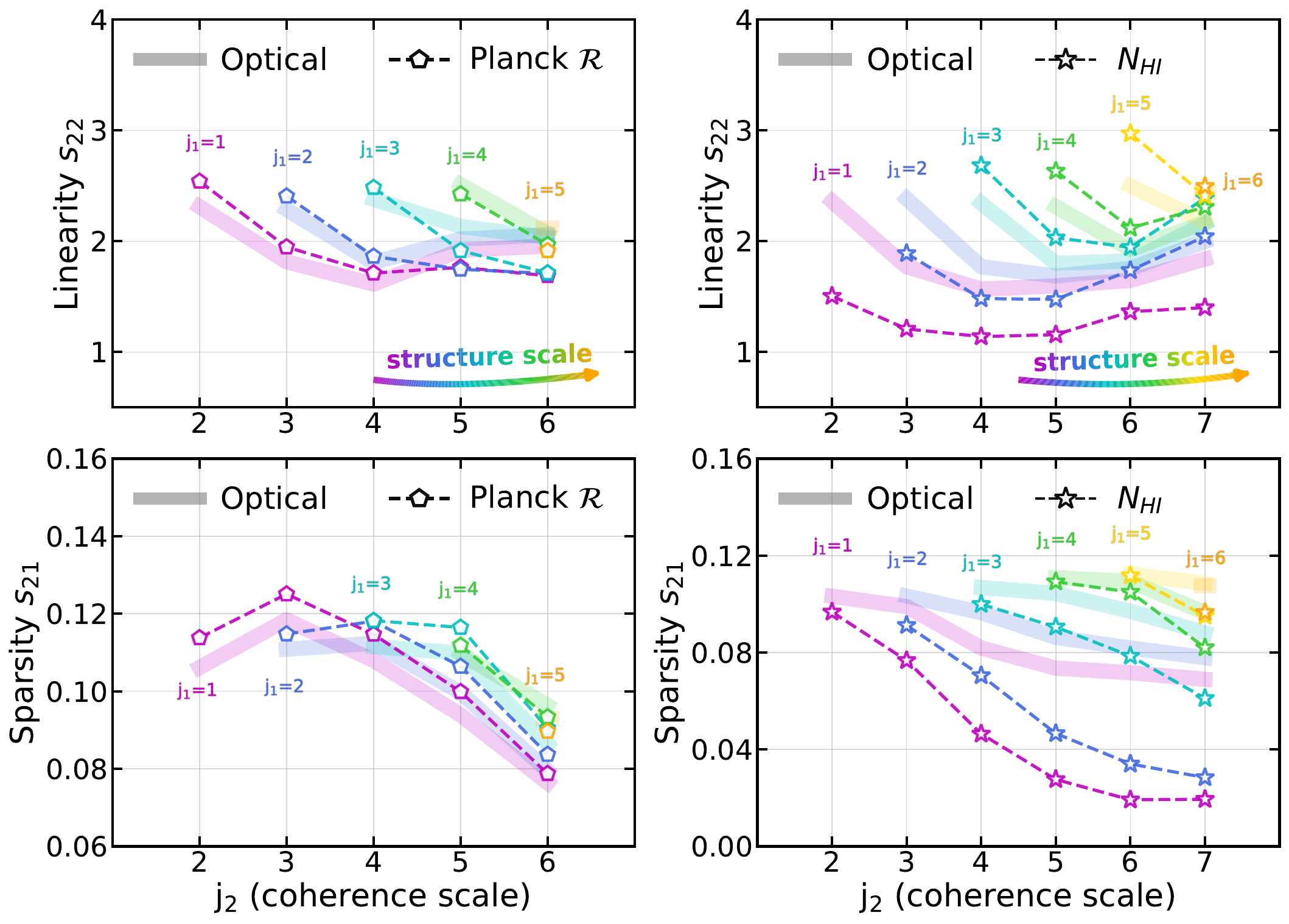}}
  \caption{WST statistics of Planck radiance map (left column) and DHIGLS $N_{{\HI}}$ map (right column) compared to those computed using the optical map with matched beam widths. The x-axis describes the coherence scale $j_2$. The structure scale is color coded by $j_1$. Only $j_2>j_1\geq1$ terms are shown because $j_1=0$ structures are significantly smaller than the beams and $j_2 \leq j_1$ terms are dependent on the wavelets. In general, $s_{22}$ and $s_{21}$ of Planck follow similar trends with those of optical, except for $s_{22}$ at large coherence scales. However, {\HI} shows deviation from the optical, which could be caused by different processing of {\HI} data and possibly indicate the decoupling between gas and dust.}
\label{fig:wst_stats_planck+HI}
\end{figure*}

The left two panels of Figure~\ref{fig:wst_stats_planck+HI} show $s_{22}$ and $s_{21}$ of the Planck radiance map compared to the matched optical map. Because the first structure $j_1=0$ is much smaller than the beam, we only show terms with $j_1\geq1$. However, we note $j_1=1$ terms also mildly suffer from smoothing by the large beam. In general, $s_{22}$ and $s_{21}$ of the Planck map follow trends similar to those of the optical map, suggesting that the dust morphology traced by radiance derived from FIR to sub-mm thermal emission is close to that traced by dust scattering. 
Optical cirrus appears to be more filamentary on larger coherence scales, and more evenly distributed (diffuse) for small-scale structures. The difference might be caused by the variation of the scattering phase function or dust temperature, however, further investigation on a much wider sky area will be needed to characterize the pattern differences and the possible causes.

The right two panels of Figure~\ref{fig:wst_stats_planck+HI} show the two summary statistics for the DHIGLS {\HI} map and the matched optical map. Similarly, we do not plot$j_1=0$ terms due to the large beam width, and note that $j_1=1$ terms are comparable to the beam width. Values for the {\HI} map shows clear deviation from the optical map at several scale combinations. At small scales ($j_1\leq 2$), the difference (especially in sparsity) can be caused by the complex noise properties of {\HI} data (\citealt{2017ApJ...834..126B}), which is partially manifested in the complex noise template in the power spectrum that increases at high $k$. The reasons for the differences at large scales, however, remain in question. One possible explanation is that large dust grains traced by optical data become decoupled from gas flows and tend to be more concentrated than gas. This is qualitatively consistent with results from hydrodynamical simulations in denser regions (\citealt{2016MNRAS.456.4174H}, \citealt{2019MNRAS.483.5623M}). \cite{2024A&A...681A...1A} used a denoising algorithm on the {\HI} velocity data cube prior to constructing the {\HI} image to create a noise-free mock dust map. It would be useful to apply similar approaches in further analysis to investigate whether instrumental effects or the physical decoupling of dust and gas is the major cause of the observed difference in their morphologies as captured in the WST statistics.

\subsection{Division of the field into subregions}
\label{appendix:spider_patch}

\begin{figure}[!htbp]
\centering
  \resizebox{0.78\hsize}{!}{\includegraphics{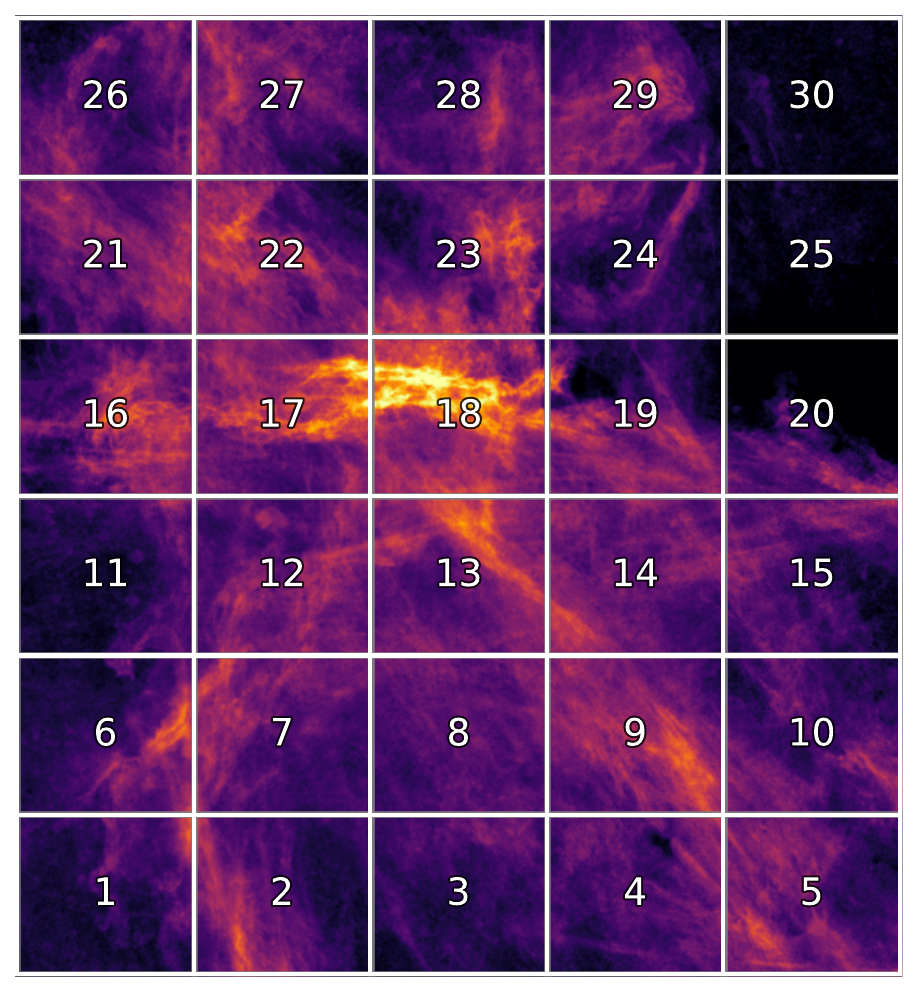}}
  \caption{Subregion division to study the ensemble property of cirrus morphology. Each cutout has a size of $45\arcmin \times 40 \arcmin$. 
  Subregion \#18 might be affected by optical depth effect (see Appendix~\ref{appendix:diffuse}). The two subregions at the upper right corner, \#25 and \#30, are marked as ``cirrus-free'' in the excise described in Section \ref{sec:discuss_tidal_inj}.}
\label{fig:spider_patch}
\end{figure}

Figure~\ref{fig:spider_patch} shows the division of the Spider field into subregions to study the ensemble properties of the WST coefficients. The two subregions, \#25 and \#30, correspond to the two rightmost points in Fig.~\ref{fig:wst_pca_tidal} and the two ``cirrus-free'' subregions in Fig.~\ref{fig:wst_lda_tidal}.

\subsection{Scattering covariance and ensemble synthesis of cirrus}
\label{appendix:wst_syn}

The WST coefficients probe only the scale interactions within a given wavelet in Fourier space. A natural extension of the WST approach is to compute the cross-correlations of the different wavelet-convolved footprints to capture the coupling between wavelets, which is significant when the sequence is dyadic \citep{2021arXiv211201288C}.
However, it is necessary to introduce non-linearities in the operation for the characterization of interactions between different scales due to phase misalignment of the fluctuations (\citealt{2020PhRvD.102j3506A}). 
To compute this correlation, one can take the modulus of the footprint, and use a non-linear Phase Harmonic operator to tune the phase information between footprints. This approach is called the Wavelet Phase Harmonic (WPH) approach. Details of the mathematical framework can be found in \cite{2020PhRvD.102j3506A}. The WPH approach has been applied to ISM and cosmology (e.g., \citealt{2020PhRvD.102j3506A}, \citealt{2021A&A...649L..18R}, \citealt{2024A&A...691A.269M}, \citealt{2024A&A...681A...1A}). The cost of a richer description is a larger set of coefficients by one or two orders of magnitude than the WST coefficients. 


With similar objectives as the WPH approach, the WST covariance of an input field $I$ down to the second order can be computed from the convolved footprints:\footnote{The covariances are calculated for wavelet-convolved fields in the Fourier domain, and therefore contain real parts and imaginary parts. Note the covariance between two complex fields ($X$, $Y$) is defined as: 
$Cov\,(X,Y)=E(X{Y^*})-E(X)E({Y^*})$, where $E$ denotes the mean.
Here we employ the realization in the \texttt{scattering} package \citep{Cheng2023}.}
\begin{eqnarray}
\begin{aligned}
    &C_{01}^{j_1,j_2,\theta_1,\theta_2} = Cov \,(I \ast \psi^{j_2,\theta_2}, |I \ast \psi^{j_1,\theta_1}| \ast \psi^{j_2,\theta_2})\,, \\
    &C_{11}^{j_1,j_2,j_3,\theta_1,\theta_2,\theta_3} = Cov \,(|I \ast \psi^{j_1,\theta_1}| \ast \psi^{j_3,\theta_3}, |I \ast \psi^{j_2,\theta_2}| \ast \psi^{j_3,\theta_3})\,.
\end{aligned}
\end{eqnarray}
The covariance depends on the amplitude of the power spectrum and so is normalized by the autocorrelation of the terms.

\begin{figure*}[!htbp]
\centering
  \resizebox{0.95\hsize}{!}{\includegraphics{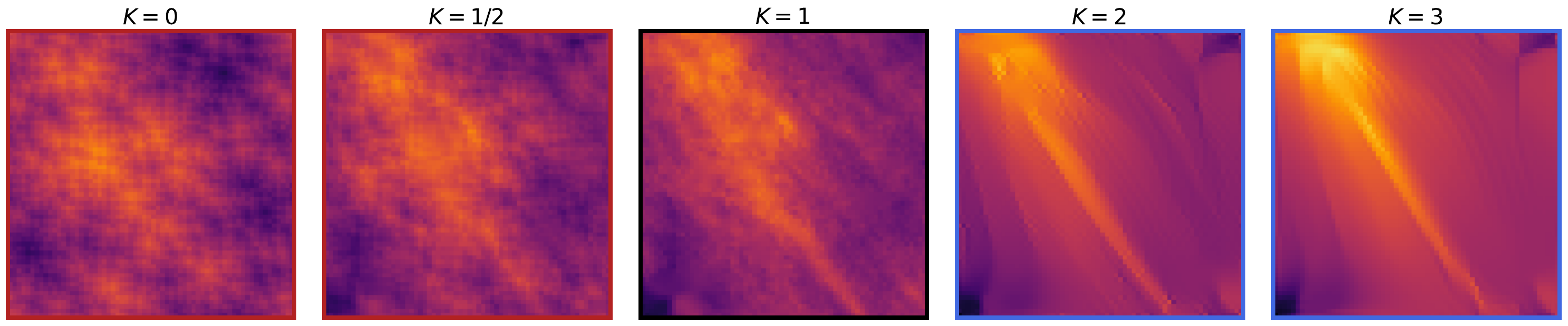}}
  \caption{Illustration of the morphological variation of cirrus structures by artificially increasing or decreasing scattering coefficients. The third panel is a $25\arcmin\times 25\arcmin$ cutout of the optical cirrus map in the example dataset, which is used as the base image. The amplitudes of scattering covariance $C_{01}$ and $C_{11}$ are scaled by a factor $K$ that is reduced in the left two panels and increased in the right two panels. From left to right, the sparsity and linearity of the field increase.}
\label{fig:st_coeff_syn}
\end{figure*}

Figure~\ref{fig:st_coeff_syn} illustrates how variations in WST covariance alter the morphology of structures through image synthesis based on a square cutout of the optical cirrus (shown in the middle) from the example dataset. The cutout is $25\arcmin\times 25\arcmin$, binned [4x4], 
and is run with $J=5$ and $\Theta = 4$.
The amplitudes of both the real and imaginary parts of $C_{01}$ and $C_{11}$ were multiplied by a scaling factor $K$ increasing from left to right. The original image (not synthesized) is displayed in the middle panel. As the covariances increase, the cirrus morphology appears more filamentary and less diffuse. This corresponds to moving from the lower left to the upper right of Fig.~\ref{fig:lin_vs_spa_mosaic}.

\begin{figure*}[!htbp]
\centering
  \resizebox{0.8\hsize}{!}{\includegraphics{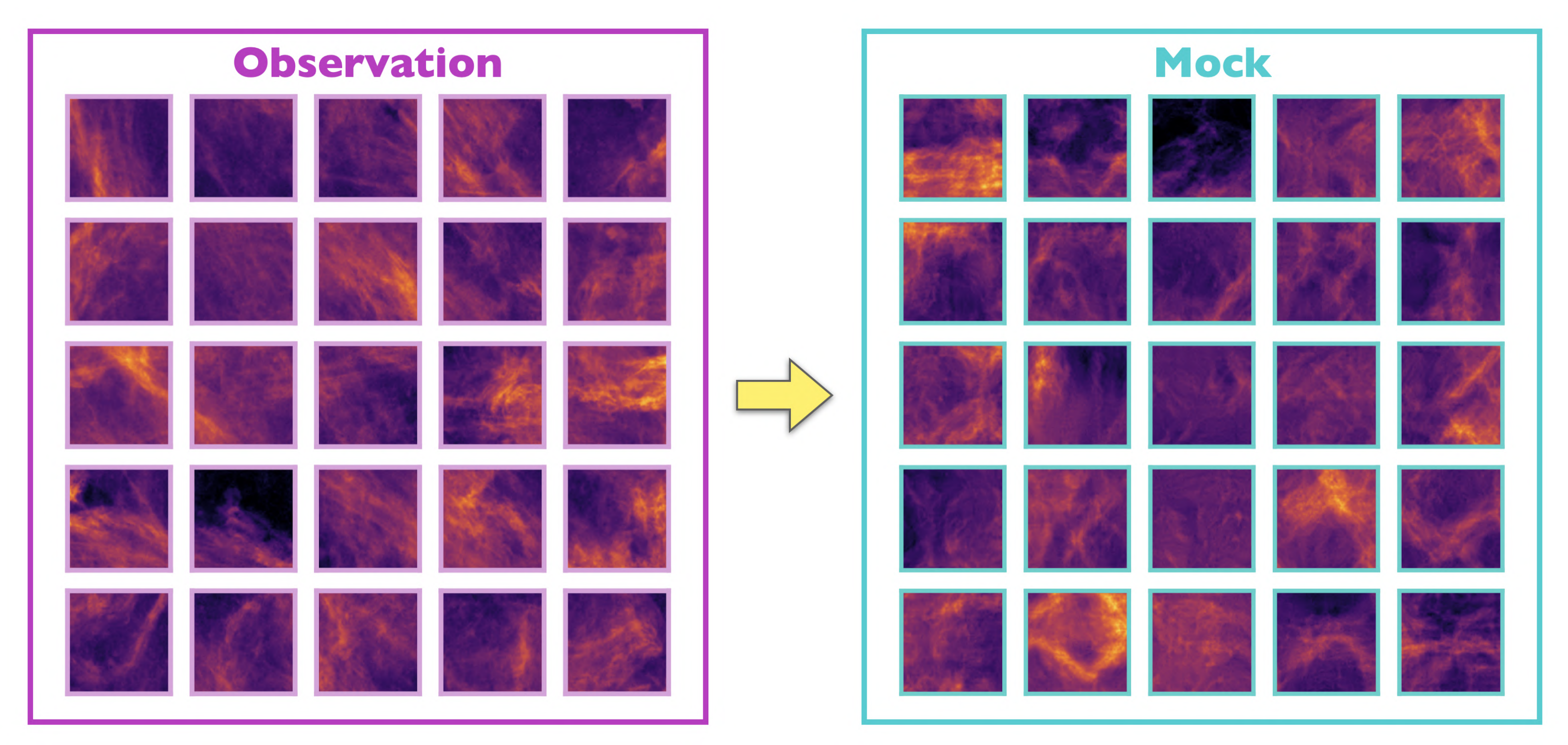}}
  \caption{Illustration of ensemble synthesis using a set of cirrus subregions. 
  The initial fields are random Gaussian fields following the same power spectra as the data in the left panel, which are
  the subregions in Fig.~\ref{fig:spider_patch} excluding faint regions \#1, \#11, \#25, \#30, and the optically thick region \#18.
  The mock cirrus cutouts are generated  by matching the averaged statistics of the input subregions. 
  }
\label{fig:spider_ensemble_syn}
\end{figure*}

The synthesis process aims to generate fields with similar texture to the input field(s) by reproducing the summary statistics (here the scattering covariances). For the synthesis, we employ the utilities of the \texttt{scattering} package (\citealt{Cheng2020}). This is done by using a gradient descent in pixel space to match the statistical estimators of the input field. The initial field is a random Gaussian field with the same power spectrum as the input image. The example of one-to-one synthesis has been displayed in Fig.~\ref{fig:spider_syn}, which was run with $J=6$ and $\Theta = 4$.
on binned cutouts of $40\arcmin\times 40\arcmin$ for computational efficiency.
Such synthesis can be generalized to a multiple-to-multiple synthesis 
that aims to reproduce
the average statistics of an ensemble of input images. 
Figure~\ref{fig:spider_ensemble_syn} illustrates the ensemble synthesis for mock cirrus based on a set of subregions of the Spider data, showing the diversity of cirrus morphology that can be reproduced through the characterization by the WST approach.

\section{Injection of simulated tidal features into cirrus regions}
\label{appendix:tidal}

\subsection{Procedures of injecting tidal tail features}
\label{appendix:inj_tidal}
We inject realistic tidal features into the example cirrus field. The test galaxy with tidal tails was generated from the TNG 50 simulation of the IllustrisTNG project \citep{2019MNRAS.490.3196P}. Details about generating mock images from the simulation are referred to \cite{Miro2025}. 
In \cite{Miro2025}, the mock images are generated as appropriate for the DECam instrument at 70 Mpc. For illustrative purposes, below we describe the injection with one example mock (we have also performed experiments using a suite of mock galaxies and cirrus patches).

To match the optical data obtained by Dragonfly while covering a range of spatial scales of the typical cirrus structure characterized in this work, we convolved the mock DECam image to the Dragonfly PSF and resampled the image which equivalently placed the mock at a closer distance of 15 Mpc.\footnote{Because cirrus structures are \textit{fractal}, the same confusion would arise for galaxies at further distance but embedded in cirrus at smaller angular scales. Those would be harder to identify with the resolution of Dragonfly, but other facilities could benefit from the methodology here at these scales.}  

Because the central galaxy is much brighter than the tidal features and cirrus, it needs to be masked to avoid the power of the galaxy dominating over tidal structures. We masked the central galaxy using a combination of an aperture mask and surface brightness mask where pixels from the central galaxies with $\mu_r < $ 25.3 mag/arcsec$^2$ were masked. Through visual inspection, we further masked background sources and nearby faint satellite galaxies that have not been stripped or merged into the central galaxy. We then replaced the masked pixels using an iterative filling process. In brief, in each iteration the nearest $N$ pixels in the unmasked portion of the image are refilled by convolving the previous image with a Gaussian kernel with increasing sizes until all the masked area is filled. This ensures the continuity of the infilled pixel values and smoothness in the center. The remaining faint tidal features have a mean surface brightness of $\left< \mu_r \right>=26.2$ mag/arcsec$^2$.\footnote{The exact values of the mean surface brightness and masking threshold are not critical because what affects the level of confusion for single-band analysis is the \textit{contrast} between the tidal and cirrus structures. As a survey goes deeper to be able to detect fainter features, more diffuse cirrus structures also emerge.} 
Their surface brightness is converted to the same unit of the cirrus map and then injected into subregions (patches) of the $g+r$ intensity image following the division shown in Fig.~\ref{fig:spider_patch} and placed in the center of the patch.

Steps in the injection process are illustrated in Figure~\ref{fig:tidal_inj}. 
The bottom right panel, which shows the injection after removing the central galaxy, is used as the input for the WST analysis. 

In Sect.~\ref{sec:discuss_tidal_inj} the investigation was performed with this single mock. We have tested with several different mocks and obtained similar results. We noticed that some geometries and projections of features make them easier to distinguish than others. Since this is a proof-of-concept experiment, we defer a more comprehensive analysis to future work.

\subsection{WST statistics of tidal tails}
\label{appendix:wst_tidal}
\begin{figure}[!htbp]
\centering
  \resizebox{0.95\hsize}{!}{\includegraphics{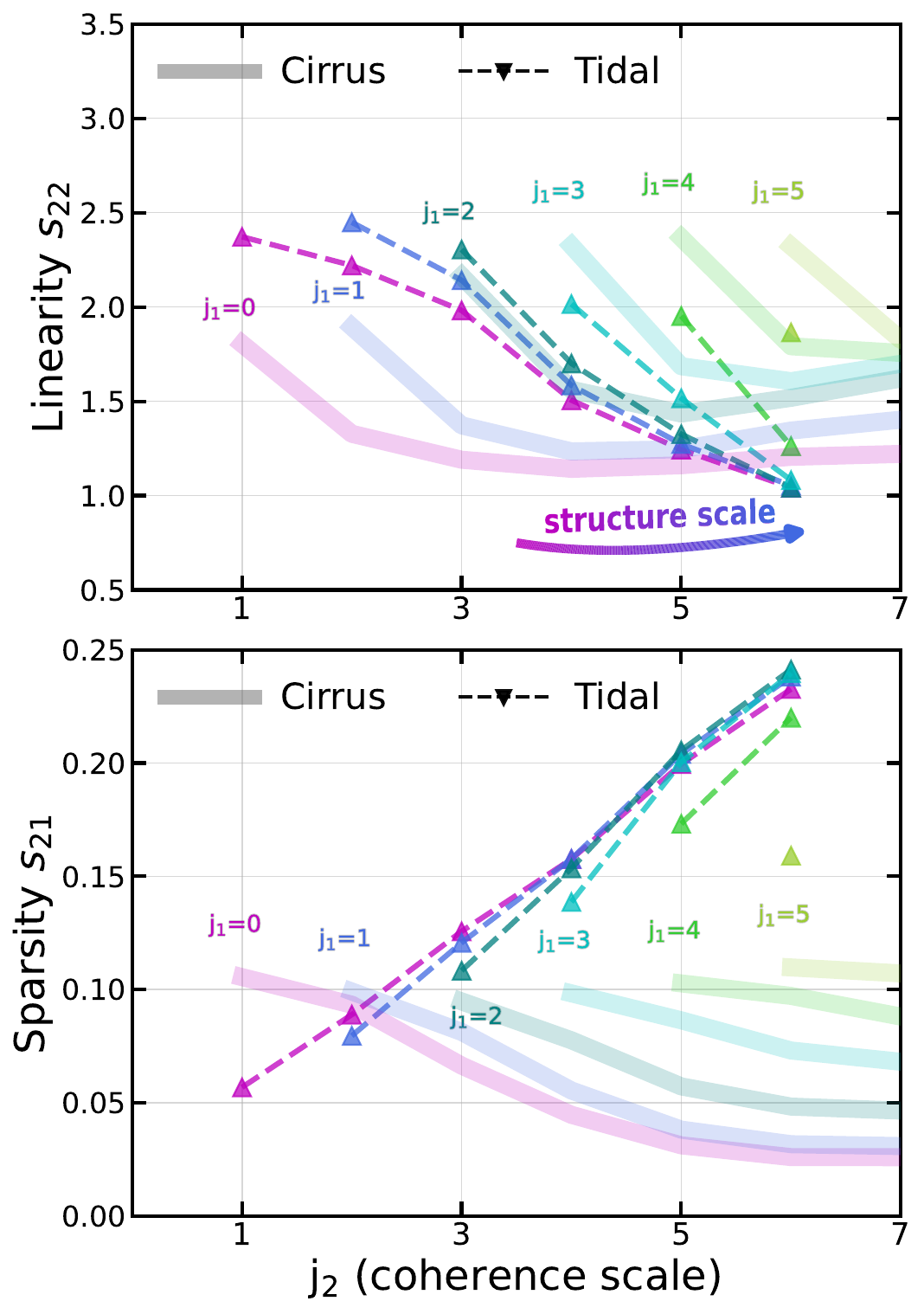}}
  \caption{WST summary statistics of the simulated tidal features (dashed lines) compared to results of optical cirrus shown in Fig.~\ref{fig:wst_stats}. 
  From small to large coherence scales, $s_{22}$ of tidal features consistently decreases, while $s_{21}$ shows opposite trends. Note that the range of the y-axis of $s_{21}$ here is larger than that in Fig.~\ref{fig:wst_stats}. The results here quantify the differences in the morphology of optical Galactic cirrus and tidal features, which lead to their differentiation shown in Figs.~\ref{fig:wst_pca_tidal} and \ref{fig:wst_lda_tidal}.}
\label{fig:wst_stats_tidal}
\end{figure}

Here we show the comparison of the WST summary statistics derived from simulated tidal features with the results from optical Galactic cirrus. Fig.~\ref{fig:wst_stats_tidal} shows the linearity and sparsity of the tidal features in Fig.~\ref{fig:tidal_inj} (bottom middle panel) computed with $J=7$ and $\Theta=8$. The host galaxy has been removed from that image. Both metrics exhibit trends that diverge markedly from those for optical cirrus. Tidal features appear more filamentary at small coherence scales, but their linearity consistently decreases as the scale increases. Conversely, sparsity rises substantially at larger scales. The difference trends can be ascribed to fundamentally different mechanisms governing the morphologies: tidal stripping is a localized gravitational process centering on the host galaxy, whereas turbulence and other physical processes in the ISM act over a wide range of angular scales. 

Because these differences persist in the summary statistics, it can be inferred that they are also manifest in the original set of WST coefficients, which leads to the differentiation of the two phenomena in the WST-PCA and WST-LDA analyses in Sect.~\ref{sec:discuss_tidal_inj}. Interestingly, such differences in morphology can be discerned by experienced observers. The quantitative descriptions introduced here now provide a rigorous basis for that perceptual knowledge.

\end{appendix}

\end{document}